\documentclass[preprint,12pt]{elsarticle}

\usepackage{graphicx}
\usepackage{amssymb}
\usepackage{amsmath}
\usepackage{float}
\usepackage{subfigure}
\usepackage{lineno,hyperref}
\usepackage{amsmath,amsfonts,amsthm,amssymb,graphicx,subfigure,mathtools,tikz,hyperref,bm,blindtext}
\usepackage{diagbox}
\usepackage{geometry}
\usepackage[acronym]{glossaries}
\usepackage{multirow}
\usepackage{algorithm}

\graphicspath{{Figs/}}
\usepackage{algpseudocode}
 \usepackage{bibentry}
 
\journal{Journal of Computational Physics}

\begin{document}

\begin{frontmatter}


\title{Central-moment discrete unified gas-kinetic scheme for incompressible two-phase flows with large density ratio}



\author[nkd]{Chunhua Zhang \fnref{1}}
\author[nkd,nkd2]{Lian-Ping Wang\fnref{2}}
\author[hzd]{Hong Liang}
\author[hust]{Zhaoli Guo}

\fntext[1]{Corresponding author: zhangch6@sustech.edu.cn.}
\fntext[2]{Corresponding author: wanglp@sustech.edu.cn.}
\address[nkd]{Guangdong Provincial Key Laboratory of Turbulence Research and Applications, Department of Mechanics and Aerospace Engineering, Southern University of Science
 and Technology, Shenzhen 518055, Guangdong, China}
\address[nkd2]{ Center for Complex Flows and Soft Matter Research, Southern University of Science  and Technology, Shenzhen 518055, Guangdong, China}
 \address[hzd]{Department of Physics, Hangzhou Dianzi University, Hangzhou 310018, China}
 \address[hust]{State Key Laboratory of Coal Combustion, Huazhong University of Science and Technology, Wuhan 430074, China}


\begin{abstract}
In this paper, we proposed a central moment discrete unified gas-kinetic scheme (DUGKS)  for multiphase flows with large density ratio and high Reynolds number. Two sets of kinetic equations with central-moment-based  multiple relaxation time collision operator
are employed to approximate the incompressible Navier-Stokes equations and a conservative phase field equation for interface-capturing.
In the framework of DUGKS,  the first moment of the distribution function for the hydrodynamic equations is defined as velocity instead of momentum. Meanwhile, the zeroth moments of the distribution function and external force are also suitably defined such that  a artificial pressure evolution equation can be recovered. Moreover, the Strang splitting technique for time integration is employed to  avoid the calculation of spatial derivatives in the force term at cell faces. For the interface-capturing equation, two equivalent DUGKS methods that deal with the diffusion term differently using a source term as well as a modified equilibrium distribution function are presented.
Several benchmark tests that cover a wide a range of density ratios (up to 1000) and Reynolds numbers (up to $10^5$) are subsequently carried out to demonstrate the capabilities of the  proposed scheme. Numerical results are in good agreement with the reference and experimental data.
\end{abstract}


\end{frontmatter}



\section{Introduction}

Numerical simulation of multiphase flows is of interest in a variety of natural  and engineering processes.
Many numerical methods, such as the front tracking method~\cite{UNVERDI199225}, volume of fluid~\cite{hirt1981volume}, level set method~\cite{chang1996level},  diffuse interface method~\cite{jacqmin1999calculation,anderson1998diffuse} and kinetic-theory-based method~\cite{kruger2017lattice}, have been proposed. In particular, the kinetic-theory-based method, such as lattice Boltzmann equation (LBE) method ~\cite{he1997theory,chen1998lattice},
gas-kinetic scheme~\cite{xu2014direct,liu2016unified}, multiphase lattice Boltzmann flux solver~\cite{shu2014development,wang2015multiphase,yang2020lattice}, discrete unified gas kinetic scheme method~\cite{guo2013discrete,guo2015discrete,guo2021progress}, have been developed into an alternative and promising numerical scheme to simulate complex multiphase flows.
The most popular kinetic method for multiphase flow is the LBE method due to its locality of the streaming and collision algorithm and ease of implementation. Based on the used strategy of dealing with the intermolecular interactions for non-ideal gas effects, various multiphase LBE methods have been  developed. To date, a variety of successful applications for multiphase flows have been achieved. Recent and comprehensive review of LBE models for multiphase flows can be found in the literature~\cite{petersen2021lattice,li2016lattice}.

However,  most of the previously proposed LBE methods are restricted to two-phase fluids with small or moderate density differences (less than $100$) due to the numerical instability~\cite{swift1996lattice,he1999lattice,zu2013phase,liang2014phase}.
Although some improved methods have been developed, numerical modeling of multiphase flows with complex interface morphology at large density contrast still remains a challenging subject in multiphase community.
In the framework of LBE,  the adopted strategies to improve the numerical stability could be roughly divided into four categories.
The first strategy is to improve the free-divergence condition. For example,
Inamuro et al.~\cite{inamuro2004lattice} employed one particle velocity distribution function for the calculation of a predicted velocity of the two-phase fluid without the pressure gradient. Then, the current velocity is determined by iteratively solving the Poisson equation at each time step.
The second strategy is to adopt a stable discretization scheme. For example, Lee et al~\cite{lee2005stable}. proposed a mixing difference scheme for the surface tension force in the pre-streaming collision step while using the standard central difference scheme for the surface tension force in the post-streaming, which is able to improve the numerical stability. However, it was proved that such treatment violates the conservation of mass~\cite{lou2011some}.
Similarly, Gong and Cheng~\cite{gong2012numerical} proposed a mixed interparticle interaction force term in the single-component multiphase pseudo-potential model to improve the numerical stability.
The third strategy is to couple Navier-Stokes (NS) equations with an improved interface-tracking equation, such as the level set equation, the volume of fluid equation and the corrected phase field equation. In particular,
the conservative Allen-Cahn equation (ACE) that is a second order partial differential equation has attracted much attention~\cite{chiu2011conservative}.
By employing the ACE for interface-capturing, several LBE models for multiphase flow with large density ratios have been developed~\cite{fakhari2017improved,liang2018phase,ren2016improved}. The main difference in these models is the definitions of the equilibrium distribution function and source term for the hydrodynamic equations.
Moreover, some modified Cahn-hilliard equations with interface correction terms are also presented to track the interface with improved accuracy such that LBE is able to simulate two phase flows with large density ratios~\cite{zhang2019fractional,zhang2017flux}.
The fourth strategy is to employ improved collision operators, such as  multiple entropic collision model~\cite{dorschner2016entropic,mazloomi2015entropic}, central moment-based collision model~\cite{geier2006cascaded} and cumulant collision model~\cite{geier2015cumulant} instead of the classical single-relaxation time and multiple-relaxation -time models. For example,
Gruszczynski et al.~\cite{gruszczynski2020cascaded} and Hajabdollahi et al.~\cite{hajabdollahi2021central} separately presented central-moment-based lattice Boltzmann model for the incompressible two-phase flows. Hosseini and Karlin et al.~\cite{hosseini2022entropic} proposed a multiple relaxation time entropic LBE model.  Sitompul and Aoki et al.~\cite{sitompul2019filtered} presented a filtered cumulant LBE method to simulate violent two-phase flow problems.
It's worth pointing out that the above strategies are not independent of each other. In most models, several strategies are usually adopted together.

On the other hand,  most of existing LBE methods  suffer from substantial drawbacks, such as the uniform grid and fixed CFL condition.
Recently, the discrete unified gas kinetic scheme  for multiscale flows based on kinetic model has been applied to simulate incompressible two phase flows~\cite{guo2013discrete,guo2021progress,zhang2018discrete}. Due to its finite-volume formulation and decoupling of the time step and mesh size, DUGKS becomes a competitive tool in comparison with LBE ~\cite{guo2021progress}.  Zhang et al.~\cite{zhang2018discrete} first extended the DUGKS to two-phase flows based on a quasi-incompressible NS equations associated with the Cahn-Hilliard equation for interface capturing.  Yang et al.~\cite{yang2019phase} further extended the DUGKS to two fluids with large density ratios  by solving the incompressible NS equations coupled with the conservative ACE.  However, multiphase flow problems with complex interface deformation are missing in their work. Recently, Zhang~\cite{zhang2022discrete} proved that
a linear reconstruction in DUGKS for the distribution function at cell face fails to recover the kinetic equation with second-order accuracy when the source term and the collision operator without the constraint of momentum conservation are involved, e.g., DUGKS is used to solve ACE. They instead proposed a piecewise-parabolic reconstruction for the distribution function at cell face. In particular, the weighted essentially non-oscillatory scheme for the reconstruction of the distribution function at the cell face can significantly improve the accuracy of interface capturing when ACE is employed to capture the interface.

In this paper, we proposed a robust and accurate DUGKS for two-phase flows with large density ratios and high Reynolds number. The features of the present model are summarized as follows. First, the central moment-based collision operator is utilized to improve the Galilean invariance, in which the central moment is derived from the continuous Maxwell equilibrium distribution function instead of a discrete one. Second, for the hydrodynamic equations, a velocity-based particle distribution function with a constant density is adopted  such that a differentiation-by-part error proportional to the density ratio in the recovered equations is removed~\cite{kim2015lattice}. Meanwhile, the zeroth order moments of the distribution function and external force are suitably modified to match an artificial compressibility model for the pressure evolution equation. Third,  the Strang splitting method is used to  reduce computational cost and avoid  calculations of various spatial derivatives in the source term at cell interfaces, which further contributes to improve numerical stability. Finally, for the conservative phase field equation,  two equivalent DUGKS methods for the conservative ACE are presented. A key step is that the moments of the equilibrium distribution are obtained from the continuous Maxwell equilibrium distribution function. Through the multiscale expansion analysis, we proved  that the correct macroscopic equations can be recovered from two sets of kinetic equations. 

The rest of the paper is organized as follows. The governing equations for incompressible and immiscible multiphase fluids are introduced in Sec.~\ref{sec:1}.
The kinetic models  with central-moment-based collision operator for the hydrodynamic equations and phase field equation are introduced in  Sec.~\ref{sec:3} and Sec.~\ref{sec:6}, respectively. Numerical results of several classical multiphase flow problems are presented to validate the capabilities of the proposed DUGKS in Sec.~\ref{sec:7}. A brief summary is given in Sec.~\ref{sec:8}.

\section{Governing equations}\label{sec:1}
The Navier-Stokes equations  for an immiscible  incompressible two phase system under isothermal condition can be written as
\begin{align}\label{eq:NS1}
\nabla\cdot \bm u &=0, \\
\label{eq:NS2}
\rho\left(\frac{\partial \bm u}{\partial t}+\nabla\cdot(\bm u\otimes \bm u) \right)&=-\nabla p+\nabla \cdot\mu(\nabla \bm u+\nabla^T\bm u)+ \bm F_b +\bm F_{sf},
\end{align}
where $\rho$ is the mixture density, $\bm u$ is the average velocity, $p$ is the pressure, $\mu$ is the dynamic viscosity, $\bm F_b$ is the body force, $\bm F_{sf}$ is the surface tension force.
For interface-capturing, the following conservative Allen-Cahn equation~\cite{chiu2011conservative} is employed,
\begin{equation}\label{eq:CACE}
\partial_t\phi+\nabla\cdot (\phi \bm u)=\nabla\cdot
M\left(\nabla\phi-\Theta \bm n\right), \qquad \Theta=-\frac{4(\phi-\phi_h)(\phi-\phi_l)}{W(\phi_h-\phi_l)}
\end{equation}
where $M$ is the mobility, $W$ is the width of interface, and $\bm n=\nabla\phi/|\nabla\phi|$ is the unit outward-normal vector of the interface.

In the phase-field-based multiphase flow model, the NS equations and the interface-capturing equation are coupled by the density, the dynamic viscosity and the surface tension force.
The density is usually defined as
\begin{align}\label{eq}
\rho &=\rho_l+ \frac{\phi-\phi_l}{\phi_h-\phi_l}(\rho_h-\rho_l),
\end{align}
where $\phi$ is a maker function (usually called the order parameter) distinguishing two fluids, i.e, $\phi=\phi_h$  denotes the heavy fluid and $\phi=\phi_l$ denotes the light fluid. The subscripts $h$ and $l$ indicate the physical properties of the heavy and light fluids, respectively.
The dynamic viscosity can be obtained by $\mu=\rho\nu$, where $\nu$ is the kinetic viscosity, defined as
\begin{equation}\label{eq}
 \nu=\nu_l+\frac{\phi-\phi_l}{\phi_h-\phi_l}(\nu_h-\nu_l).
\end{equation}
As for the surface tension force, many formulations of the surface tension force are available in the literature. Comparisons among different surface tension formulas have been made~\cite{kim2005continuous,zhang2021formulations}. Here,  the density-scaled continuum surface  tension force is used~\cite{brackbill1992continuum}
\begin{equation}\label{eq:sf}
\bm F_{sf}=-\sigma (\nabla\cdot \bm n)\nabla\phi \frac{2\rho}{(\rho_h+\rho_l)(\phi_h-\phi_l)},
\end{equation}
where $\sigma$ is the surface tension coefficient and $\nabla\cdot \bm n$ denotes the interface curvature. It is noted that
$\sigma$ may vary along the interface and its gradient tangent to the interface. For simplicity, $\sigma$ is treated as constant.
It can be seen that the acceleration produced by the surface tension force  $\bm F_{sf}/\rho$ is independent of density. Such symmetric acceleration can present a good numerical stability~\cite{sitompul2019filtered}.  The curvature of the interface  can be  obtained by
\begin{equation}\label{eq}
\begin{aligned}
\nabla\cdot \bm n=\nabla\cdot\left(\frac{\nabla\phi}{|\nabla \phi|}\right)=\frac{\phi_x^2\phi_{yy} -2\phi_x \phi_y \phi_{xy}+\phi_y^2 \phi_{xx}}{ (\phi_x^2+\phi_y^2)^{3/2}}.
\end{aligned}
\end{equation}

To update the pressure, the pressure Poisson equation often needs to be solved iteratively that is the most time-consuming process in simulation. Another approach is to adopt a so-called artificial compressibility method where a pressure evolution equation is postulated. Following the latter strategy, we employ the following pressure equation instead of Eq.~(\ref{eq:NS1})~\cite{toutant2017general}
\begin{equation}\label{eq:ACM}
\partial_t p+ c_0\rho \nabla\cdot\bm u=0,
\end{equation}
where $c_0$ is a artificial compressibility coefficient and its value is usually chosen from the range $0.1\leq c_0\leq 10$.

\section{DUGKS for incompressible  Navier-Stokes equations}\label{sec:2}
\subsection{Continuous kinetic model for the  Navier-Stokes equations}\label{kinetic-NS}
The  kinetic equations for solving the hydrodynamic equations can be expressed as~\cite{he1998discrete}
\begin{equation}\label{eq:NS_kinetic}
\frac{\partial f}{\partial t}+\bm \xi\cdot \nabla f+ \frac{\bm F_{ext}}{\rho_0}\cdot \nabla_{\bm \xi} f=\Omega(f,f^{eq}) ,
\end{equation}
where $f=f(\bm x,\bm \xi, t)$ is the particle distribution functions with its velocity $\bm \xi$ at position $\bm x$ and time $t$, $f^{eq}$ is the corresponding particle equilibrium distribution function,  $\bm F_{ext}$ is the external force, including the surface tension force $\bm F_{sf}$, the gravitational force $\bm F_g$ and other correction terms to be determined,
 $\Omega(f,f^{eq})$ represents the collision operator satisfying the conservation of mass and momentum.
For Maxwellian molecules, the  equilibrium distribution function $f^{eq}$ is defined as
\begin{equation}\label{eq:maxwellian-NS}
f^{eq}(\bm x,\bm \xi,t)=\frac{\rho_0}{2\pi c_{s,f}^2}\exp\left[-\frac{(\bm \xi-\bm u)^2}{2c_{s,f}^2} \right],
\end{equation}
where $\rho_0$ is a reference density, $c_{s,f}=\sqrt{RT_f}$, where $R$ and $T_f$ are the specific gas constant and the temperature. $c_{s,f}$ is
also known as the sound speed at temperature $T_f$. It should be stressed that the temperature $T_f$ has no physical significance here since we focus on  an isothermal system.
The fluid density $\rho_0$ and velocity $\bm u$ are determined from the zeroth and first moments of the distribution function
\begin{equation}\label{eq:macro_from_dugks_NS}
\begin{aligned}
\rho_0=\int fd\bm \xi, \quad  \rho_0 \bm u=\int\bm \xi fd\bm \xi.
\end{aligned}
\end{equation}
As $f^{eq}$ is the leading part of the distribution function $f$, the forcing term can be approximated by~\cite{guo2002discrete}
\begin{equation}\label{eq}
\mathit{F}(\bm\xi)=-\frac{\bm F_{ext}}{\rho_0}\cdot \nabla_{\bm \xi} f\approx \frac{\bm F_{ext}}{\rho_0}\cdot \frac{(\bm\xi-\bm u)}{c_{s,f}^2}f^{eq}.
\end{equation}

In order to solve the kinetic equation numerically,  discretization of the  velocity space of the distribution function is required in addition to temporal and spatial discretizations. This means that the continuous integrals must be changed into discrete sums evaluated at certain points in velocity space, namely,
\begin{equation}\label{eq}
\begin{aligned}
\int \bm \xi^{k} f^{eq}d\bm \xi &=\sum_i \omega_i \bm \xi_i^{k} f^{eq}(\bm \xi_i),\\
\int \bm \xi^{k} fd\bm \xi &=\sum_i \omega_i \bm \xi_i^{k} f(\bm \xi_i),  \\
\int \bm \xi^{k} \mathit{F}(\bm \xi)& =\sum_i \omega_i \bm \xi_i^{k} \mathit{F}(\bm \xi_i)
\end{aligned}
\end{equation}
for $0\leq k\leq 1$, where $f_i$ and $f_i^{eq}$ are the corresponding discrete velocity distribution function, $\omega_i$ is  the weight
of the numerical quadrature at the discrete velocity $\bm \xi_i$.
In particular, the moment integrals of the distribution function can be effectively evaluated by a Gauss-Hermite quadrature, where the abscissae of the quadrature comprises a set of discrete velocities~\cite{shan2006kinetic}. For nearly compressible flow or incompressible flow, the three-point Gauss-Hermite quadrature is used to obtain the one-dimensional discrete particle velocities and associated weights, taking one-dimensional velocity as a example,
\begin{equation}\label{eq}
\begin{aligned}
  \xi_{-1}=-c_f, \qquad \xi_{0}=0,\qquad \xi_{1}=c_f \\
  \omega_{0}=2/3, \qquad  \omega_{\pm1}=1/6.
\end{aligned}
\end{equation}
In two-dimensional space, the discrete velocities  and  the corresponding weights  can be calculated by using the tensor product method, leading to~\cite{shan2006kinetic}
\begin{equation}\label{eq}
\begin{aligned}
\vert \xi_{ix} \rangle=& \left(0,1,0,-1,0,1,-1,-1,1 \right)c_f, \\
\vert \xi_{iy} \rangle=& \left(0,0,1,0,-1,1,1,-1,-1 \right)c_f,\\
\vert W_i \rangle =& \left(\frac{4}{9},\frac{1}{9},\frac{1}{9},\frac{1}{9},\frac{1}{9},\frac{1}{36},\frac{1}{36},\frac{1}{36},\frac{1}{36} \right),
\end{aligned}
\end{equation}
where $c_f=\sqrt{3}c_{s,f}$ and $\vert \ldots \rangle$ represents a nine-dimensional row vector.
As a result,  the discrete Boltzmann equation becomes
\begin{equation}\label{eq:DBE_f}
\frac{\partial f_i}{\partial t}+\bm \xi_i\cdot \nabla f_i=\Omega(f_i,f_i^{eq}) + \mathit{F}_i.
\end{equation}
where $\mathit{F}_i=(\bm \xi_i-\bm u)\cdot \bm F_{ext} f_i^{eq}/(c_{s,f}^2\rho_0)$.
The above discrete Boltzmann equation will be discretized by DUGKS later.

Through multiscale analysis, the following continuity and momentum equations can be recovered~\cite{qian1993lattice}
\begin{equation}\label{eq:continue_0}
\partial_t \rho_0+\nabla\cdot(\rho_0\bm u)=0,
\end{equation}
\begin{equation}\label{eq:momentum_0}
\partial_t(\rho_0 \bm u)+\nabla\cdot(\rho_0\bm u\bm u)=-\nabla p_0+\nabla \cdot(\rho_0\nu(\nabla \bm u+\nabla\bm u^T)+\rho_0\eta\nabla\cdot \bm u)+\bm F_{ext},
\end{equation}
where $p_0=c_{s,f}^2\rho_0$, $\nu$ and $\eta$ are the kinetic viscosity and bulk viscosity related to the relaxation time in the collision model.
It is clear that Eq.~(\ref{eq:continue_0}) is different from Eq.(\ref{eq:ACM}). To recover the correct pressure equation,  the zeroth moment of the distribution function should be suitably  modified. To do this, we redefine the distribution function and force term as
\begin{equation}\label{eq:new_feq_F}
  \begin{aligned}
  \bar{\bar{f}} &=f + (p-\rho_0)\exp\left(-\frac{|\bm \xi|^2}{2c_{s,f}^2}\right)\delta(\xi_x)\delta(\xi_y),\\
  \bar{\bar{f}}^{eq} &=f^{eq}+ (p-\rho_0)\exp\left(-\frac{|\bm \xi|^2}{2c_{s,f}^2}\right)\delta(\xi_x)\delta(\xi_y),\\
  \bar{\bar{F}} &=F(\bm\xi)+ (\rho_0-c_0\rho)\nabla\cdot \bm u\exp\left(-\frac{|\bm \xi|^2}{2 c_{s,f}^2}\right) \delta(\xi_x)\delta(\xi_y),
  \end{aligned}
\end{equation}
where $\delta $ denotes the one-dimensional Dirac delta function.
As shown in Eq.(\ref{eq:rawmoment_NS}),  the integration of the redefined distribution function and force term has no effect on all higher moments. However, the zeroth moments of the distribution function and force term become
\begin{equation}\label{eq}
p=\int \bar{\bar{f}}d\bm \xi=\int \bar{\bar{f}}^{eq} d\bm\xi, \qquad  (\rho_0-c_0\rho)\nabla\cdot\bm u=\int \bar{\bar{F}}d\bm \xi.
\end{equation}
For convenience, we omit the notation $\bar{\bar{}}$ in what follows. The novel distribution function  obeys Eq.(\ref{eq:NS_kinetic}) as well.

By comparing Eqs.(\ref{eq:NS2}) and (\ref{eq:momentum_0}), the external force $\bm F_{ext}$ can be defined as $ \bm F_{ext} =\frac{\rho_0}{\rho}\left[\bm F_{sf}+\bm F_g+\bm F_p +\bm F_{\mu}\right]$, where $\bm F_p$ and $\bm F_{\mu}$ are introduced correction terms to match the pressure gradient  and the viscous stress force in Eq.(\ref{eq:NS2}),namely,
\begin{equation}\label{eq:redefined-forceNS}
\bm F_p=-\nabla p,\qquad
\bm F_{\mu}=\left(\nu(\nabla \bm u+\nabla\bm u^T)+\eta\nabla\cdot \bm u\right)\cdot \nabla\rho.
\end{equation}
Inserting Eq.(\ref{eq:redefined-forceNS}) into Eq.(\ref{eq:momentum_0}), the above momentum equation can be rewritten as
\begin{equation}\label{eq:momentum}
\partial_t\bm u+\nabla\cdot(\bm u\bm u)=-\frac{\nabla p}{\rho}+ \nabla \cdot(\nu(\nabla \bm u+\nabla\bm u^T)+\eta\nabla\cdot \bm u)
+\frac{\left(\nu(\nabla \bm u+\nabla\bm u^T)+\eta\nabla\cdot \bm u\right)\cdot \nabla\rho}{\rho}
+\frac{\bm F_{sf}+\bm F_b}{\rho},
\end{equation}
which is consistent with Eq.(\ref{eq:NS2}) in the incompressible limit.

\subsection{Central-moment-based collision operator}
Geier~\cite{geier2006cascaded} insists that an insufficient degree of Galilean invariance of the relaxation-type collision operator leads to numerical instability when small viscosity (or high Reynolds number) is involved. To this end, they performed collision operator in a frame of reference shifted by the macroscopic fluid velocity instead of the previous collision operator in a rest frame of reference. The raw moments in a rest frame and the central moments in a moving frame are respectively defined as
\begin{equation}\label{eq:rawmoment-centralmoment}
\bm {\Upsilon}_{pq}=\int_{-\infty}^{+\infty} \int_{-\infty}^{+\infty} \xi_x^p \xi_y^q f d\xi_x d\xi_y,
\quad
\bm{\widehat{\Upsilon}}_{pq}=\int_{-\infty}^{+\infty} \int_{-\infty}^{+\infty}  c_x^pc_y^q f \xi_x d\xi_y,
\end{equation}
where $\vert c_x\rangle=\vert \xi_x-u_x \rangle$ and $\vert c_y\rangle=\vert \xi_y- u_y\rangle$ are relative velocities,
 $(u_x,u_y)$ are  components of the macroscopic velocity $\bm u$ in Cartesian coordinates.
The equilibrium counterparts can be defined analogously by replacing $f$ with $f^{eq}$. As recommended in~\cite{premnath2009incorporating},
the following non-orthogonal basis vectors obtained from the combinations of the monomials  $\xi_x^p\xi_y^q$ and  $c_x^pc_y^q$ in an ascending order
\begin{align}\label{eq:rawvelocity}
\bm M &=\left[ 1,\xi_x,\xi_y,\xi_x^2+\xi_y^2,\xi_x^2-\xi_y^2,\xi_x\xi_y,\xi_x^2\xi_y,\xi_x\xi_y^2,\xi_x^2\xi_y^2 \right],\\
\label{eq:centralvelocity}
\bm {{K}} &=\left[ 1,c_x,c_y,c_x^2+c_y^2,c_x^2-c_y^2,c_x c_y,c_x^2 c_y,c_xc_y^2,c_x^2c_y^2 \right]
\end{align}
The relationship between raw moments and central moments can be expressed in the following matrix form,
\begin{equation}\label{eq}
\bm {K}=\bm N \bm M,
\end{equation}
where the transformation matrix $\bm M$ is independent of the macroscopic velocity and the shift matrix $\bm N={\bm {K}}\bm M^{-1}$ is obtained by matrix operations, which can be given as~\cite{fei2017consistent,lycett2014multiphase}
\begin{equation}\label{eq:M_matrix}
\bm M=\left[
\begin{matrix}
1 &  1 &   1&    1&    1&    1&    1&      1&  1    \\  
0 &  c_f &   0&   -c_f&    0&    c_f&   -c_f&     -c_f&  c_f    \\  
0 &  0 &   c_f&    0&   -c_f&    c_f&    c_f&     -c_f& -c_f    \\  
0 &c_f^2 & c_f^2&  c_f^2&  c_f^2& 2c_f^2& 2c_f^2&   2c_f^2& 2c_f^2  \\  
0 &c_f^2 &-c_f^2&  c_f^2& -c_f^2&    0&    0&      0&  0    \\  
0 &  0 &   0&    0&    0&  c_f^2& -c_f^2&    c_f^2& -c_f^2  \\  
0 &  0 &   0&    0&    0&  c_f^3&  c_f^3&   -c_f^3& -c_f^3  \\  
0 &  0 &   0&    0&    0&  c_f^3& -c_f^3&   -c_f^3&  c_f^3  \\  
0 &  0 &   0&    0&    0&  c_f^4&  c_f^4&    c_f^4&  c_f^4  \\  
\end{matrix}
\right]
\end{equation}
and,
\begin{equation}\label{eq:N_matrix}
\bm N=\left[
\begin{matrix}
1             &       0   &    0      &               0&              0&         0&    0&    0&  0  \\  
-u_x          &       1   &    0      &               0&              0&         0&    0&    0&  0  \\  
-u_y          &       0   &    1      &               0&              0&         0&    0&    0&  0  \\  
u_x^2+u_y^2   & -2u_x     &-2u_y      &               1&              0&         0&    0&    0&  0  \\  
u_x^2-u_y^2   & -2u_x     &2u_y       &               0&              1&         0&    0&    0&  0  \\  
u_x u_y       & -u_y      &-u_x       &               0&              0&         1&    0&    0&  0  \\  
-u_x^2 u_y    & 2u_x u_y  &u_x^2      & -\frac{u_y}{2}         &-\frac{u_y}{2}         & -2u_x    &    1&    0&  0  \\  
-u_y^2 u_x    & u_y^2     &2u_xu_y    & -\frac{u_x}{2}         &\frac{u_x}{2}          & -2u_y    &    0&    1&  0  \\  
u_x^2 u_y^2   &-2u_x u_y^2&-2u_yu_x^2 & \frac{u_x^2}{2}+\frac{u_y^2}{2}&\frac{u_y^2}{2}-\frac{u_x^2}{2}& 4u_x u_y &-2u_y&-2u_x&  1  \\  
\end{matrix}
\right]
\end{equation}
It is worth noting that the formulations for $\bm M$ and $\bm N$ are not unique. Another formulations of  $\bm M$ and $\bm N$ can be found in~\cite{2017Modeling}. The key difference is the definition of the combined raw moment set. With the above moments, the collision operator is given by
\begin{equation}\label{eq}
\Omega_f=-\bm M^{-1}\bm N^{-1}\bm \Lambda_f \bm N\bm M(f_i-f_i^{eq})=-\bm K^{-1}\bm \Lambda_f \bm K(f_i-f_i^{eq}),
\end{equation}
where the diagonal relaxation matrix is given by
\begin{equation}\label{eq}
\bm \Lambda_f=diag\left(
 \lambda_{f,0}^{-1},\lambda_{f,1}^{-1},\lambda_{f,2}^{-1}
 ,\lambda_{f,3}^{-1},\lambda_{f,4}^{-1},\lambda_{f,5}^{-1}
 ,\lambda_{f,6}^{-1},\lambda_{f,7}^{-1},\lambda_{f,8}^{-1} \right).
\end{equation}
The fluid kinematic viscosity and bulk viscidity are defined as
\begin{equation}\label{eq}
  \nu=c_{s,f}^2\lambda_{f,4}=c_{s,f}^2\lambda_{f,5}, \qquad  \nu_b=2c_{s,f}^2\lambda_{f,3},
\end{equation}

From the definition of the raw moment, namely, Eq.(\ref{eq:rawmoment-centralmoment}), integrating $f^{eq}$ and $F(\bm \xi)$ in Eq.(\ref{eq:new_feq_F}), we can obtain the raw moments of the  equilibrium distribution function  and the source term
\begin{equation}\label{eq:rawmoment_NS}
\bm m_f^{eq}=\left( \begin{array}{c}
p,\\
\rho_0 u_x,\\
\rho_0 u_y,\\
\rho_0(u_x^2+u_y^2)+2c_{s,f}^2\rho_0,\\
\rho_0(u_x^2-u_y^2),\\
\rho_0 u_x u_y,\\
\rho_0 u_y(c_{s,f}^2+u_x^2),\\
\rho_0 u_x(c_{s,f}^2+u_y^2),\\
\rho_0 (c_{s,f}^2+u_y^2)(c_{s,f}^2+u_x^2),\\
\end{array}
\right),
\quad
\bm S_f= \left( \begin{array}{c}
(\rho_0-c_0\rho)\nabla\cdot\bm u,\\
F_x,\\
F_y,\\
2(F_x u_x+F_y u_y),\\
2(F_x u_x -F_y u_y),\\
F_x u_y+F_y u_x,\\
2F_x u_x u_y+F_y(c_{s,f}^2+u_x^2),\\
2F_y u_x u_y+F_x(c_{s,f}^2+u_y^2),\\
2F_xu_x(c_{s,f}^2+u_y^2)+2F_y u_y(c_{s,f}^2+u_x^2),\\
\end{array}
\right),
\end{equation}
where $(F_x,F_y)$ are components of the external force $\bm F_{ext}$.
It is noted that we do not  employ the well-known second-order truncated equilibrium distribution function to obtain the raw moments.

\subsection{Operator splitting for the discrete Boltzmann equation}
When high density ratios are present in multiphase flows, the discrete  error from the calculation of the gradients of the density across the interface is inevitably introduced, often causing numerical instability.
An convenient strategy is to employ the Strang splitting method to deal with the force term without affecting the second order accuracy of DUGKS.
In Strang splitting method, the discrete Boltzmann equation without the force term is discretized by DUGKS  while the external force term is applied twice for a half time step  before and after the solution of the DUGKS procedure. This procedure can be represented as
\begin{align}\label{eq}
\text{Pre-forcing}:&  &\frac{\partial f_i}{\partial t}&=\mathit{F}_i. &\quad  [t, t+\Delta t/2] \\
\text{DUGKS without force}:&  & \frac{\partial f_i}{\partial t}+\bm \xi\cdot\nabla f_i &=\Omega_f, &\quad [t, t+\Delta t] \label{SS_DUGKS}\\
\text{post-forcing}:&  &\frac{\partial f_i}{\partial t} &=\mathit{F}_i. &\quad  [t, t+\Delta t/2]
\end{align}
The pre-forcing step is realized as
\begin{equation}\label{eq}
f_i^*=f_i^n+\frac{\Delta t}{2}\left(F_i^{n}(\phi^n,p^n,\bm u^n)+\mathit{F}_i^{*}(\phi^n,p^*,\bm u^*)\right),
\end{equation}
where $p^*$ and $u^*$ are values from the prediction step at $t=\Delta t/2$. The post-forcing step is performed by analogy.

\subsection{Discrete unified gas-kinetic scheme without force term}
In what follows, we shall give the detail of the DUGKS based on Eq.(\ref{SS_DUGKS}).
The integration of the above equation over a control volume  yields
\begin{equation}\label{eq:dbe-ns}
\frac{\partial \bar{f}_i}{\partial t}+\frac{1}{|V_j|} \oint_{V_j}\bm \xi_i f_id\bm A=\bar{\Omega}_f ,
\end{equation}
where $|V_j|$ is the volume of the control volume $V_j$ and $A$ is the cross-sectional area of the control volume face, $\bar{f}_i$ is the cell-averaged values of the distribution function located in  $V_j$, i.e.,
\begin{equation}\label{eq}
\bar{f}_i=\frac{1}{|V_j|}\int_{V_j} f_i (\bm x_j,\xi_i,t_n)d\bm x
\end{equation}
For brevity, we omit the overbar from the notation.
Time integration of  Eq.(\ref{SS_DUGKS}) from time step $n\Delta t$ to the next time step $(n+1)\Delta t$ gives
\begin{equation}\label{eq:time_integrating_ns}
f_i^{n+1}-f_i^{n}+\frac{\Delta t}{|V_j|}\oint(\bm \xi_i \cdot \bm n) f_i^{n+1/2}dA=\frac{\Delta t}{2}(\Omega_f^{n+1}+\Omega_f^{n}),
\end{equation}
where the midpoint rule for the integration of the convection term and trapezoidal rule for the
collision term.
To remove the implicity, the following auxiliary distribution functions are introduced,
\begin{equation}\label{eq:auxxiliary_first_ns}
\begin{aligned}
\widetilde{f}_i &=f_i-\frac{\Delta t}{2}\Omega_{f}
=K^{-1}\left[\frac{2  I+\Lambda_f \Delta t}{2} Kf_i-\frac{\Delta t\Lambda_f}{2}Kf_i^{eq}\right],
 \\
\widetilde{f}_i^{+} &=f_i+\frac{\Delta t}{2}\Omega_{f}
=K^{-1}\left[
\frac{2I-\Delta t \Lambda_f}{2I+\Delta t \Lambda_f} K\widetilde{f}_i  +\frac{2\Delta t \Lambda_f}{2I+\Delta t\Lambda_f} K f_i^{eq}\right], \\
\end{aligned}
\end{equation}
Substituting Eq.(\ref{eq:auxxiliary_first_ns}) into Eq.(\ref{eq:time_integrating_ns}), one can obtain
\begin{equation}\label{eq:evolution_ns}
\widetilde{f}_i=\widetilde{f}_i^{+} -\frac{\Delta t}{|V_j|}\int_{\partial V_j}(\bm \xi_i \cdot \bm n)f_i^{n+1/2}d\partial V_j.
\end{equation}
Then, the conserved variables are updated by
\begin{equation}\label{eq}
\begin{aligned}
\rho_0 \bm u =\sum_i \bm \xi_i \widetilde{f}_i,  \qquad  p = \sum_i \widetilde{f}_i.
\end{aligned}
\end{equation}
Therefore, we only need to track the distribution function $\widetilde{f}_i$ instead of $f_i$.

The only unknown parameter is the interface flux in Eq.~(\ref{eq:evolution_ns}). To obtain the interface flux, we integrate Eq.~(\ref{SS_DUGKS}) within a half time step $h=\Delta t/2$ along the characteristic line assuming the end point located at the cell interface $\bm x_b=\bm x_j+\bm \xi_i h$, leading to
\begin{equation}\label{eq:half_time_integrate_ns}
f_i^{n+h}(\bm x_b, \bm \xi_i)-f_i^n(\bm x_b-\bm \xi_i h,\bm \xi_i)=
\frac{h}{2}\left[\Omega_f^{n+h}+\Omega_f^{n} \right].
\end{equation}
Similarly, to eliminate the implicity, another two auxiliary distribution functions $\widehat{f}_i$ and $\widehat{f}_i^{+}$  are introduced,
\begin{equation}\label{eq:auxxiliary_second_ns}
\begin{aligned}
\widehat{f}_i &=f_i-\frac{h}{2}\Omega_{f}
=K^{-1} \left[
\frac{2  I +h\Lambda_f }{2}Kf_i-\frac{h\Lambda_f}{2}Kf^{eq}_i \right],
\\
\widehat{f}_i^+ &=f_i + \frac{h}{2}\Omega_{f} =
K^{-1}\left[
\frac{2  I-h\Lambda_f}{2I+\Delta t\Lambda_f} K\widetilde{f}_i
+\frac{3h\Lambda_f}{2I+\Delta t\Lambda_f} Kf^{eq}_i\right]
\end{aligned}
\end{equation}
Then, Eq.(\ref{eq:half_time_integrate_ns}) can be rewritten as
\begin{equation}\label{eq:flux_ns}
\begin{aligned}
\widehat{f}_i^{n+1/2}(\bm x_b, \bm \xi_i)=\widehat{f}_i^{+,n}(\bm x_b-\bm \xi_i h,\bm \xi_i).
\end{aligned}
\end{equation}
By employing the linear reconstruction around the cell face $\bm x_b$, the above equation can be approximated as
\begin{equation}\label{eq:halt_time_widehatPhi_ns}
\widehat{f}_i^{n+1/2}(\bm x_b-\bm \xi_i h, \bm \xi_i)=\widehat{f}_i^{+,n}(\bm x_b,\bm \xi_i)- h \bm \xi_i \cdot \nabla \widehat{f}_i^{+,n}(\bm x_b,\bm \xi_i).
\end{equation}
The calculation of the values of $\widehat{f}_i^{+,n}$ has an important effect on the results~\cite{zhang2022discrete}. Here, the fifth-order targeted essentially non-oscillatory scheme (TENO) is employed~\cite{shu1998essentially}. The gradient $\nabla\widehat{f}_i^{+,n}$ is approximated by the linear interpolation. The macroscopic quantities at cell faces are updated by,
\begin{equation}\label{eq}
\begin{aligned}
\rho_0 \bm u^{n+1/2} &=\sum_i \bm \xi_i \widetilde{f}_i^{n+1/2}, \\
p^{n+1/2} &= \sum_i \widetilde{f}_i^{n+1/2}.
\end{aligned}
\end{equation}

With the above conserved parameters at the cell face, the original distribution function can be calculated by
\begin{equation}\label{eq:original_FD_NS}
f_i=K^{-1}\left(
\frac{2I}{2I+h\Lambda_f}K\widehat{f}_i+\frac{h\Lambda_f}{2I+h\Lambda_f} K f_i^{eq}\right).
\end{equation}
In addition, based on Eqs.~(\ref{eq:auxxiliary_first_ns}) and (\ref{eq:auxxiliary_second_ns}), the following relation is found
\begin{equation}\label{eq:relation_auxxiliary_ns}
\begin{aligned}
\widetilde{f}_i^{+} &=\frac{4}{3}\widehat{f}_i^{+} -\frac{1}{3}\widetilde{f}_i,
\end{aligned}
\end{equation}

In DUGKS, the time step $\Delta t$ is an adjustable variable determined by Courant-Friedricbs-Lewy (CFL) condition,
\begin{equation}\label{eq}
\Delta t=\text{CFL}\frac{\text{min}\left(\Delta x, \Delta y\right)}{|\bm u|_{max}+c_{s,f}},
\end{equation}
where $\text{CFL}$ lies between $0$ and $1$, $\Delta x_{min}$ and $\Delta y_{min}$ denote the minimal grid spacing in the x- and y-directions respectively.

\section{DUGKS for the conservative ACE}\label{sec:3}
\subsection{Discrete Boltzmann equation  for the conservative ACE}\label{sec:4}
The kinetic equation with a source term for the conservative ACE can be written as
\begin{equation}\label{eq:kinetic-AC}
\frac{\partial g}{\partial t}+\bm \xi\cdot \nabla g=\Omega(g,g^{eq})+\mathit{G}(\bm \xi),
\end{equation}
where $g=g(\bm x,\bm \xi, t)$ is the distribution function with particle velocity $\bm \xi$ at position $\bm x$ and time $t$ for the order parameter.  $g^{eq}$ is the corresponding equilibrium distribution function, $\mathit{G}$ is the source term.  $\Omega_g$ represents the effect of the binary particle collisions only satisfying the conservation of zeroth moment. The  equilibrium distribution function  $g^{eq}$ in analogy with $f^{eq}$ by  replacing the density with the order parameter is defined as
\begin{equation}\label{eq:dugks_AC}
g^{eq}(\bm x,\bm \xi,t)=\frac{\phi}{2\pi c_{s,g}^2}\exp\left[{-\frac{(\bm \xi-\bm u)^2}{2c_{s,g}^2}}\right],
\end{equation}
where $c_{s,g}=\sqrt{RT_g}$ and $T_g$ represents the temperature whose value can be different from $T_f$. The source term is given by
\begin{equation}\label{eq}
\mathit{G}(\bm \xi)=-\frac{\bm G}{\phi}\cdot\nabla_{\xi} g\approx \frac{\bm G\cdot (\bm \xi-\bm u)}{c_{s,g}^2\phi}g^{eq},
\end{equation}
 The order parameter is defined as
\begin{equation}\label{eq}
  \phi=\int gd\bm \xi.
\end{equation}
Similar to the velocity-space discretization in the previous  section,
the three-point Gauss-Hermite quadrature is used to obtain the one-dimensional discrete particle velocities and associated weights for the distribution function for the order parameter. The resulting discrete velocities and weights in two-dimensional space is calculated by the tensor product method, namely,
\begin{equation}\label{eq}
\begin{aligned}
\vert \xi_{ix} \rangle=& \left(0,1,0,-1,0,1,-1,-1,1 \right)c_g, \\
\vert \xi_{iy} \rangle=& \left(0,0,1,0,-1,1,1,-1,-1 \right)c_g,\\
\vert W_i \rangle =& \left(\frac{4}{9},\frac{1}{9},\frac{1}{9},\frac{1}{9},\frac{1}{9},\frac{1}{36},\frac{1}{36},\frac{1}{36},\frac{1}{36} \right),
\end{aligned}
\end{equation}
where $c_g=\sqrt{3}c_{s,g}$.  Note that the value of $c_g$ does not need to be equal to $c_f$.
Then, the discrete continue kinetic equations with the source term for the conservative ACE is written as
\begin{equation}\label{eq:dbe-AC}
\frac{\partial g_i}{\partial t}+\bm \xi_i\cdot \nabla g_i=\Omega(g_i,g_i^{eq})+G_i,
\end{equation}
where $G_i=\frac{(\bm \xi_i-\bm u)\cdot \bm G}{c_{s,g}^2\phi} g_i^{eq}$.
Accordingly, the collision operator is defined as
\begin{equation}\label{eq}
\Omega_g=-\bm M_g^{-1}\bm N^{-1}\bm{\Lambda}_g\bm N\bm M_g(g_i-g_i^{eq})=-\bm K_g^{-1}\bm{\Lambda}_g\bm K_g(g_i-g_i^{eq}),
\end{equation}
where $\bm M_g$ has the same form of $\bm M$ but $c_f$ is replaced by $c_g$, and $\bm K_g=\bm N\bm M_g$. For brevity, we still denote $\bm M_g$ by $\bm M$ and $\bm K_g$ by $\bm K$.
The diagonal relaxation matrix $\bm {\Lambda}_g$ is defined as
\begin{equation}\label{eq}
\bm \Lambda_g=diag\left(
 \lambda_{g,0}^{-1},\lambda_{g,1}^{-1},\lambda_{g,2}^{-1}
 ,\lambda_{g,3}^{-1},\lambda_{g,4}^{-1},\lambda_{g,5}^{-1}
 ,\lambda_{g,6}^{-1},\lambda_{g,7}^{-1},\lambda_{g,8}^{-1} \right).
\end{equation}

To recover the correct Eq.(\ref{eq:CACE}), $\bm G$ should be defined as $\bm G=c_{s,g}^2 \Theta\bm n$. Then, the order parameter can be updated by
\begin{equation}\label{eq:macro_AC_from_DUGKS}
\phi(\bm x,t)=\sum_i g_i.
\end{equation}

With the above discrete velocity set, we can obtain the raw moments of the continuum equilibrium distribution $g^{eq}$ and the source term $G_i$,
\begin{equation}\label{eq:rawmoments_continuum_AC}
\bm m_g^{eq}=\left( \begin{array}{c}
\phi,\\
\phi u_x,\\
\phi u_y,\\
\phi(u_x^2+u_y^2)+ 2c_{s,g}^2\phi,\\
\phi(u_x^2-u_y^2),\\
\phi u_x u_y,\\
\phi u_y(c_{s,g}^2+u_x^2),\\
\phi u_x(c_{s,g}^2+u_y^2),\\
\phi (c_{s,g}^2+u_y^2)(c_{s,g}^2+u_x^2),\\
\end{array}
\right),
\quad
\bm S_g=\left( \begin{array}{c}
0,\\
c_{s,g}^2G_x,\\
c_{s,g}^2G_y,\\
2c_{s,g}^2(G_x u_x+G_y u_y),\\
2c_{s,g}^2(G_x u_x -G_y u_y),\\
c_{s,g}^2(G_x u_y+G_y u_x),\\
2c_{s,g}^2G_x u_x u_y+G_yc_{s,g}^2(c_{s,g}^2+u_x^2),\\
2c_{s,g}^2G_y u_x u_y+ c_{s,g}^2G_x(c_{s,g}^2+u_y^2),\\
2c_{s,g}^2G_x u_x(c_{s,g}^2+u_y^2)+2c_{s,g}^2G_y u_y(c_{s,g}^2+u_x^2),\\
\end{array}
\right),
\end{equation}
where $(G_x,G_y)$ are  components of $\bm G$ in Cartesian coordinates.
It can be proved that the correct ACE can be recovered from the above model and details are provided in ~\ref{sec:App_B}.

In addition to employing the source term to correct the diffusion term in ACE, a relatively simple way to recover the ACE is to redefine the second and third moments of the equilibrium distribution function such that the source term is unnecessary.
To this end, the raw moments of the equilibrium distribution can be redefined as
\begin{equation}\label{eq:dugksWithoutforce}
\bm m_g^{eq}=\left( \begin{array}{c}
\phi,\\
\phi u_x+c_{s,g}^2\Theta n_x,\\
\phi u_y+c_{s,g}^2\Theta n_y,\\
\phi(u_x^2+u_y^2)+ 2c_{s,g}^2\phi,\\
\phi(u_x^2-u_y^2),\\
\phi u_x u_y,\\
\phi u_y(c_{s,g}^2+u_x^2),\\
\phi u_x(c_{s,g}^2+u_y^2),\\
\phi (c_{s,g}^2+u_y^2)(c_{s,g}^2+u_x^2),\\
\end{array}
\right),
\end{equation}
In the absence of source force term, Eq.(\ref{eq:dbe-AC}) with Eq.(\ref{eq:dugksWithoutforce}) is still able to recover the correct ACE.
In the next section, both DUGKS methods with and without source term for ACE will be validated and compared.

\subsection{Discrete unifed gas-kinetic scheme with source term}\label{sec:5}
Without loss of generality, the DUGKS scheme with source term  for ACE is presented below.
First, integration of  Eq. (\ref{eq:dbe-AC}) over a control volume  yields
\begin{equation}\label{eq:cv_ac}
\frac{\partial \bar{g}_i}{\partial t}+\frac{1}{|V_j|} \oint_{V_j}\bm \xi_ig_id\bm A=\bar{\Omega}_g + \bar{G}_i,
\end{equation}
where $\bar{g}_i$ and $\bar{G}_i$ are the cell-averaged values of the distribution function and source term located in the control volume $V_j$. In what follows, we omit the overbar from the notation again.
Integrating Eq.(\ref{eq:cv_ac}) from time step $n\Delta t$ to the next time step $(n+1)\Delta t$ gives
\begin{equation}\label{eq:time_integrating_ac}
g_i^{n+1}-g_i^{n}+\frac{\Delta t}{|V_j|}\oint(\bm \xi_i \cdot \bm n) g_i^{n+1/2}dA=\frac{\Delta t}{2}(\Omega_g^{n+1}+\Omega_g^{n})+\frac{\Delta t}{2}(G_i^{n+1} +G_i^{n}),
\end{equation}
where the midpoint rule for the integration of the convection term and trapezoidal rule for
the collision term and external force term. $g_i^{n+1/2}$ denotes a face-averaged value at the cell faces.

To remove the implicity, the following auxiliary distribution functions are introduced,
\begin{equation}\label{eq:auxxiliary_first_ac}
\begin{aligned}
\widetilde{g}_i &=g_i-\frac{\Delta t}{2}g_i-\frac{\Delta t}{2}G_i
=K^{-1}\left[\frac{2I+\Lambda_g \Delta t}{2} K g_i-\frac{\Delta t\Lambda_g}{2}Kg_i^{eq}
-\frac{\Delta t}{2} KG_i \right],
 \\
\widetilde{g}_i^{+} &=g_i+\frac{\Delta t}{2}g_i+\frac{\Delta t}{2}G_i
=K^{-1}\left[
\frac{2I-\Delta t \Lambda_g }{2I+\Delta t \Lambda_g } K\widetilde{g}_i  +\frac{2\Delta t \Lambda_g }{2I+\Delta t\Lambda_g } K g_i^{eq}
+\frac{2\Delta t}{2I+\Delta t\Lambda_g } K G_i \right]
, \\
\end{aligned}
\end{equation}
Substituting Eq.(\ref{eq:auxxiliary_first_ac}) into Eq.(\ref{eq:time_integrating_ac}), one can obtain
\begin{equation}\label{eq:evolution_ac}
\widetilde{g}_i=\widetilde{g}_i^{+} -\frac{\Delta t}{|V_j|}\int_{\partial V_j}(\bm \xi_i \cdot \bm n)g_i^{n+1/2}d\partial V_j.
\end{equation}
Based on Eq.(\ref{eq:auxxiliary_first_ac}), the conserved variables can be calculated by
\begin{equation}\label{eq}
\begin{aligned}
\phi =\sum_i \widetilde{g}_i,\\
\end{aligned}
\end{equation}
Thus, we only need to track the distribution function $\widetilde{g}_i$ instead of $g_i$.

To update the interface flux, we integrate  Eq.~(\ref{eq:dbe-AC})  within a half time step $h=\Delta t/2$ along the characteristic line assuming the end point located at the cell interface $\bm x_b=\bm x_j+\bm \xi_i h$,
\begin{equation}\label{eq:half_time_integrate_ac}
g_i^{n+1/2}(\bm x_b, \bm \xi_i)-g_i^n(\bm x_b-\bm \xi_i h,\bm \xi_i)=
\frac{h}{2}\left[\Omega_g^{n+1/2}+\Omega_g^n \right]+\frac{h}{2}\left[G_i^{n+1/2}+G_i^{n} \right],
\end{equation}
Similarly, to eliminate the implicity, another two auxiliary distribution functions $\widehat{g}_i$ and $\widehat{g}_i^{+}$  are introduced,
\begin{equation}\label{eq:auxxiliary_second_ac}
\begin{aligned}
\widehat{g}_i &=g_i-\frac{h}{2}\Omega_{g}-\frac{h}{2}G_i
=K^{-1} \left[
\frac{2 I+h\Lambda_g }{2}Kg_i-\frac{h\Lambda_g }{2}Kg_i-\frac{h}{2}Kg_i \right],
\\
\widehat{g}_i^+ &=g_i-\frac{h}{2}\Omega_{g}-\frac{h}{2}G_i=
K^{-1}\left[
\frac{2I-h\Lambda_g }{2I+\Delta t\Lambda_g } K\widetilde{g}_i
+\frac{3h\Lambda_g }{2I+\Delta t\Lambda_g } Kg^{eq}+\frac{3 h}{2I+\Delta t\Lambda_g }KG_i
\right]
\end{aligned}
\end{equation}
Then, Eq.(\ref{eq:half_time_integrate_ac}) can be rewritten as
\begin{equation}\label{eq}
\begin{aligned}
\widehat{g}_i^{n+1/2}(\bm x_b-\bm \xi_i h, \bm \xi_i)=\widehat{g}_i^{+,n}(\bm x_b-\bm \xi_i h,\bm \xi_i),
\end{aligned}
\end{equation}
As $\widehat{g}_i^{+,n}(\bm x_b-\bm \xi_i h,\bm \xi_i)$ is not exactly located at the centre of the control volume, the above equation is approximated as~\cite{PhysRevE.105.045317}
\begin{equation}\label{eq:halt_time_widehatPhi_ac}
\widehat{g}_i^{n+1/2}(\bm x_b, \bm \xi_i)\approx \widehat{g}_i^{+,n}(\bm x_b,\bm \xi_i)- h \bm \xi_i \cdot \nabla \widehat{g}_i^{+,n}(\bm x_b,\bm \xi_i)-\frac{h^2}{2}\bm \xi_i \bm \xi_i :\nabla\nabla\widehat{g}_i^{+,n}(\bm x_b,\bm \xi_i) ,
\end{equation}
The calculation of the values of $\widehat{g}_i^{+,n}$   at the cell faces is approximated by fifth-order TENO scheme and the gradients of $\widehat{g}_i^{+,n}$   at the cell faces are obtained by the linear interpolation. Once $\widehat{g}_i^{n+1/2}$  is obtained,
the macroscopic quantities at the cell face can be obtained,
\begin{equation}\label{eq}
\begin{aligned}
\phi^{n+1/2} &=\sum_i \widetilde{g}_i^{n+1/2}.
\end{aligned}
\end{equation}
From Eq.(\ref{eq:auxxiliary_second_ac}), the original distribution function is calculated by
\begin{equation}\label{eq}
g_i=K^{-1}\left[
\frac{2I}{2I+\Delta t \Lambda_g} K\widehat{g}_i+\frac{\Delta t\Lambda_g}{2I+\Delta t \Lambda_g}  Kg_i^{eq}+\frac{\Delta t}{2I+\Delta t\Lambda_g}KG_i\right].
\end{equation}
Finally, the interfacial flux in Eq.(\ref{eq:evolution_ac}) can be updated.
In addition, based on Eqs.~(\ref{eq:auxxiliary_first_ac}) and (\ref{eq:auxxiliary_second_ac}), the following relation is built,
\begin{equation}\label{eq:relation_ac}
\begin{aligned}
\widetilde{g_i}_i^{+} &=\frac{4}{3}\widehat{g_i}^{+} -\frac{1}{3}\widetilde{g_i},
\end{aligned}
\end{equation}

\section{Approximation of spatial derivatives in the force terms}\label{sec:6}
The first derivatives in the force terms, such as $\nabla\phi$, $\nabla p$, $\nabla\cdot\bm u$, must be calculated carefully to obtain satisfactory results. We take the first derivative of the order parameter as an example. For a uniform mesh, $\partial_x \phi$ can be evaluated by
\begin{equation}\label{eq}
\frac{\partial \phi}{\partial x}=\frac{\phi_{i+\frac{1}{2},j}-\phi_{i-\frac{1}{2},j}}{\Delta x},
\end{equation}
with
\begin{equation}\label{eq}
\begin{aligned}
\phi_{i+\frac{1}{2},j}&=\frac{7\phi_{i,j}+7\phi_{i+1,j}-\phi_{i-1,j}-\phi_{i+2,j}}{12}, \\
\phi_{i-\frac{1}{2},j}&=\frac{7\phi_{i,j}+7\phi_{i-1,j}-\phi_{i-2,j}-\phi_{i+1,j}}{12}.
\end{aligned}
\end{equation}
where $\phi_{i,j}=\phi(x_i,y_j)$ and $(x_i,y_j)$ represents the coordinate index in the computational domain.
The velocity derivatives can be calculated by the second-order moment of the non-equilibrium distribution in DUGKS instead of  finite difference methods
\begin{equation}\label{eq:fneq_to_velocity}
\begin{aligned}
\Pi_{xx} &=3c_{s,f}^2\partial_x u_x+c_{s,f}^2\partial_y u_y=-\sum_i \xi_{ix}\xi_{ix}(\bm M^{-1}\bm N^{-1}\Lambda_f \bm N \bm M)(f_i-f_i^{eq}), \\
\Pi_{xy} &=c_{s,f}^2(\partial_x u_y+\partial_x u_y)=-\sum_i \xi_{ix}\xi_{iy}(\bm M^{-1}\bm N^{-1}\Lambda_f \bm N \bm M)(f_i-f_i^{eq}),\\
\Pi_{yy} &=3c_{s,f}^2\partial_y u_y+c_{s,f}^2\partial_x u_x=-\sum_i \xi_{iy}\xi_{iy}(\bm M^{-1}\bm N^{-1}\Lambda_f \bm N\bm M)(f_i-f_i^{eq}),
\end{aligned}
\end{equation}
Using Eq.(\ref{eq:fneq_to_velocity}), $\bm F_{\mu}=(F_{\mu,x},F_{\mu,y})$ can be calculated by
\begin{equation}\label{eq}
\begin{aligned}
\rho_0\nabla\cdot \bm u &=\frac{\Pi_{xx}+\Pi_{yy}}{4c_{s,f}^2},\\
\rho_0F_{\mu,x} &=\left[\lambda_{f,4}\Pi_{xx}+(\lambda_{f,3}-\lambda_{f,4})\frac{\Pi_{xx}+\Pi_{yy}}{2}\right] \partial_x\rho+\lambda_{f,4} \Pi_{xy}\partial_y\rho, \\
\rho_0 F_{\mu,y} &=\lambda_{f,4}\Pi_{xy}\partial_x\rho
+\left[\lambda_{f,4} \Pi_{yy}+(\lambda_{f,3}-\lambda_{f,4})\frac{\Pi_{xx}+\Pi_{yy}}{2}
\right]\partial_y\rho, \\
\end{aligned}
\end{equation}

\section{Results and discussions}\label{sec:7}
In this section, we will present a validation study of our proposed DUGKS by a series of benchmark problems.
First, the first three tests are used to evaluate the accuracy of capturing the interface and modeling the surface tension effect. The last four tests are used to further validate the accuracy and robustness of the present model for modeling large-density-ratio flows with complex interface topologies.
For each test, in the presence of walls, the missing distribution functions are obtained by the bounce-back scheme~\cite{yang2020analysis}, which means that the velocities and  normal gradient of the order parameter remain zero at the boundary. For the free-slip boundary condition, the missing distribution functions for the hydrodynamic equations are realized by specular reflection scheme, which means that the normal velocity remains zero. For the pressure boundary, the zero normal pressure gradient at the boundary is enforced. The open boundary is realized by setting zero normal gradients of distribution function, order parameter and pressure. In the calculations of the force term, the values of the order parameter and pressure outside of the domain are required. These parameters are obtained by linear extrapolation.

In all simulations, the CFL number is fixed at $0.5$,  the interface width $W$ is set to $5$, $\rho_0=min(\rho_l,\rho_h)$ and $c_{s,f}=c_{s,g}=1$ unless other stated. The uniform grid is used.
The relaxation times in $\Lambda_g$ are set to be $0.25$ except $\lambda_{f,2}^g=\lambda_{f,3}^g$ are determined by the mobility. The relaxation times in $\Lambda_f$ are set to be $0.25$ except
$\lambda_{f,4}=\lambda_{f,5}$ are determined by the viscosity while $\lambda_{f,3}$ is set to be $1.0$. The order parameters  are $\phi_h=1$ and $\phi_l=0$. To suppress the pressure oscillations,  a fifth-order spatial filter is applied to the pressure field~\cite{kim2015lattice,zhang2019fractional}.

\subsection{Reversed single vortex}
\begin{figure}
  \centering
 \subfigure[t=T/2 ]{
  \includegraphics[width=0.5\textwidth,trim=20 0 20 0,clip]{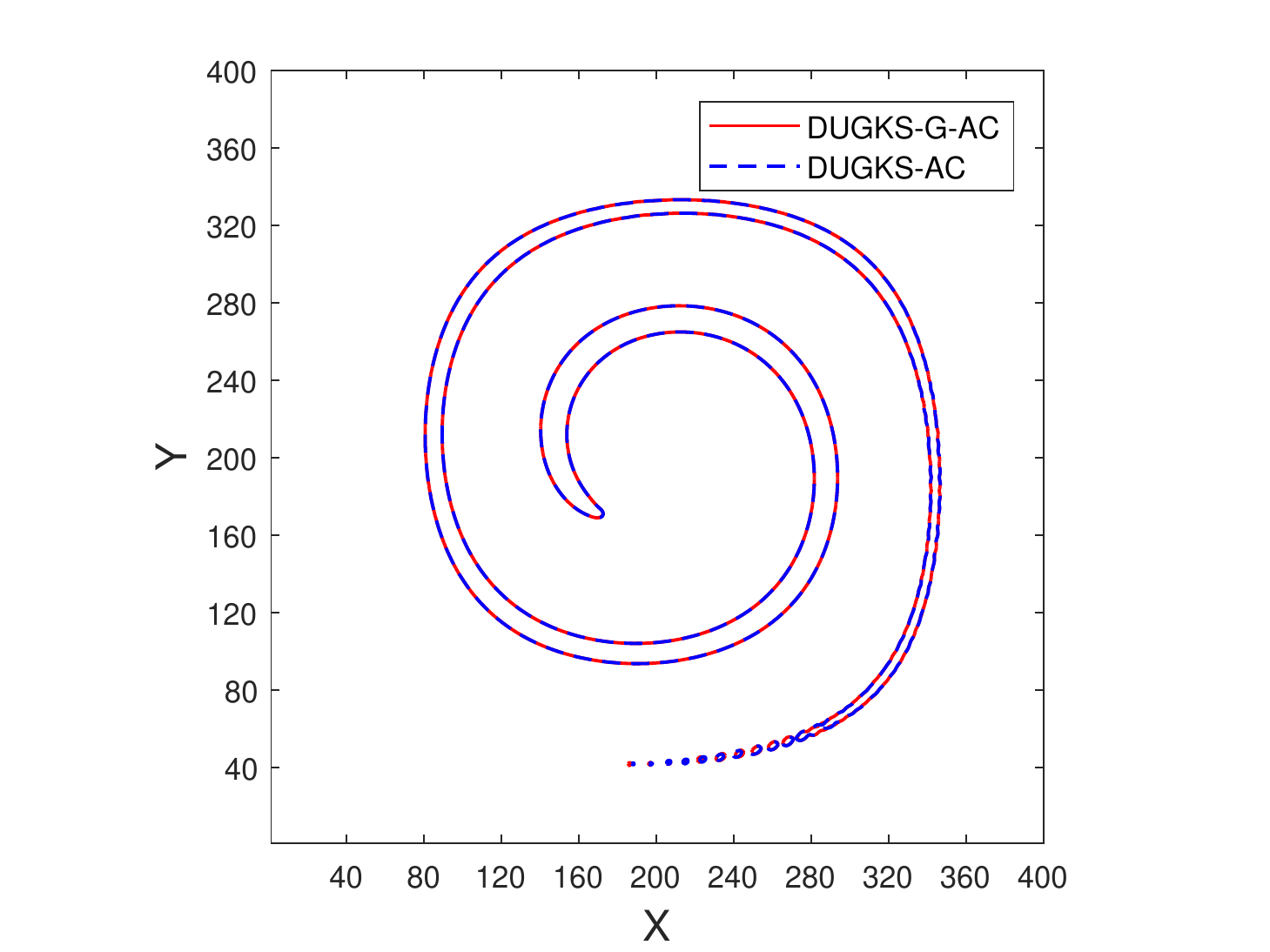}}~
 \subfigure[t=T ]{
  \includegraphics[width=0.5\textwidth,trim=20 0 20 0,clip]{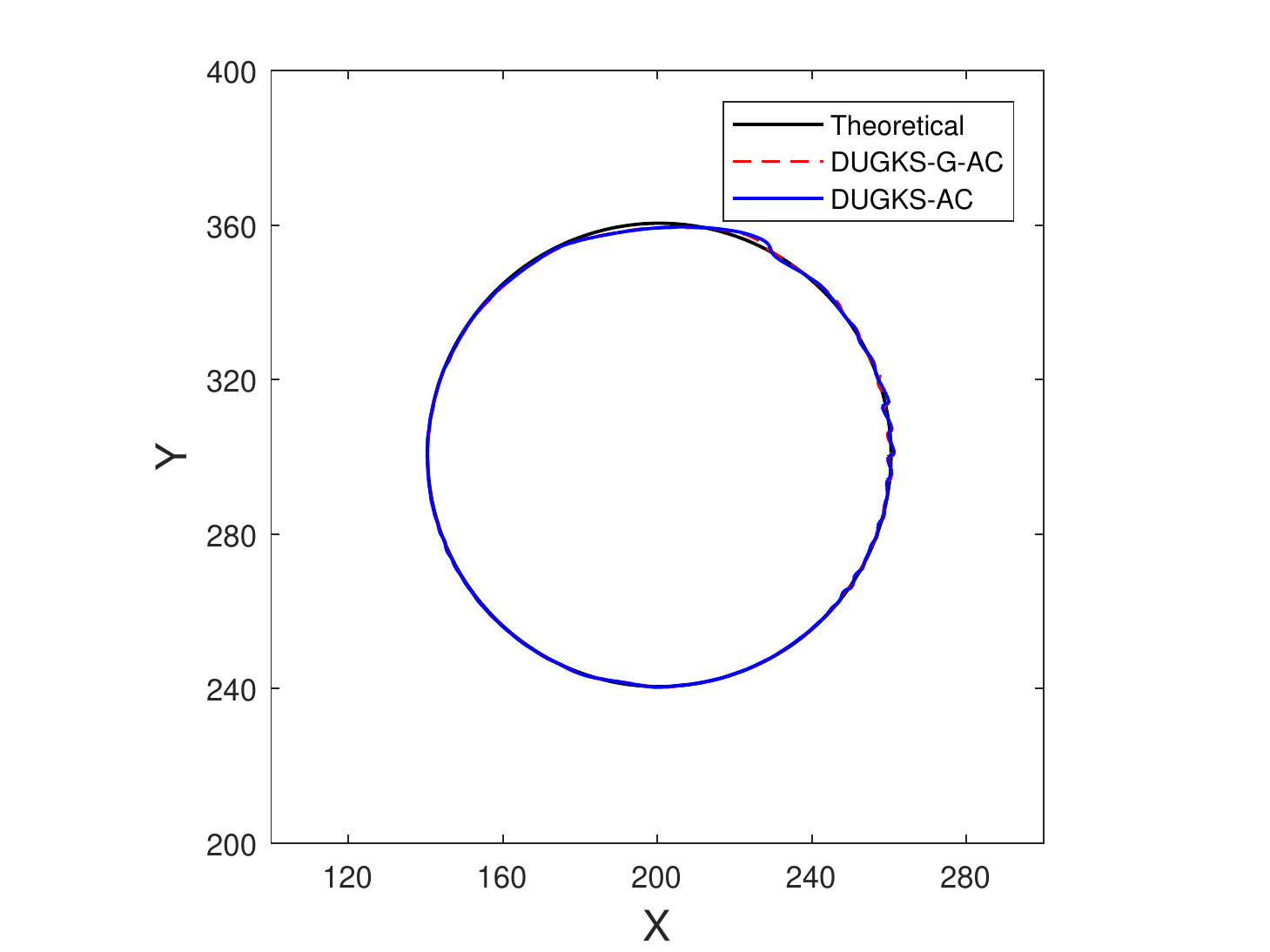}}
 \subfigure[t=T/2 ]{
 \includegraphics[width=0.5\textwidth,trim=20  0 20 0,clip]{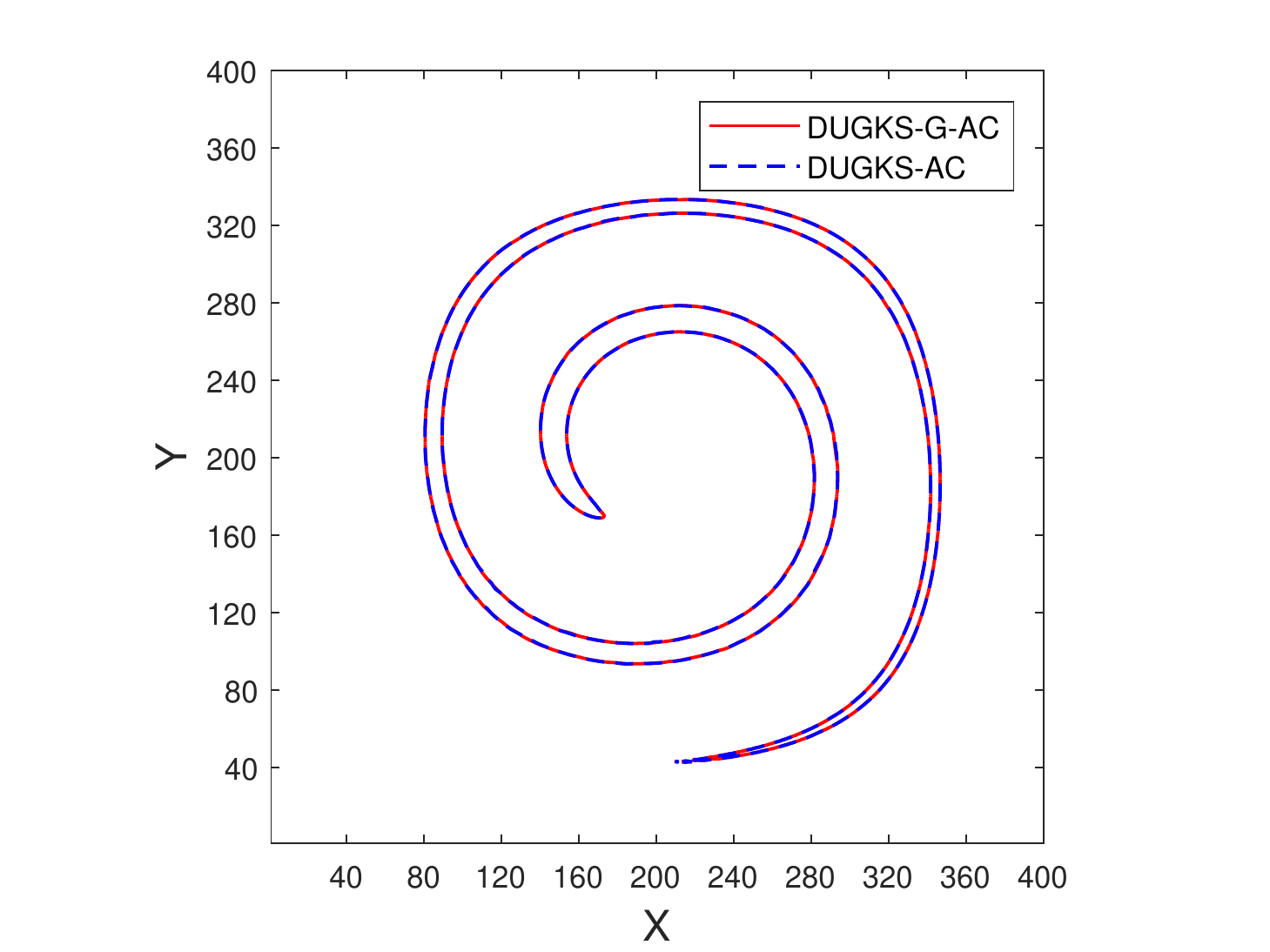}}~
 \subfigure[t=T ]{
  \includegraphics[width=0.5\textwidth,trim=20 0 20 0,clip]{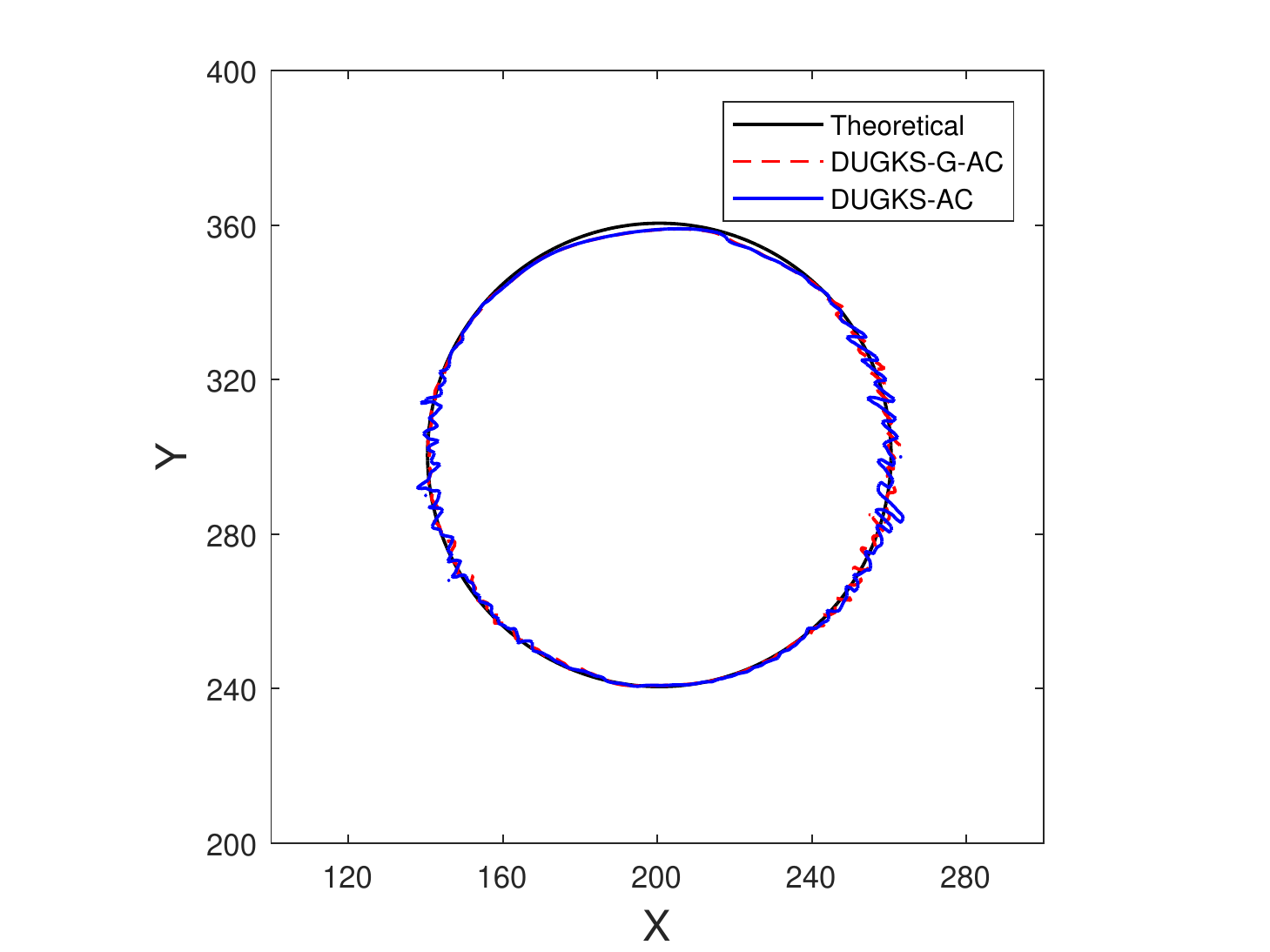}}
  \caption{The phase-field contour ($\phi=0.5$) of reversed single vortex at (a)(b) $\text{Pe}=100$ and (c)(d) $\text{Pe}=1600$.}\label{deformation}
\end{figure}

\begin{figure}
  \centering
   \includegraphics[width=0.5\textwidth,trim=10 0 20 0,clip]{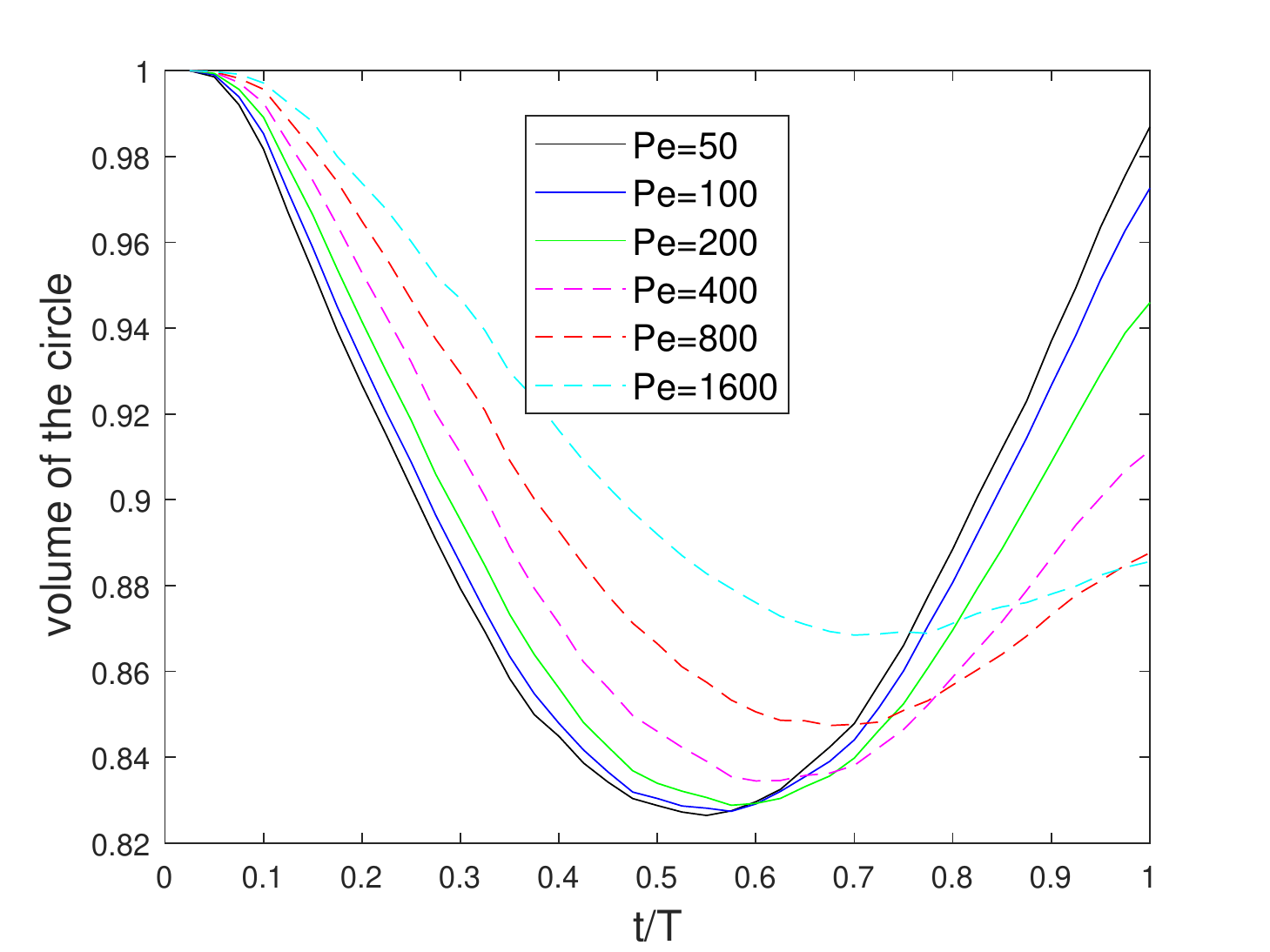}~
  \caption{Evolution of the normalized volume of the circle at different Peclet numbers predicted by DUGKS-G-AC.}\label{volume_evolution}
\end{figure}

To evaluate the performance of the central moment DUGKS for tracking interface, a reversed single vortex evolving in a shear flow is simulated.
For the sake of discussion, the DUGKS method with Eq.~(\ref{eq:rawmoments_continuum_AC}) for ACE is denoted by DUGKS-G-AC while the DUGKS method with Eq.~(\ref{eq:dugksWithoutforce}) for ACE is denoted by DUGK-AC.
Initially, a circular droplet with radius of $R=0.15L_0$ is placed at $(0.5L_0, 0.75L_0)$ in a square computational domain $L_0\times L_0$. The solenoidal velocity field is given as follows~\cite{zhang2019fractional}
\begin{equation}\label{eq}
\begin{aligned}
u(x,y,t)=&U_0\sin^2\left(\frac{\pi x}{L_0}\right)\sin\left(\frac{2\pi y}{L_0}\right)\cos\left(\frac{\pi t}{T}\right), \\
v(x,y,t)=&-U_0 \sin^2\left(\frac{\pi y}{L_0}\right)\sin\left(\frac{2\pi x}{L_0}\right)\cos\left(\frac{\pi t}{T}\right),
\end{aligned}
\end{equation}
where $u$ and $v$ are the horizontal and vertical components of velocity and $T=nL_0/U_0$ is the period.
The prescribed velocity field can produce strong vorticity at the interface and stretch and tear the interface.
The term $\cos(\pi t/T)$ renders the flow reversible at $t=T/2$, such that the circle is fully stretched at $T/2$ and comes back to its initial state at $T$.
The parameters are set as $L_0=400$, $W=5$, $U_0=0.04$ and $n=8$. The mobility is determined by the Peclet number defined as $\text{Pe}=U_0L_0/M$ and $\text{Pe}=100$.
The periodic boundary conditions are applied at all sides. As the velocity field is specified,  the hydrodynamic equations are not needed in the current simulations.
Fig.~\ref{deformation}(a) shows the reconstructed interface of the circle by DUGKS-G-AC and DUGKS-AC.
As expected, at $t=T/2$, the circle  is stretched into a thin filament that spirals toward the vortex center and small droplets of several grid sizes at the thin end are formed. At $t=T$, the circle returns back to its initial shape but with some wiggles. These behaviors are also found in the previous literature~\cite{zhang2019fractional}.
By comparison, the results given by both DUGKS methods are nearly identical at $\text{Pe}=100$.
As the Peclet number has an important effect on the interface shape, a reversed single vortex with a larger Peclet number is simulated. The results are shown in Fig.~\ref{deformation}(b).
In such case, the shapes of the calculated interface are still similar.
Furthermore, the time evolution of the  volume of the circle normalized by its initial volume is calculated. The results given by DUGKS-G-AC are presented in Fig.~\ref{volume_evolution}.
The results given by DUGKS-AC that are not shown here are the same as  Fig.~\ref{volume_evolution}.
It is observed that the volume loss calculated at $t=T$ increases as the Peclet number increases.
For quantitative comparison, the relative errors and volume loss  for a wide range of Pelect numbers at $t=T$  are calculated by
\begin{equation}\label{eq}
E_{\phi}=\sqrt{\frac{|\phi(T)-\phi(0)|^2}{|\phi(0)|}},\qquad
E_{vol}=\left|\frac{\phi(T)|_{\phi>0.5}-\phi(0)|_{\phi>0}}{\phi(0)|_{\phi>0}}\right|.
\end{equation}
The relative errors are shown in Table.~\ref{tab: Pe-comparison}.
As can be seen, the relative errors given by DUGKS-AC are slightly more accurate than the DUGKS-G-AC.
However, there are no  significant difference between them for each Pe.
The  loss of volume of the circle is given in Table.~\ref{tab: Pe_volume_loss}.
The values of volume loss for both DUGKS methods are almost identical.   In what follows, the DUGKS-AC is employed to track the interface to save computing cost.

\begin{table}[!htb]
\centering
\caption{Relative  errors of the order parameter  for Reversed single vortex  at various $\text{Pe}$.} \label{tab: Pe-comparison}
\setlength{\tabcolsep}{0.5mm}{%
\begin{tabular}{cccccccccccc}
\hline
Pe                & 50  &   & 100     &  &200     &  &400   &    &800    & &1600     \\
\hline
DUGKS-G-AC  & 0.0410   &  & 0.0635    &  &0.1155  &  &0.1706  &    &0.2065 & &0.2149   \\
DUGKS-AC  & 0.0373   & & 0.0643   &  &0.1150   &  &0.1699  &    &0.2059  & &0.2141 \\
\hline
\end{tabular}}
\end{table}

\begin{table}[!htb]
\centering
\caption{ Volume loss of the circle for Reversed single vortex  at various $\text{Pe}$.} \label{tab: Pe_volume_loss}
\setlength{\tabcolsep}{0.5mm}{%
\begin{tabular}{cccccccccccc}
\hline
Pe                & 50  &   & 100     &  &200     &  &400   &    &800    & &1600     \\
\hline
DUGKS-G-AC  &  0.0061  &  & 0.0194 &  &0.0474  &  &0.0837 &    &0.1099 & &0.1127  \\
DUGKS-AC & 0.0063  & & 0.0196 &  &0.0475  &  &0.0837 &    &0.1093 & &0.1117  \\
\hline
\end{tabular}}
\end{table}

\subsection{Static droplet}
\begin{figure}
  \centering
 \subfigure[]{
  \includegraphics[width=0.5\textwidth]{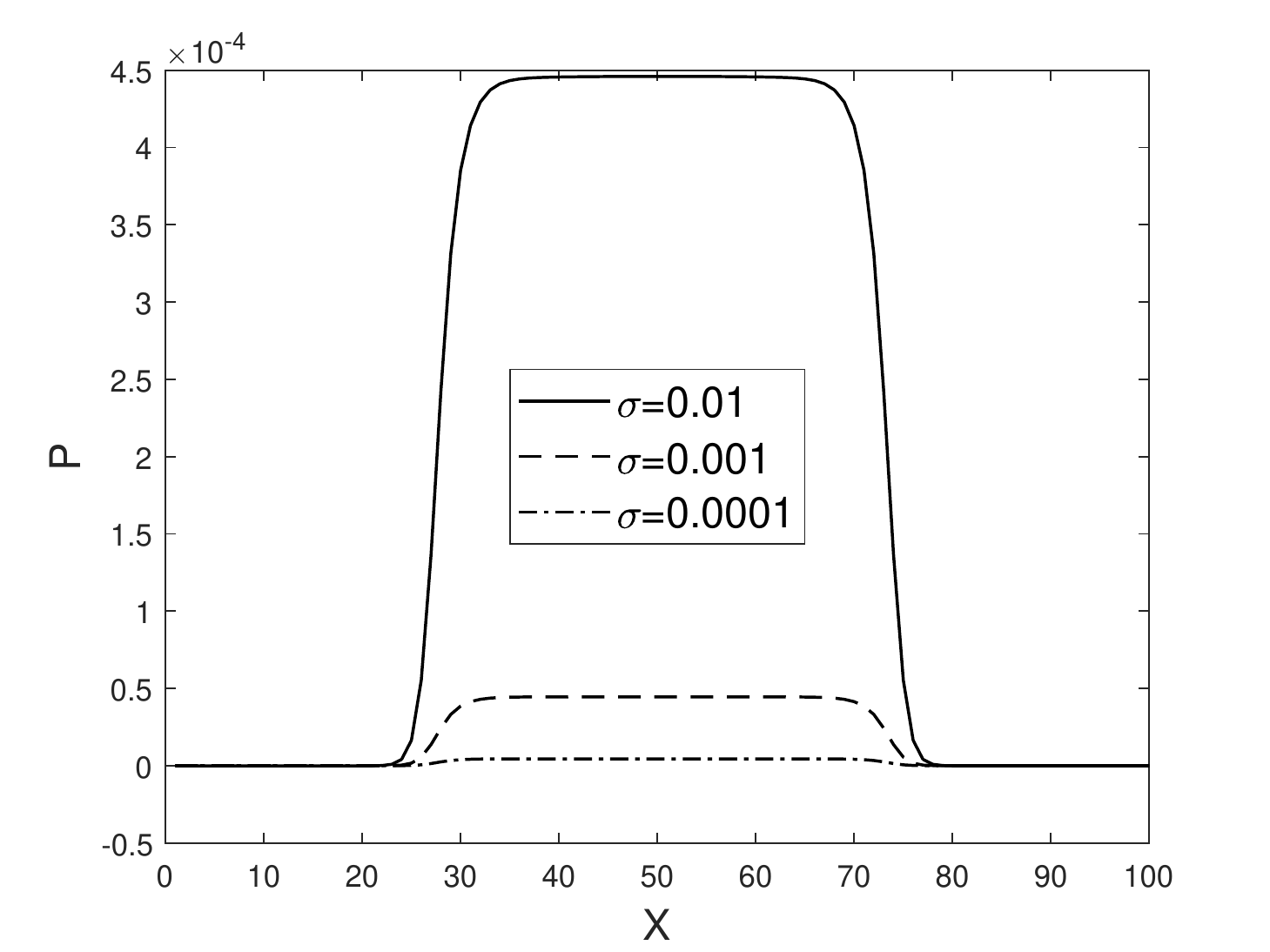}}~
 \subfigure[]{
  \includegraphics[width=0.5\textwidth]{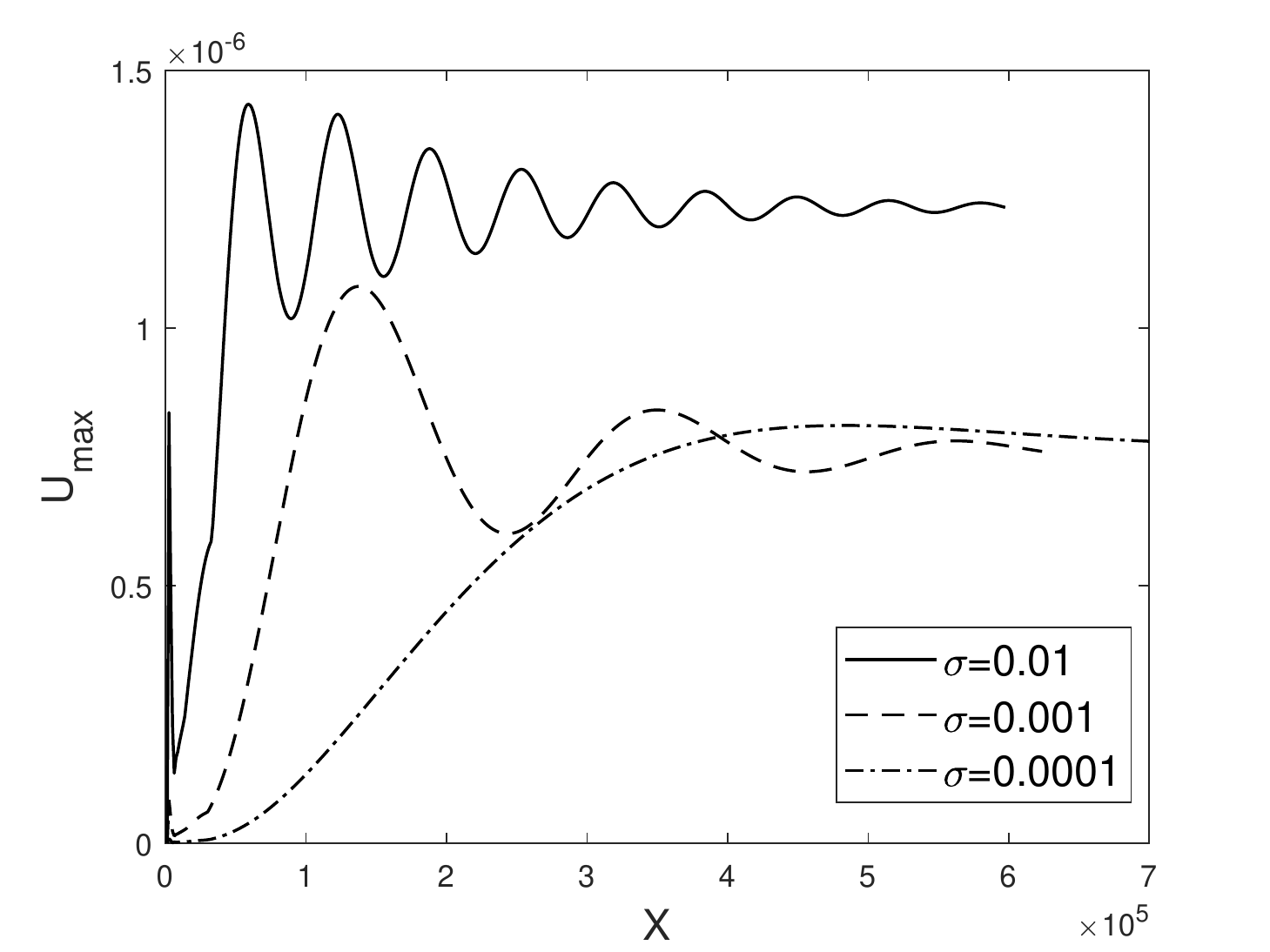}}
  \caption{ (a) The equilibrium pressure profile across the horizontal center of the droplet and (b) the evolution of the maximum velocity with time at $\rho_h/\rho_l=1000$, $\mu_h/\mu_l=100$ and $R=24$.}\label{test1:pressure_Velocity}
\end{figure}

\begin{figure}
  \centering
 \subfigure[]{
  \includegraphics[width=0.5\textwidth]{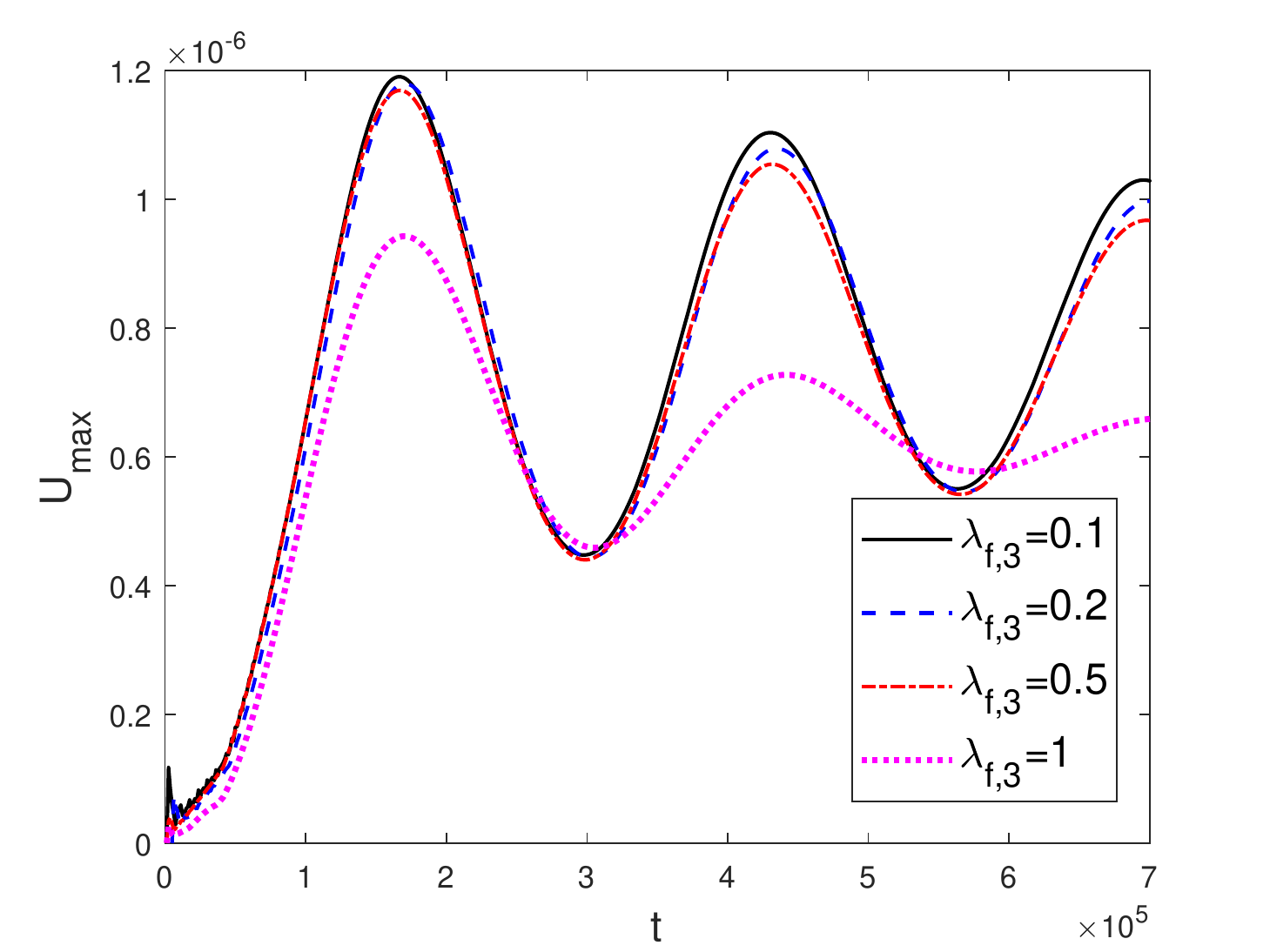}}~
 \subfigure[]{
  \includegraphics[width=0.5\textwidth]{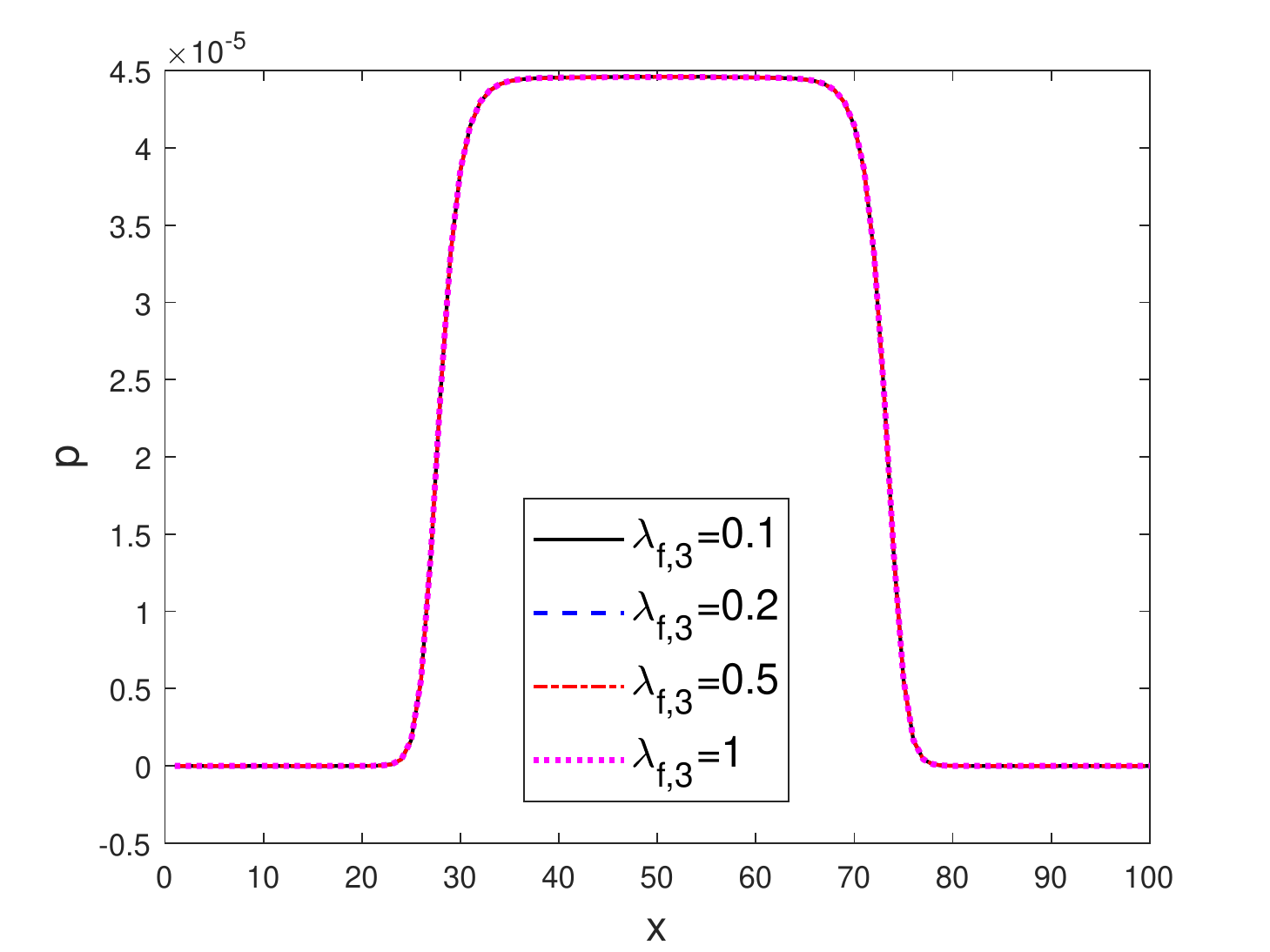}}
  \caption{ Effect of the bulk viscosity on (a) the maximum spurious velocity and   (b) the equilibrium pressure profile across the horizontal center of the droplet at $\sigma=0.001$, $R=24$.}\label{test1:bulk_viscosity}
\end{figure}

In this test, a two-dimensional static droplet is  simulated to validate Laplace's law. Initially, a circular droplet of radius $R$ is placed at the central of a $L\times L$ domain.
The gravity is neglected.
The periodic boundary condition is applied to all sides.
Both the pressure and velocity fields are initialized as zero. The order parameter is initialized as follows
\begin{equation}\label{eq}
\phi(x,y)=\frac{\phi_h+\phi_l}{2}-\frac{\phi_h+\phi_l}{2}\tanh\left(\frac{2(R-\sqrt{(x-x_c)^2+(y-y_c^2)})}{W} \right).
\end{equation}
where  $(x_c,y_c)=(0.5L,0.5L)$ is the central position of the computational domain.
In simulations, the densities are set as $\rho_h=1000, \rho_l=1$, $\mu_h=100\mu_l$, $L=100$,  $\text{CFL}=0.5\sqrt{2}$. The dynamic viscosity of the droplet is determined by the Laplace number defined as $\text{La}=\rho_h 2R\sigma/\mu_h^2=12000$.
Since $\sigma$ and $R$ are input parameters, the analytic value of $\Delta p_{exact}$ can be obtained from Laplace law ,i.e, $\Delta p_{exact}=\sigma/R$. When the relative error $|\phi_{t}-\phi_{t-1000\Delta t}|/|\Delta \phi_{t-1000\Delta t}|$ becomes less than $1.0\times 10^{-8}$, the solution is assumed converged. Then, the relative error of pressure is calculated by $|\Delta p_{exact}-\Delta p_{num}|/|\Delta p_{exact}|\times 100$.
Table~\ref{tab:test1-laplace} shows the relative error of the measured pressure difference to the analytic pressure difference with different $\sigma$ and $R$.
The maximum error is $8.9\%$. When the system is at equilibrium, the pressure profile across the horizontal center line at $R=24$ is plotted in Fig.~\ref{test1:pressure_Velocity}(a).
It can be seen that the pressure inside the droplet is higher than the one outside the droplet, and
the pressure continuously varies across the interface, which is consistent with the Laplace law.
It is well-known that spurious velocities in the vicinity of the interface  always appear in multiphase simulation.
A large spurious velocity could lead to  unphysical movement of the interfaces, even numerical instability. To examine this, the time evolution of the maximum velocity calculated by $U_{max}=|\bm u|_{max}$ in the whole domain is plotted  in Fig.~\ref{test1:pressure_Velocity}(b).
The maximum magnitude of  spurious currents at equilibrium is on the order of $10^{-6}$ for $\sigma=0.01$ and $10^{-7}$ for $\sigma=0.001$ and $0.0001$. In the previous results of multiphase simulations with large density ratios~\cite{liang2018phase,lee2005stable,li2013lattice,xu2015three}, the maximum magnitudes of the spurious velocities has the order of $10^{-2}$  in~\cite{li2013lattice,xu2015three} and $10^{-6}$ in~\cite{lee2005stable,yang2019phase}. In this study, the magnitude of the spurious velocities is relatively small and has  little effect on the interface movement in most simulations.
Finally, we examine the effect of the bulk viscosity on the pressure and velocity fields. The maximum velocity and the equilibrium pressure profiles across the horizontal center line at different bulk viscosity coefficients are shown in Fig.~\ref{test1:bulk_viscosity}.
As seen in Fig.~\ref{test1:bulk_viscosity}, increasing the bulk viscosity decreases the amplitude of the spurious velocity. while has little effect on the pressure.
We also found that a smaller or larger bulk viscosity can cause numerical instability. Generally,  numerical stability can be maintained when $\lambda_{f,3}$ is between $0.1$ and $2$.

\begin{table}[!htb]
\centering
\caption{The relative error ($\%$) of pressure difference between the inside and outside of the droplet at various $\sigma$ and $R$.} \label{tab:test1-laplace}
\setlength{\tabcolsep}{2mm}{%
\begin{tabular}{cccccccc}
\hline
$\sigma$   & R=20&      & R=24 &   & R=28 &  & R=32   \\
\hline
0.01     &  8.90   &    &7.08     &   &5.88     &    &5.04             \\
0.001    &  8.80   &    &7.00     &   &5.85     &    &5.00           \\
0.0001   &  7.71   &    &6.47     &   &5.54     &    &4.83           \\
\hline
\end{tabular}}
\end{table}

\subsection{Oscillating drops}
\begin{figure}[htp]
\centering
\subfigure[$\sigma=0.01$]{\includegraphics[width=0.5\textwidth,trim=10 20 10 10,clip]{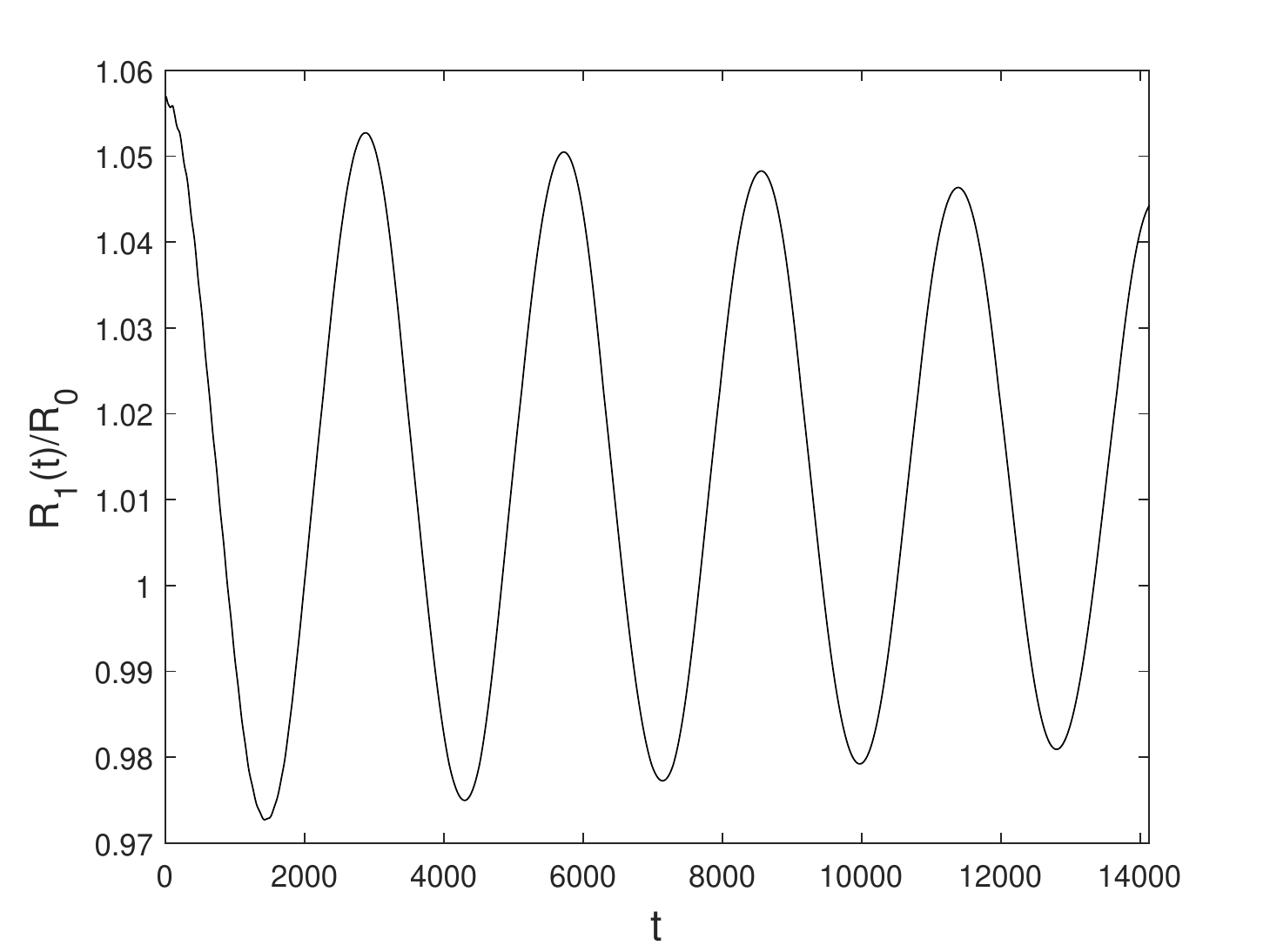}}~
\subfigure[$\sigma=0.001$]{\includegraphics[width=0.5\textwidth,trim=10 20 10 10,clip]{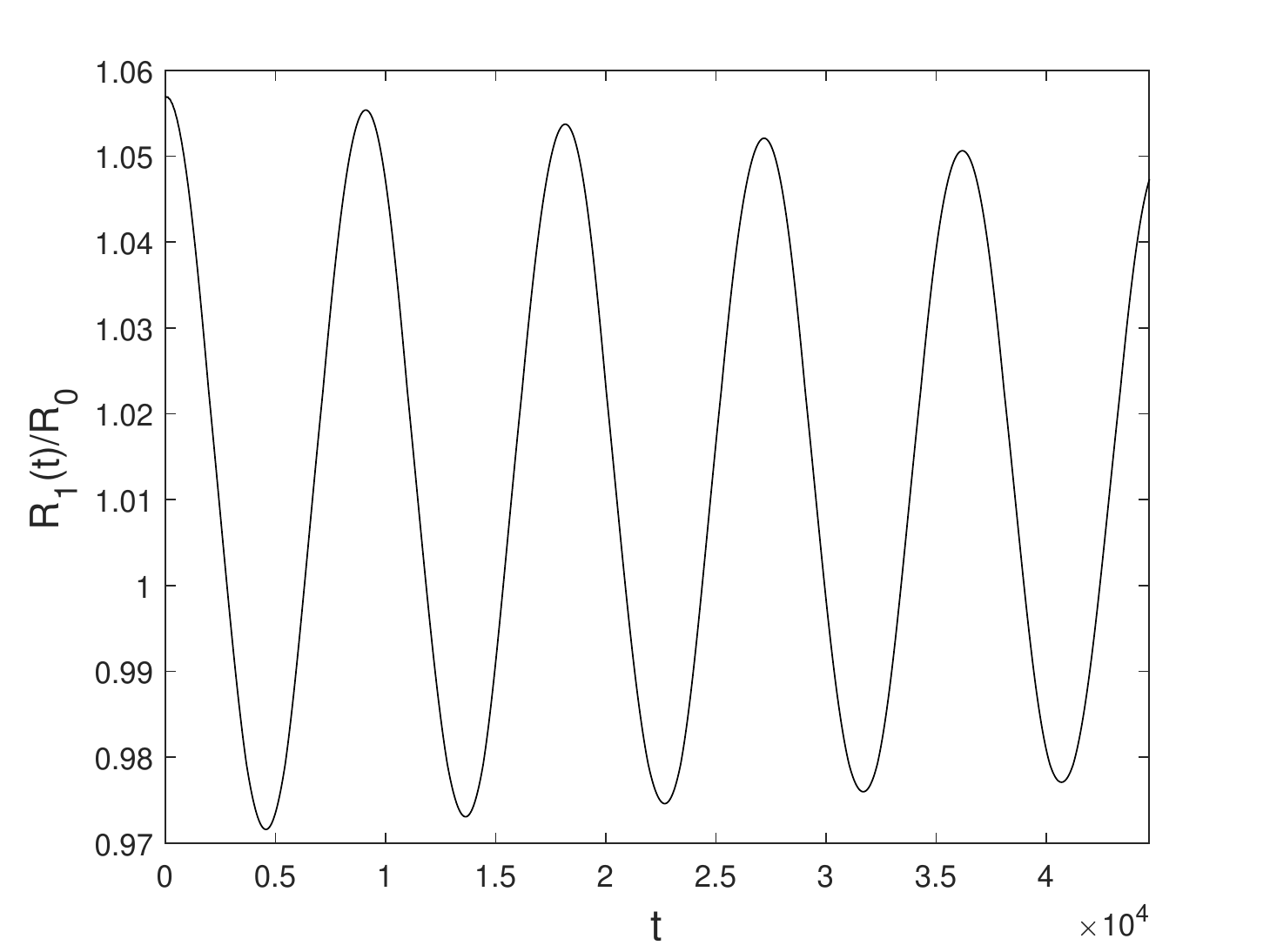}}\\
\subfigure[$\sigma=0.0001$]{\includegraphics[width=0.5\textwidth,trim=10 20 10 10,clip]{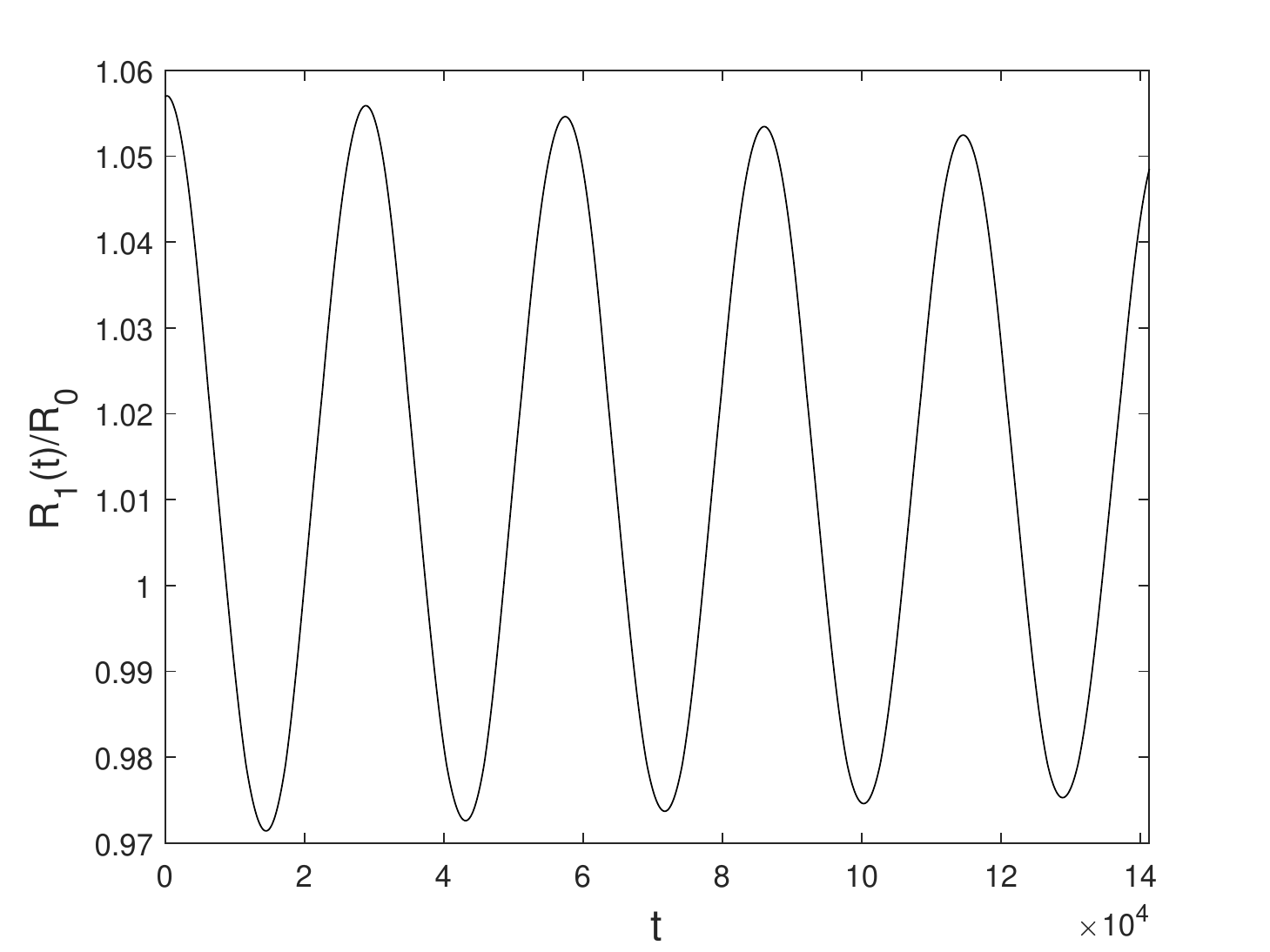}}
\caption{Oscillation of an elliptical droplet at different viscosities,
$\rho_l/\rho_g=1000$ and $\mu_l/\mu_g=40$. }
\label{test2:oscilation_periodic}
\end{figure}

Oscillating drop is  a classical benchmark test for examining the accuracy of the surface tension effect. We consider the same case as in~\cite{mukherjee2007pressure}. Initially, an elliptical droplet is given by
\begin{equation}\label{eq}
\phi(x,y)=\frac{\phi_h+\phi_l}{2}+\frac{\phi_h-\phi_l}{2}\tanh\left(\frac{2R_0(1-\sqrt{(x-x_c)^2/R_1^2+(y-y_c)^2/R_2^2})}{W} \right)
\end{equation}
where $R_1$ and $R_2$  are the major radius and minor radius respectively, and $R_0=\sqrt{R_1R_2}$ is the equilibrium droplet radius.
According to Lamb~\cite{lamb1932hydrodynamics}, the oscillation period for a two-dimensional droplet is given as follows:
\begin{equation}\label{eq:Oscilation}
T_a=2\pi\left[n(n^2-1)\frac{\sigma}{\rho_h R_0^3}\right]^{-1/2},
\end{equation}
where $n=2$ denotes the mode of oscillation. In simulations,
the parameters are set as $R_1=24$, $R_2=22$, $\rho_h=1$, $\rho_l=0.001$, $\mu_h/\mu_l=100$ and $\nu_h=1/3\times 10^{4}$.
Three different surface tension coefficients are considered: $\sigma=0.01,0.001$ and $0.0001$. The periodic boundary conditions are applied at all boundaries.
The time evolution of the major radius is shown in Fig.~\ref{test2:oscilation_periodic},
As seen, the oscillation time period increases as the surface tension coefficient decreases, which is consistent with the theoretical prediction.
The predicted oscillation periods are $2839.7$, $9045.3$ and $28621$ for $\sigma=0.01, 0.001$ and $0.0001$, respectively.
Based on Eq.~(\ref{eq:Oscilation}), the corresponding relative errors are $0.51\%$, $1.24\%$ and $1.30\%$ respectively. This implies that accurate modeling of the surface tension effect can be achieved by the present model.

\subsection{Rayleigh-Taylor instability}
\begin{figure}
  \centering
\subfigure[$t^*=1$]{\includegraphics[width=0.2\textwidth,trim=150 20 150 20,clip]{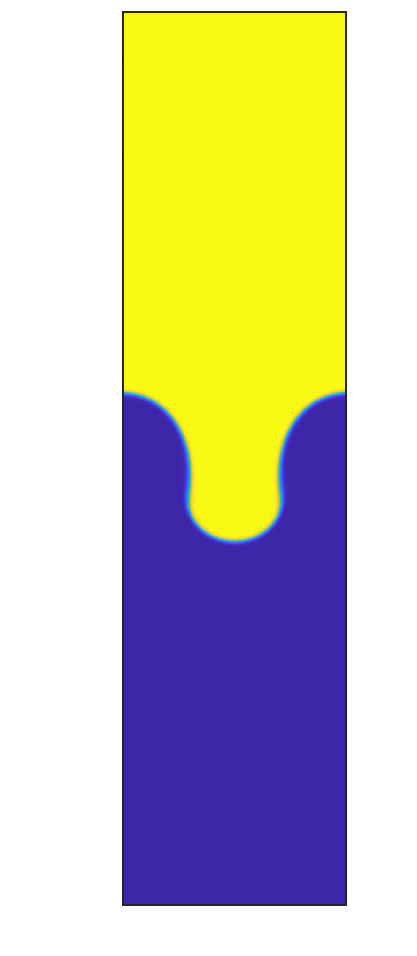}}~
\subfigure[$t^*=1.5$]{\includegraphics[width=0.2\textwidth,trim=150 20 150 20,clip]{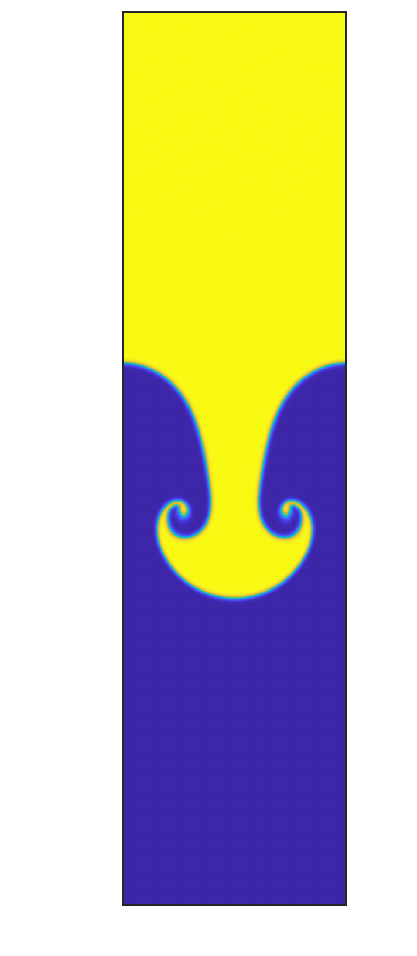}}~
\subfigure[$t^*=2$]{\includegraphics[width=0.2\textwidth,trim=150 20 150 20,clip]{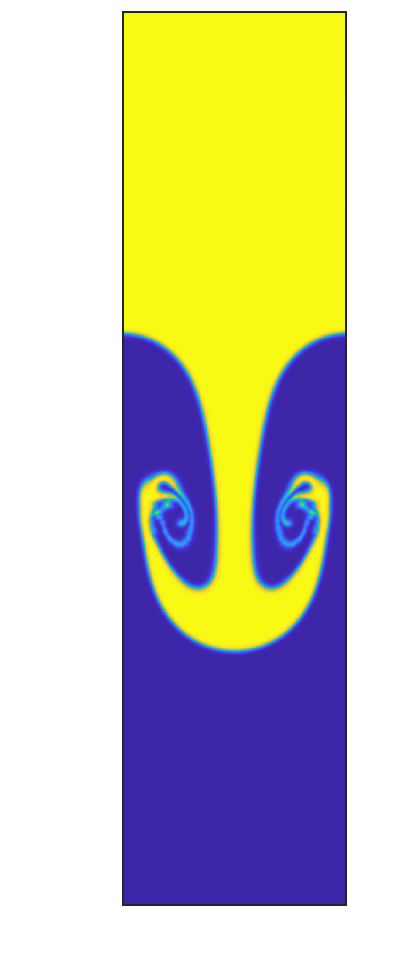}}~
\subfigure[$t^*=2.5$]{\includegraphics[width=0.2\textwidth,trim=150 20 150 20,clip]{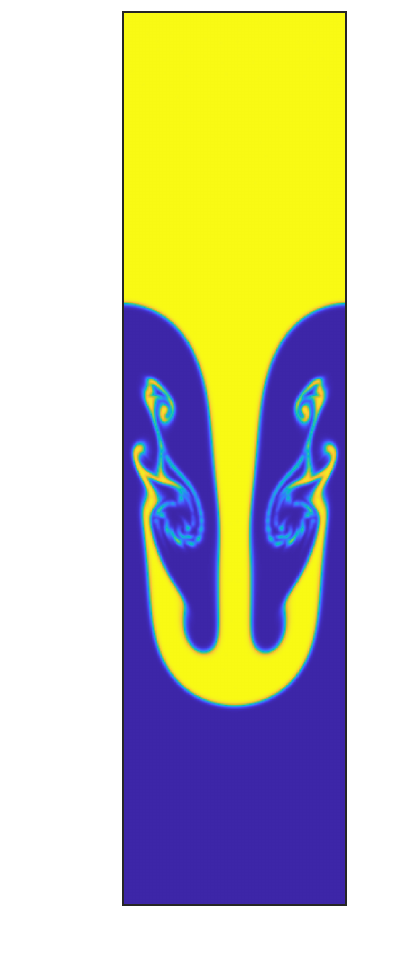}}~
\subfigure[$t^*=3$]{\includegraphics[width=0.2\textwidth,trim=150 20 150 20,clip]{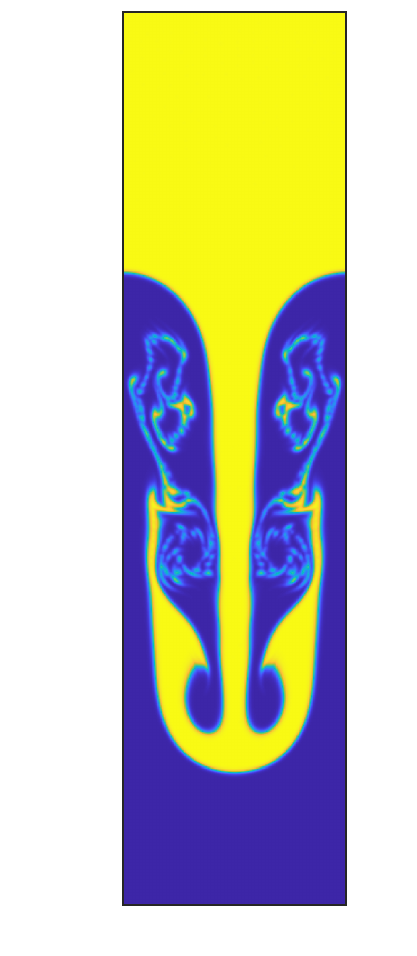}}\\
  \caption{Evolution of the fluid interface for the Rayleigh-Taylor instability at $\text{Re}=3000$, $A_t=0.5$, $\text{Pe}=1000$ and $\mu_h/\mu_g=1$.}\label{test3:At05}
\end{figure}

\begin{figure}[htp]
\centering
\subfigure[]{\includegraphics[width=0.5\textwidth]{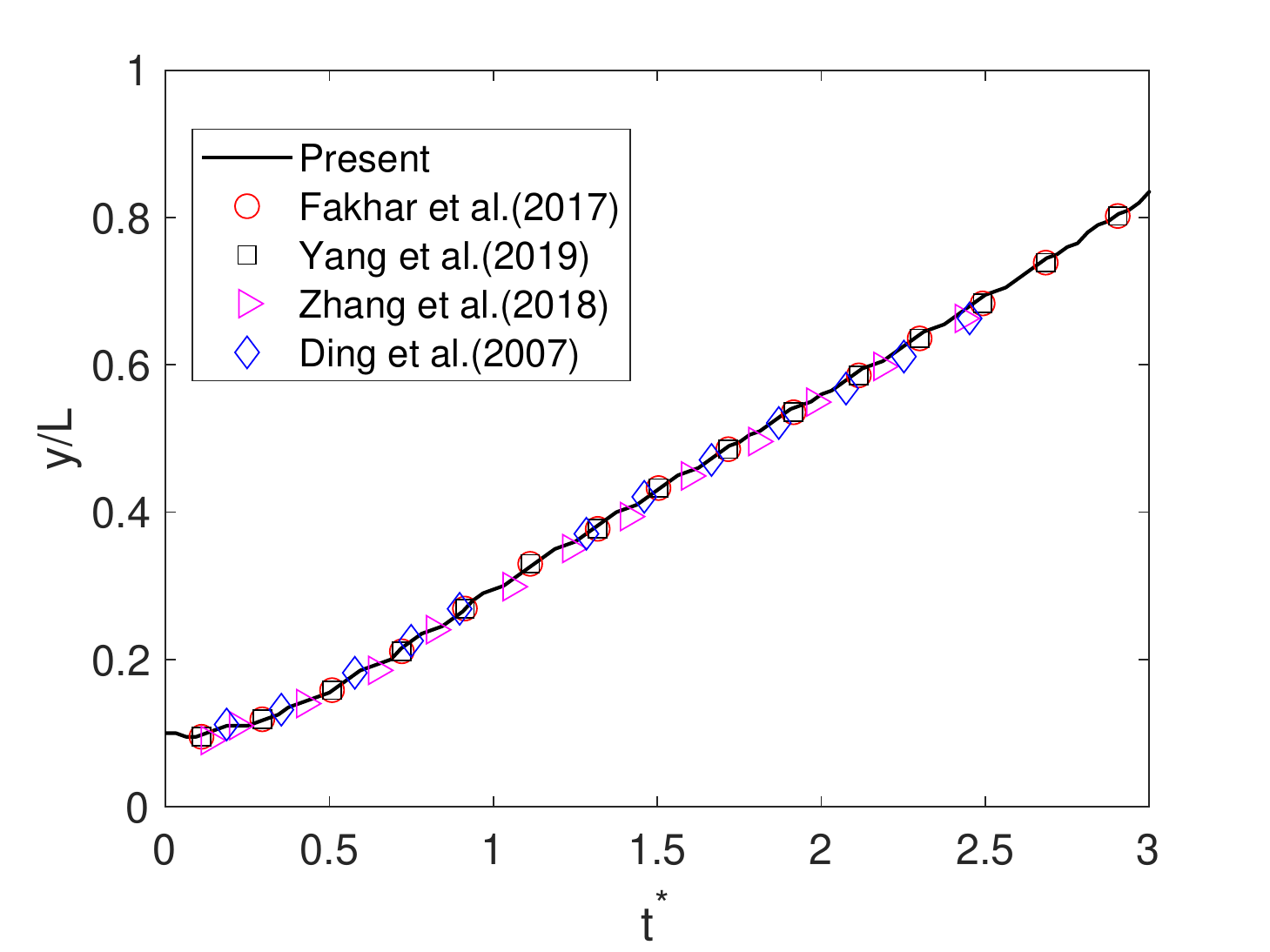} }~
\subfigure[]{\includegraphics[width=0.5\textwidth]{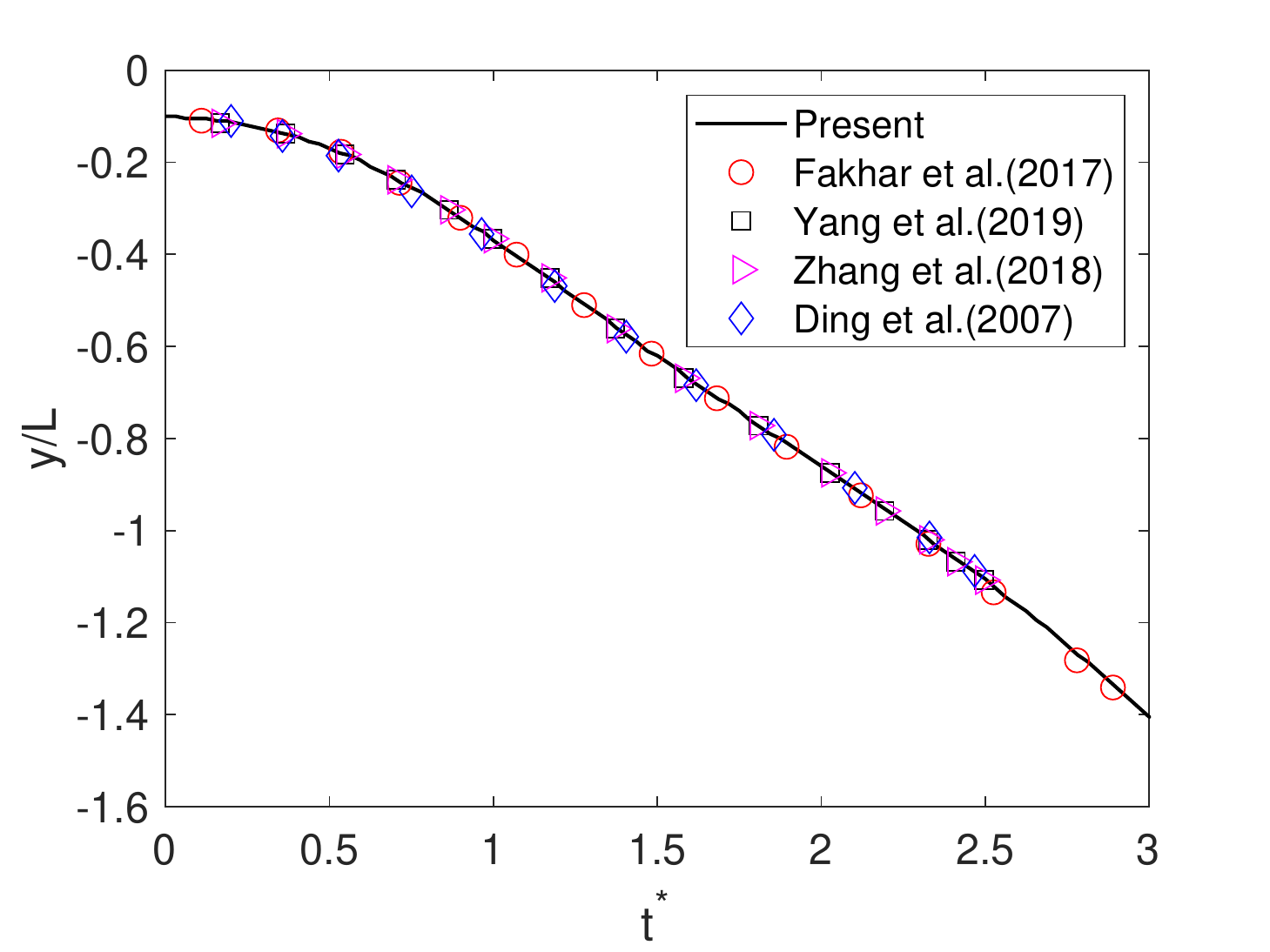}}
\caption{Evolution of (a) bubble front and (b) spike tip positions at $\text{Re}=3000$, $A_t=0.5$, $\text{Pe}=1000$ and $\mu_h/\mu_g=1$. }
\label{test3:At05postion}
\end{figure}

\begin{figure}
  \centering
\subfigure[$t^*=0.5$]{\includegraphics[width=0.25\textwidth,trim=150 20 150 20,clip]{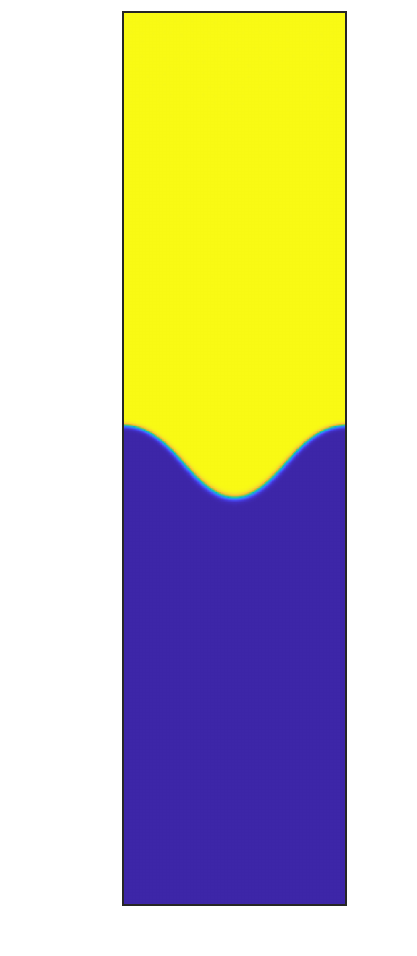}}~
\subfigure[$t^*=1$]{\includegraphics[width=0.25\textwidth,trim=150 20 150 20,clip]{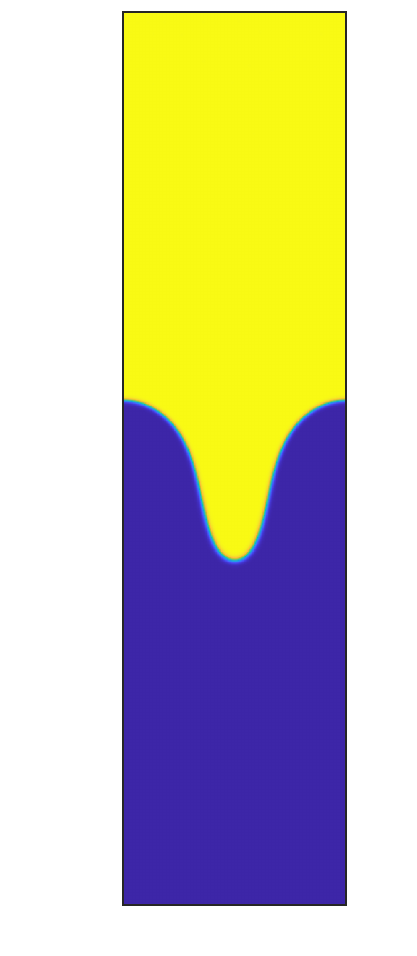}}~
\subfigure[$t^*=1.5$]{\includegraphics[width=0.25\textwidth,trim=150 20 150 20,clip]{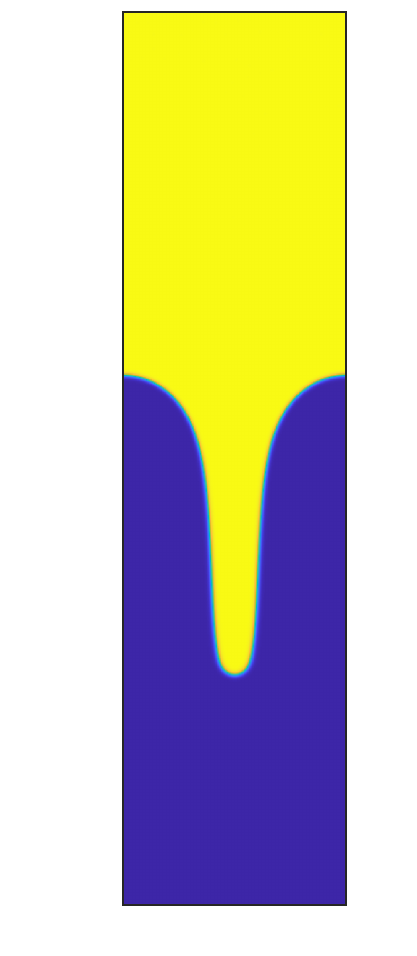}}~
\subfigure[$t^*=2$]{\includegraphics[width=0.25\textwidth,trim=150 20 150 20,clip]{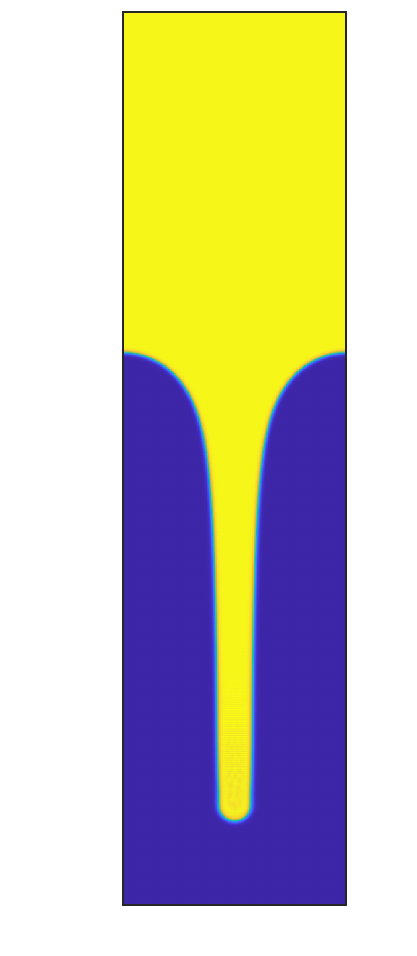}}\\
  \caption{Evolution of the fluid interface for the Rayleigh-Taylor instability at $\text{Re}=3000$, $A_t=0.998$, $\text{Pe}=200$ and $\mu_h/\mu_g=100$.}\label{test3:At0998}
\end{figure}

The Rayleigh-Taylor instability of binary fluids under gravity involves very interesting phenomenon and has been investigated extensively by many researchers~\cite{he1999lattice,zu2013phase,wang2015multiphase,ren2016improved,fakhari2017improved,zhang2018discrete}. 
This problem with high Reynolds number is considered to assess the performance of the  present model. Initially, the both fluids are filled in a rectangular domain of $[0,L]\times[-2L,2L]$.
The heavy fluid is placed on the top of the light one. The interface has a slight perturbation and the order parameter is given by
\begin{equation}\label{eq}
\phi(x,y)=\frac{\phi_h+\phi_l}{2}+\frac{\phi_h-\phi_l}{2}\tanh\left(\frac{2(y-y_0)}{W}\right),\quad
y_0(x)=0.1L\cos\left(\frac{2\pi x}{L}\right),
\end{equation}
The bounce-back boundary conditions are applied to the bottom and top boundaries and periodic boundary conditions are imposed on the lateral boundaries.
The interface behavior is determined by two major dimensionless numbers: the Atwood number $\text{$A_t$}=(\rho_h-\rho_l)/(\rho_h+\rho_l)$ and the Reynolds number $\text{Re}=\rho_h\sqrt{gL}/\mu_h$.
First,  the dimensionless numbers are set as $A_t=0.5$ and $\text{Re}=3000$ for comparison with the previous literature results~\cite{yang2019phase,zhang2018discrete,2017Improved,ding2007diffuse}.
This means that the density ratio is $3$.  The other parameters are set as $L=200$, $\text{CFL}=0.25$ and $\text{Pe}=1000$. The surface tension is given by setting the Weber number $\text{We}=\rho_h L\sqrt{gL}/\sigma=50000$. This produces a small surface tension value so that the surface tension effect can be ignored.
The time evolution of interface shape is shown in Fig.~\ref{test3:At05}.
It can be seen that the heavy fluid symmetrically penetrates the lighter fluid, and two counter-rotating vortices behind the heavy liquid front are generated, which are similar to the previous studies~\cite{zhang2018discrete,2017Improved}.
For the quantitative evaluation of the results,
the time evolutions of both the bubble front and spike tip are shown in Fig.~\ref{test3:At05postion}. It is clear that the results of the present method are in  good agreement with  those from the previous studies~\cite{zhang2018discrete,2017Improved,ding2007diffuse}.
Then, we consider a case with a large Atwood number $A_t=0.998$ and $Re=3000$, which means the density ratio is $1000$. The Peclet number is set as $\text{Pe}=200$. The predicted interface shapes are shown in Fig.~\ref{test3:At0998}.  The interface shapes given by the present model  agree very well with those of Fakhari et al.~\cite{2017Improved}. In contrast, the previous DUGKS model proposed by Yang et al.~\cite{2019Phase} fails to simulate such problem with high Reynolds number and large density ratio.

\subsection{Bubble rising}
\begin{figure}
\centering
 \includegraphics[width=0.5\textwidth,trim=0 0 0 10,clip]{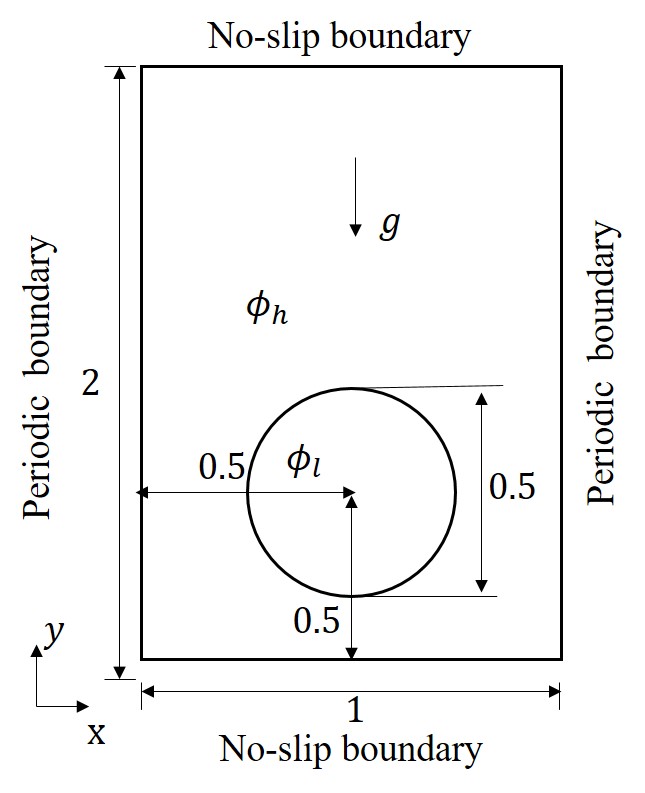}
\caption{Initial configuration for the rising bubble.}
\label{fig:Rising_initial}
\end{figure}

\begin{figure}[htp]
\centering
\subfigure[]{\includegraphics[width=0.5\textwidth,trim=10 0 10 20,clip]{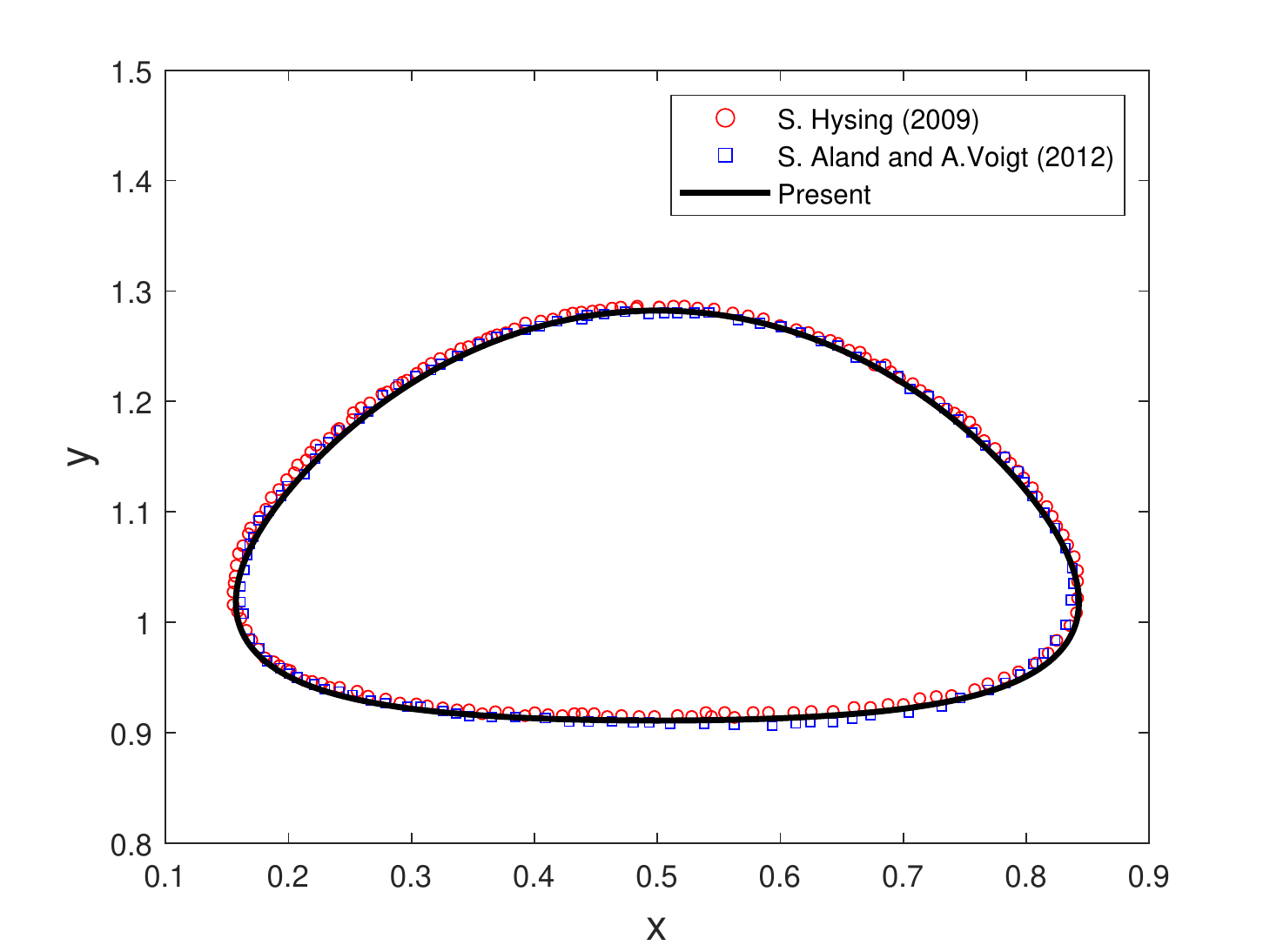} }~
\subfigure[]{\includegraphics[width=0.5\textwidth,trim=10 0 10 20,clip]{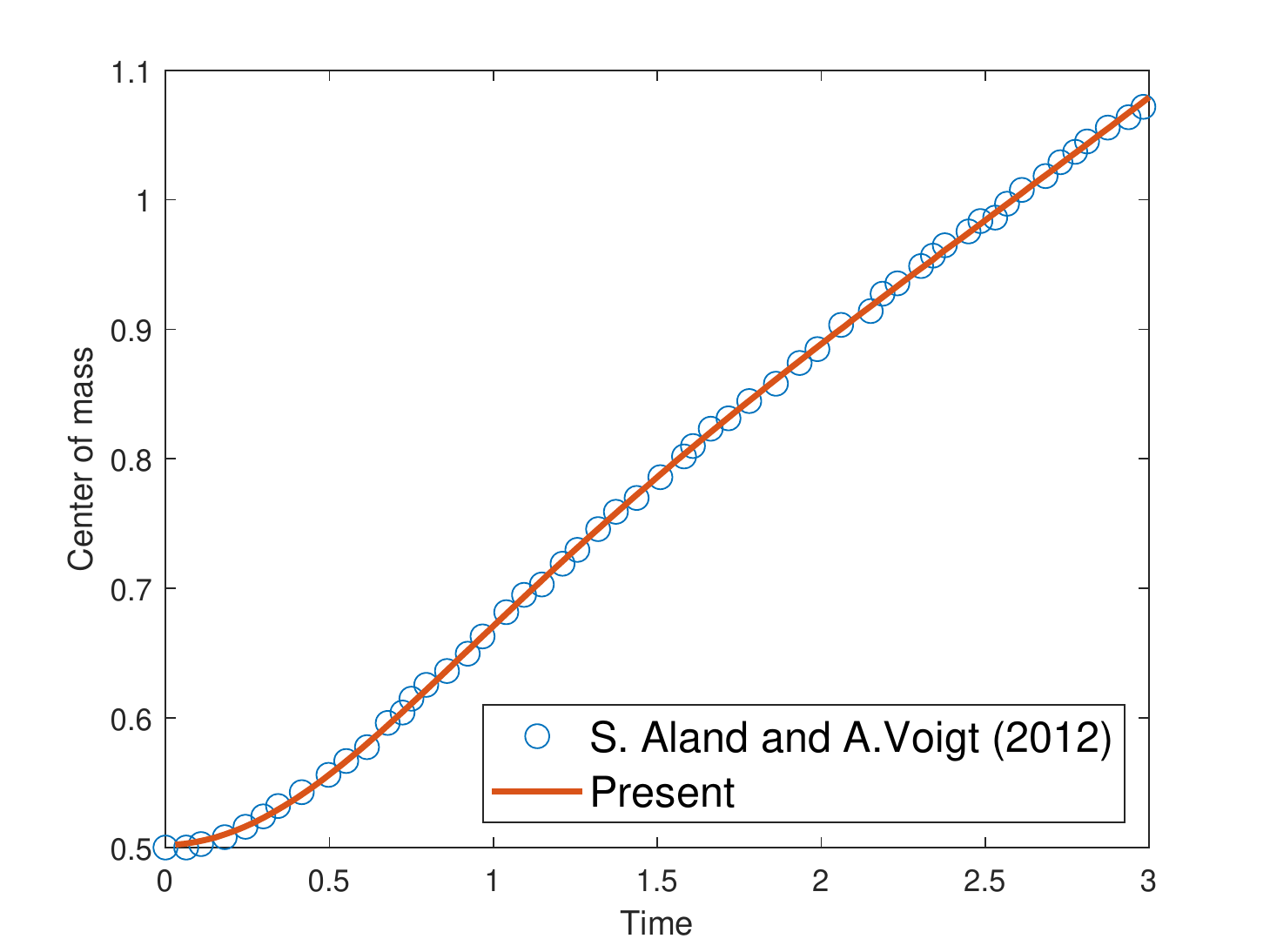}}
\subfigure[]{\includegraphics[width=0.5\textwidth,trim=10 0 10 20,clip]{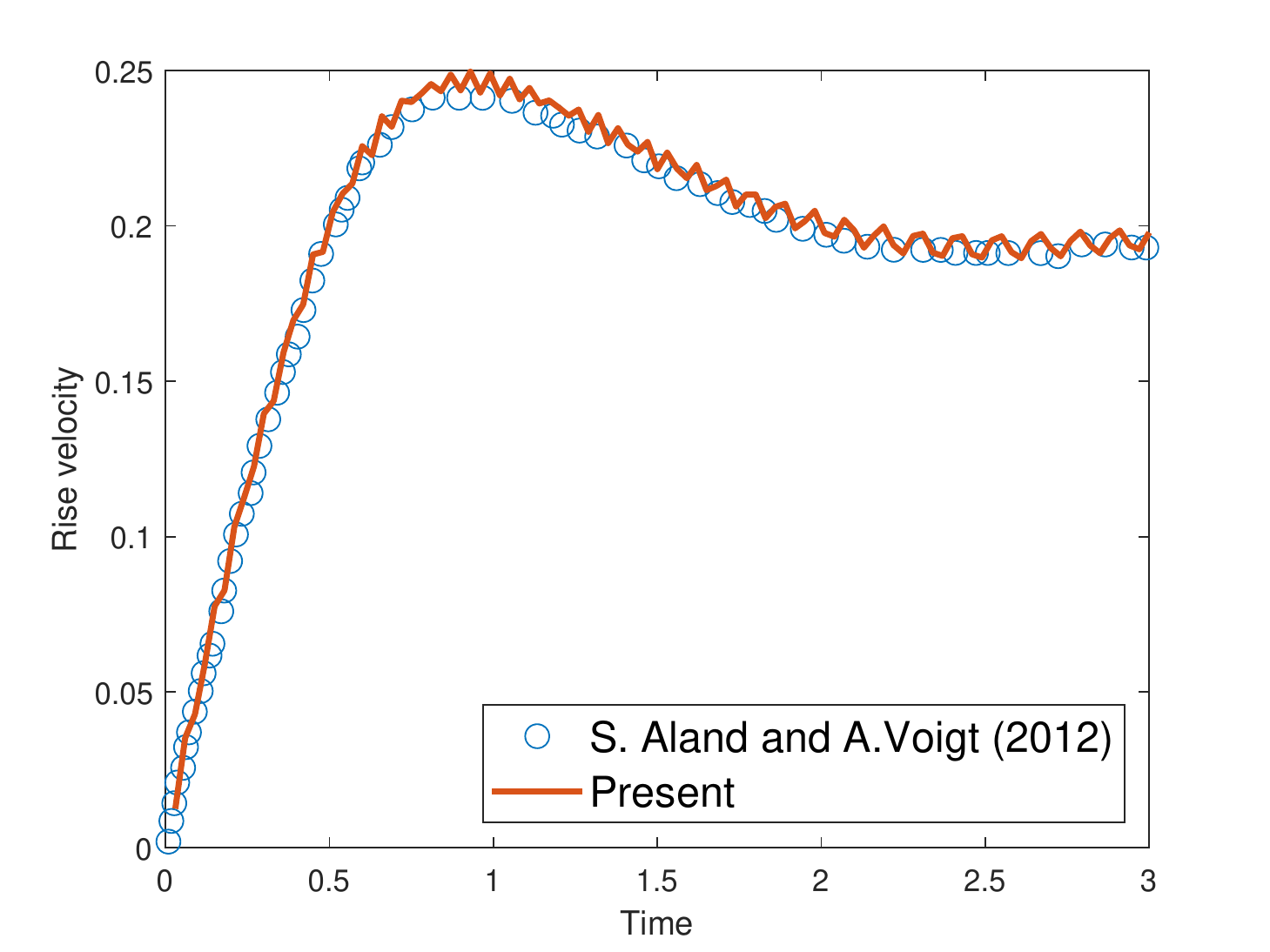}}
\caption{ Quantitative comparison of calculated results with numerical data from~\cite{hysing2009quantitative} for case 1. (a) Bubble shapes at  $t=3s$, (b) time evolution of  center of mass of the bubble and  (c)  time evolution of mean rising velocity of the bubble.}
\label{test4:case1}
\end{figure}

\begin{figure}[htp]
\centering
\subfigure[]{\includegraphics[width=0.5\textwidth,trim=10 0 10 20,clip]{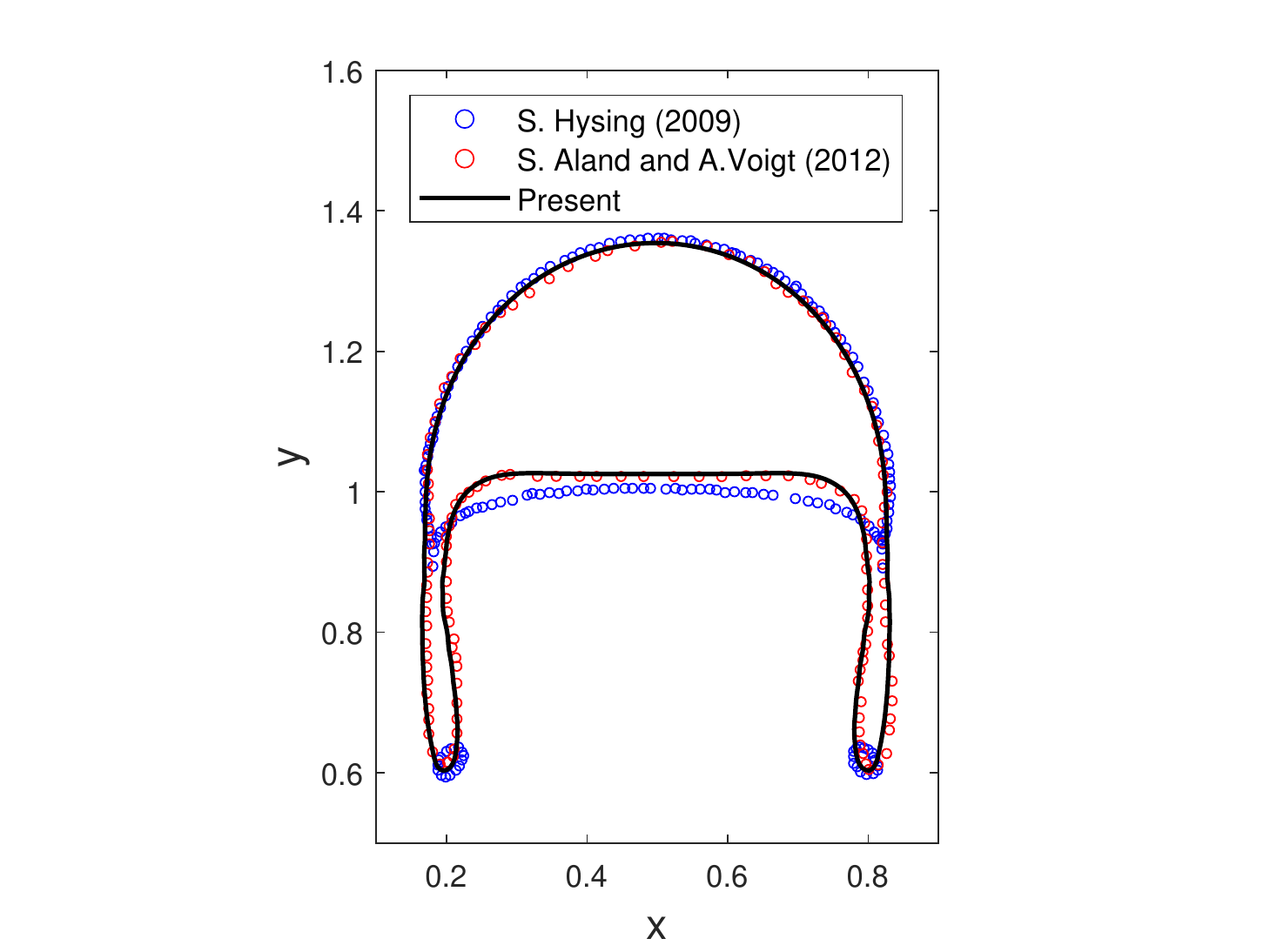} }~
\subfigure[]{\includegraphics[width=0.5\textwidth,trim=10 0 10 20,clip]{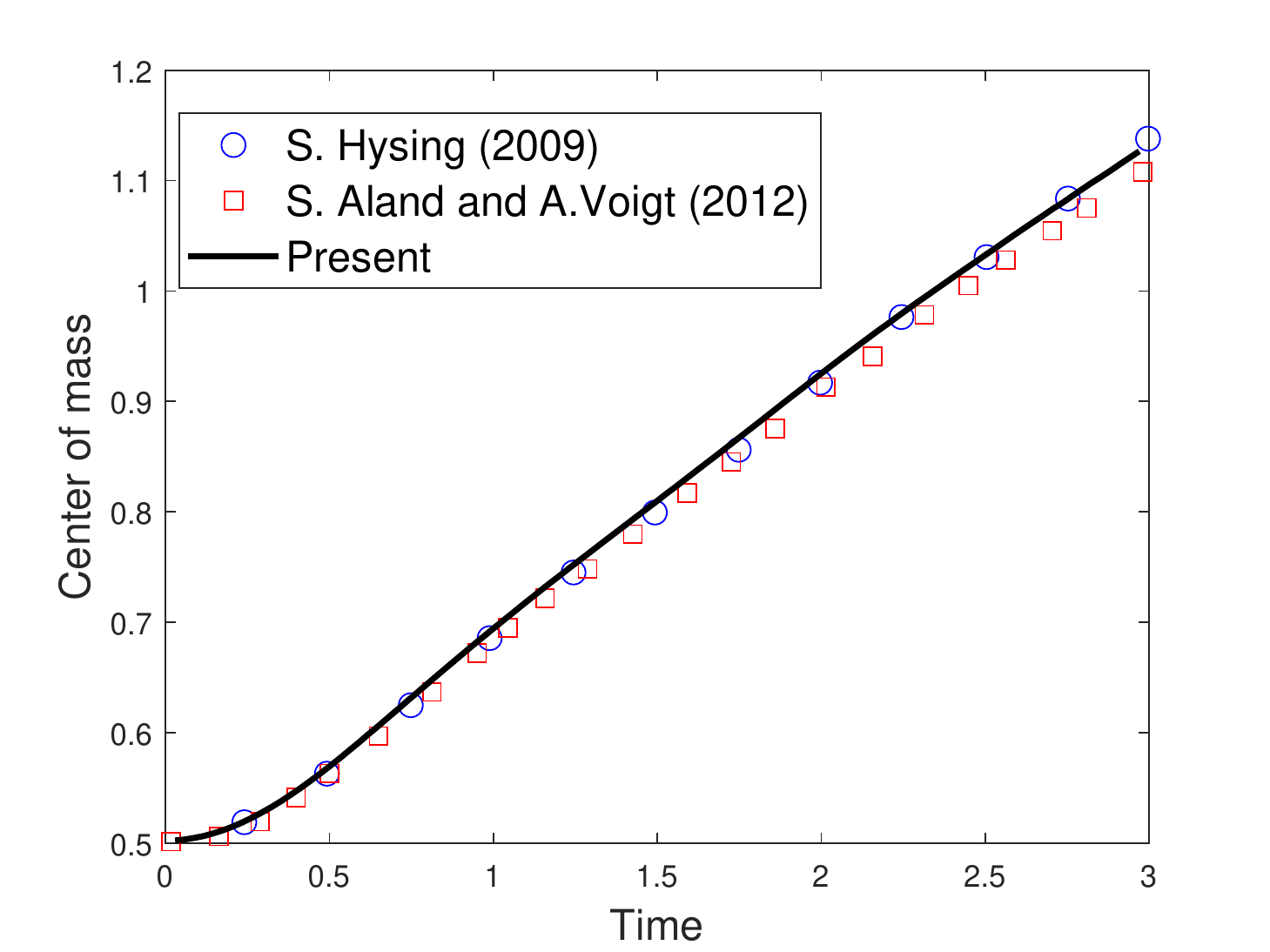}}
\subfigure[]{\includegraphics[width=0.5\textwidth,trim=10 0 10 20,clip]{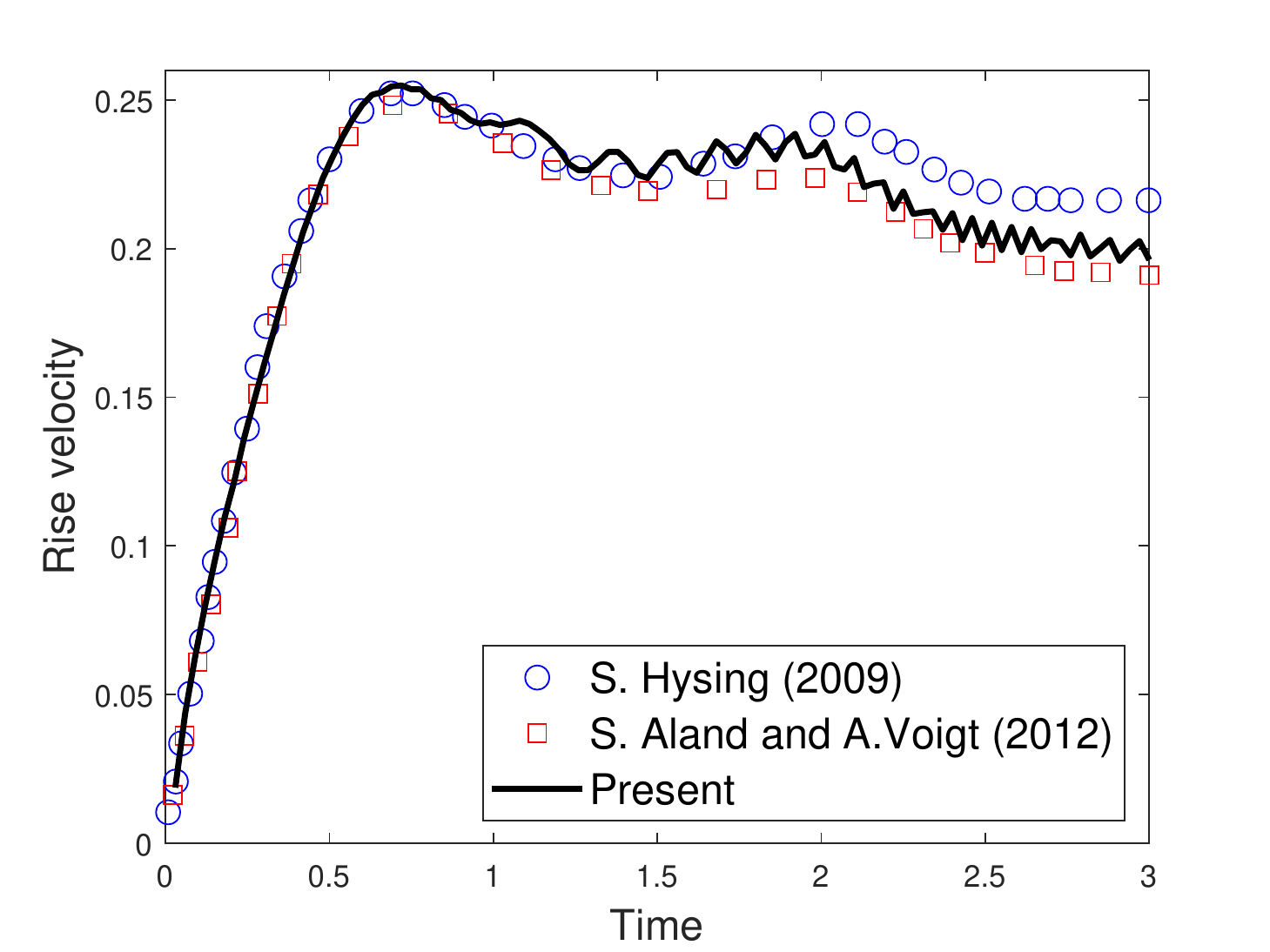}}
\caption{Quantitative comparison of calculated results with numerical data from ~\cite{hysing2009quantitative} and ~\cite{aland2012benchmark} for case 2. (a) Bubble shapes at  $t=3s$, (b) time evolution of  center of mass of the bubble and  (c)  time evolution of mean rising velocity of the bubble.}
\label{test4:case2}
\end{figure}

In this subsection, a single bubble rising in a quiescent water channel is considered. For such problem, numerical benchmark data have been provided by the finite element method together with level set and Cahn-Hilliard for interface tracking~\cite{aland2012benchmark,hysing2009quantitative}.
The sketch of this problem is shown in Figure~\ref{fig:Rising_initial}. The computational domain is $L\times 2L$.
Initially, a bubble of diameter $D$ is placed at the location of $(D,D)$. Periodic condition is applied at the left and right boundaries and no-slip boundary conation is used at the upper and lower walls.
Two key nondimensional parameters including the Reynolds number and Eotvos number can be defined as
\begin{equation}\label{eq}
\text{Re}=\frac{\rho_h D\sqrt{gD}}{\mu_h}, \quad  \text{Eo}=\frac{\rho_h g D^2}{\sigma},
\end{equation}
In order to quantify the dynamics of the bubble during rising, the rising velocity and center of mass of the bubble are measured by
\begin{equation}\label{eq}
y_c=\frac{\int_{\phi\leq 0.5(\phi_h+\phi_l)}\phi d\bm x}{\int_{\phi\leq0.5(\phi_h+\phi_l)}1 d\bm x}, \quad u_c=\frac{\int_{\phi\leq 0.5(\phi_h+\phi_l)} u_y(\bm x) d\bm x}{\int_{\phi\leq 0.5(\phi_h+\phi_l)}1 d\bm x}
\end{equation}
where $u_y$ is the velocity component in the vertical direction.
Following the same parameters in ~\cite{aland2012benchmark,hysing2009quantitative}, the fluid and physical parameters are listed in Table~\ref{tab:risingbubble}.
In the simulations, the parameters are set as $L=240$, $U_0=0.001944$ and $\text{Pe}=40$. The surface tension and viscosity are determined by $\text{Eo}$ and $\text{Re}$.
In case 1, the surface tension effect is strong such that the bubble will be less deformation as it rises.
Fig.~\ref{test4:case1} shows  the interface shape, center of mass and the mean rising velocity of the bubble for case 1. The benchmark results for case1 are plotted as well.
It can be observed that the bubble shapes and central of mass from the present method agree very well with the references. The calculated rise velocity slightly oscillates around a mean value. This oscillation may be caused by the compressibility effects where pressure waves travel with the speed proportional to the sound speed. Apart from this artifact, the mean rising velocity still agrees well with the benchmark data.
In case 2, the effect of the surface tension is lower such that the bubble shall produce severe deformations.
Fig.~\ref{test4:case2} shows   the interface shape, center of mass and the mean rising velocity of the bubble for case 2. The bubble shapes at $t=3s$ from the proposed method match the benchmark data with minor differences in the tails of the bubble.
This differences are largely caused by the grid resolution. The reference results are obtained based on a sharp interface method while the present results are obtained by the diffusion interface method where an appropriate thickness of the interface is required.  However, good agreement with the benchmark results can be observed in terms of the central of mass  and the mean rising velocity of the bubble. These demonstrate that the present model is able to simulate multiphase flow with large density ratio.

\begin{table}[!htb]
\centering
\caption{Physical parameters and dimensionless numbers in 2D rising bubble problem.} \label{tab:risingbubble}
\setlength{\tabcolsep}{1.8mm}{%
\begin{tabular}{ccccccccccc}
\hline
Case   &  $\rho_h$  & $\rho_l$ &   $\mu_h$ &     $\mu_l$ &   $g$ &  $\sigma$ &  $\text{Re}$ &  $\text{Eo}$ &   $\rho_h/\rho_l$ &  $\mu_h/\mu_l$ \\
\hline
1   & 1000  & 100  &  10   & 1    & 0.98 & 24.5  & 35    &  10 & 10    & 10       \\
2   & 1000  &   1  &  10   & 0.1  & 0.98 & 1.96  & 35    &  125& 1000  &100            \\
\hline
\end{tabular}}
\end{table}

\subsection{Droplet splashing}\label{sec:Erroranalysis}
\begin{figure}
\centering
\includegraphics[width=0.5\textwidth,trim=10 0 10 10,clip]{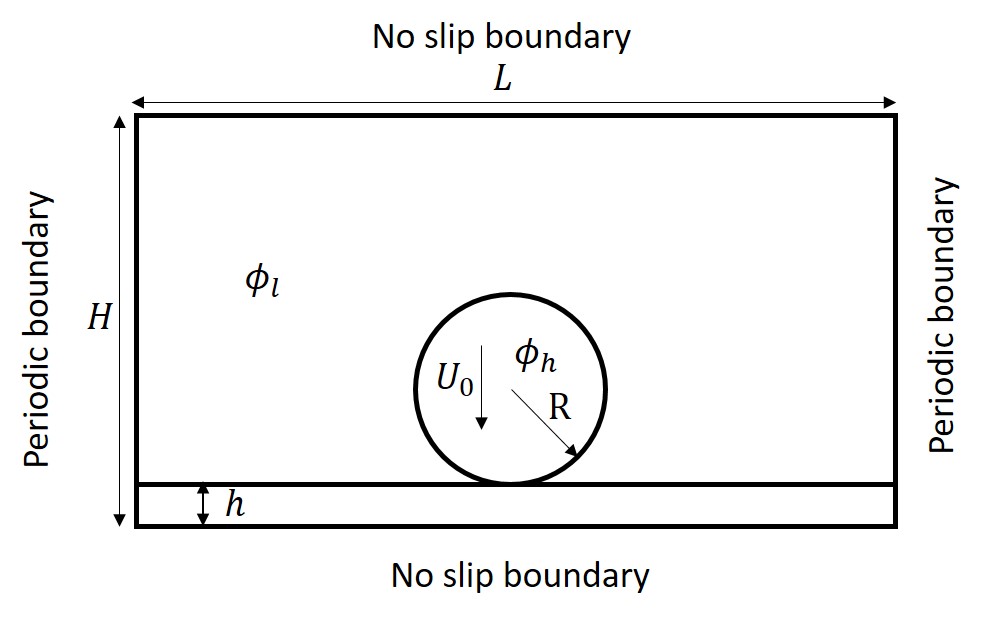}
\caption{Schematic of the initial setup for the droplet impact on an thin liquid film.}
\label{fig:splashingsetup}
\end{figure}

In this subsection, we consider the impact of a circular drop on a thin layer of fluid that widely exits in nature sciences and industrial applications, such as
a raindrop splashing on the ground and ink-jet printing.
Due to the presence of the numerical singularity at the impact point and large density and viscosity differences, it is a great challenging for numerical methods to maintain numerical stability.
Up to now, many experimental~\cite{cossali1997impact,rioboo2003experimental} and numerical research~\cite{josserand2003droplet,yarin2006drop,lee2005stable,wang2015multiphase,
liang2018phase} on such problem have been carried out.
Previous studies show that the impact radius that can be defined as the radius of the point where the interfacial velocity of the fluid is maximal at a given instant  obeys a power law at short time after the impact. This power law will be used to validate the proposed scheme.
The dynamics and the impact outcomes of this problem is mainly determined by the following non-dimensional parameters: the Reynolds number $\text{Re}$ and the Weber number $\text{We}$,
\begin{equation}\label{eq}
\begin{aligned}
\text{Re}=\frac{2\rho_h U_0 R}{\mu_h}, \qquad  \text{We}=\frac{2\rho_h U_0^2 R}{\sigma},
\end{aligned}
\end{equation}
where $U_0$ is the velocity of the droplet at the instant of impact and $R$ is the radius of the droplet.
Schematic representation of the simulation setup is shown in Fig.~\ref{fig:splashingsetup}.
The order parameter is initialized by
\begin{equation}\label{eq}
\phi(x,y)=\left\{
\begin{aligned}
&\frac{\phi_h+\phi_l}{2}+\frac{\phi_h-\phi_l}{2}\tanh\frac{2(R-\sqrt{(x-0.5L)^2+(y-R-H_w)^2}}{W}, \quad y\geq H_w  \\
&\frac{\phi_h+\phi_l}{2}+\frac{\phi_h-\phi_l}{2}\tanh\frac{2(H_w-y)}{W}, \quad y<H_w \\
\end{aligned}
\right.
\end{equation}
The velocity field is initialized as
\begin{equation}\label{eq}
(u,v)=\begin{cases}
(0,-\frac{\phi-\phi_l}{\phi_h-\phi_l} U_0),  &  y\ge H_w \\
(0,0),  &  u< H_w
\end{cases}
\end{equation}
The periodic boundary condition is applied at the left and right boundaries while the no-slip boundary condition is
enforced at the bottom and upper boundaries. The computational domain is resolved with $L\times H=1500\times 500$ meshes. The height of the liquid film on the bottom wall is $H_w=0.1H$ and the radius of the circular droplet is $R=0.2H$.
The  parameters are set as $U_0=0.004$, $CFL=0.1$, $\rho_h=1$, $\rho_l=0.001$.
The Weber number is fixed $8000$ for all cases while the Reynolds numbers are changed from $20$ to $1000$.
Following the treatment of  Lee~\cite{2010Lattice}, the viscosity of the vapor is kept constat such that the Reynolds number of the droplet is determined by the viscosity of the liquid.
The viscosity of the vapor is given by $\text{Re}=500$ and $\mu_h/\mu_l=40$.

\begin{figure}[htp]
\centering
\subfigure[$t^*=0.1$]{\includegraphics[width=0.5\textwidth,trim=40 100 40 100,clip]{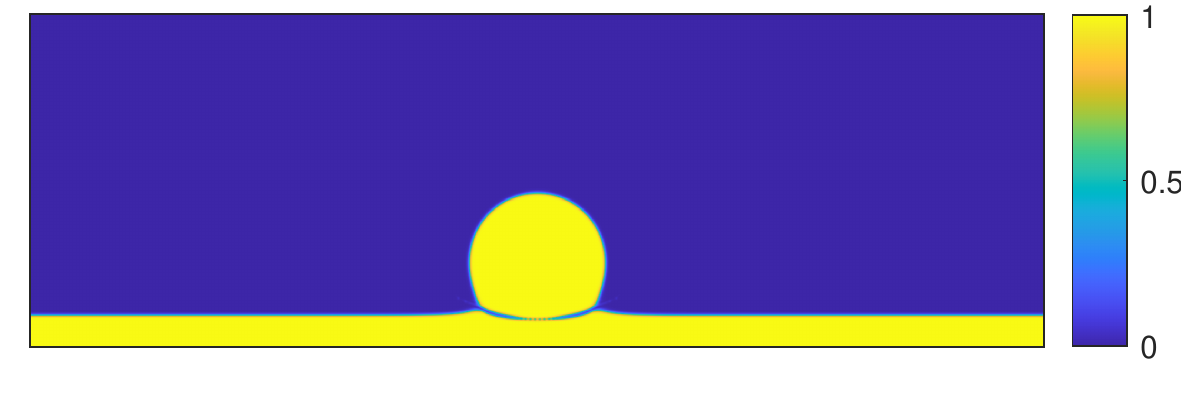}}~
\subfigure[$t^*=0.2$]{\includegraphics[width=0.5\textwidth,trim=40 100 40 100,clip]{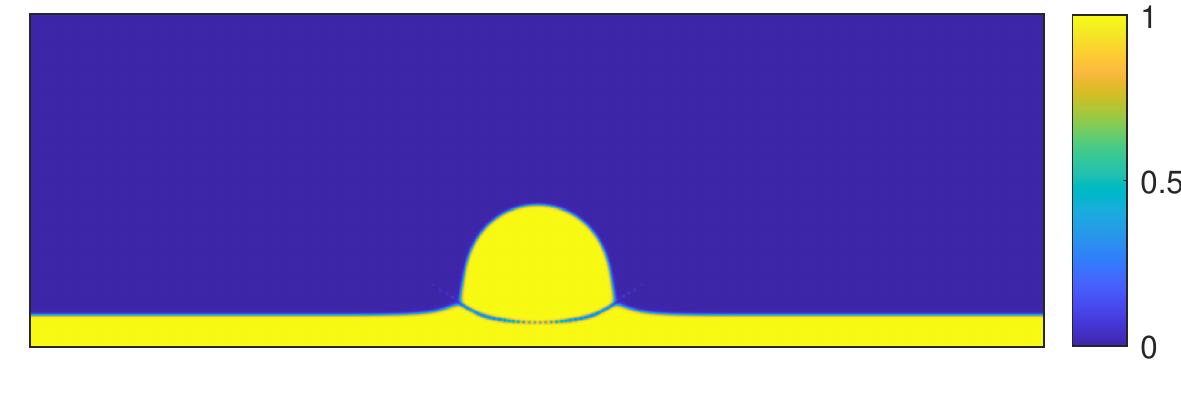}}\\
\subfigure[$t^*=0.4$]{\includegraphics[width=0.5\textwidth,trim=40 100 40 100,clip]{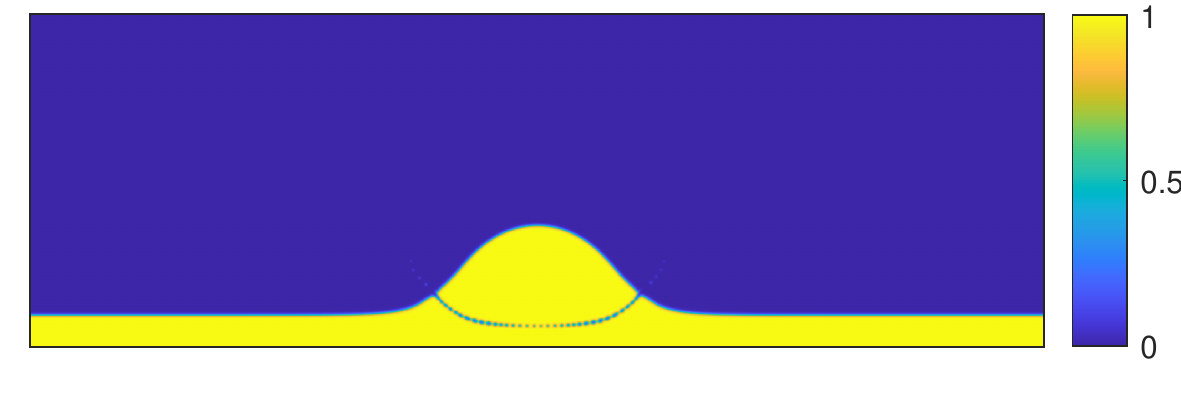}}~
\subfigure[$t^*=0.8$]{\includegraphics[width=0.5\textwidth,trim=40 100 40 100,clip]{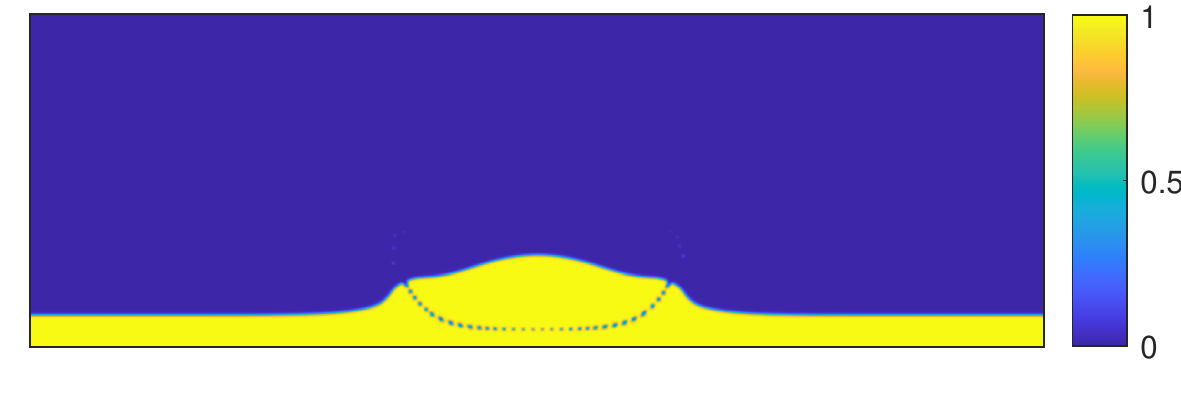}}\\
\subfigure[$t^*=1.2$]{\includegraphics[width=0.5\textwidth,trim=40 100 40 100,clip]{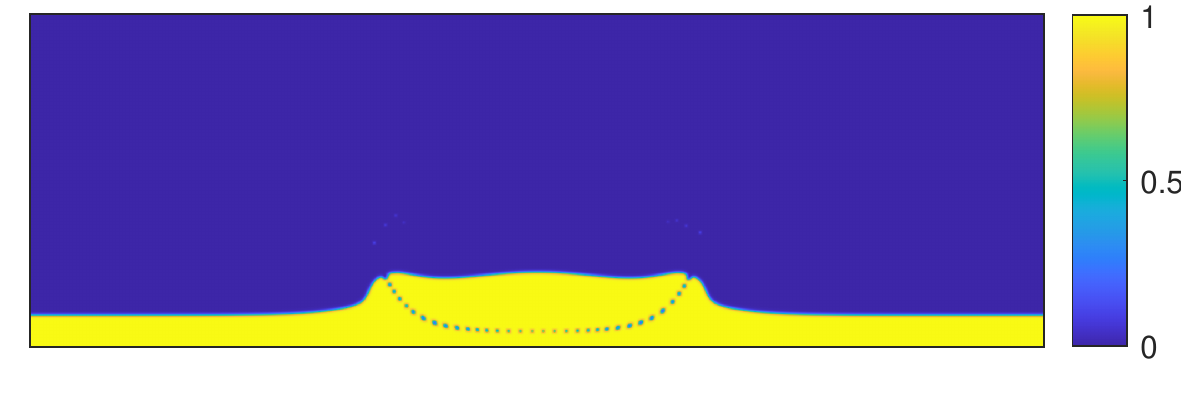}}~
\subfigure[$t^*=1.6$]{\includegraphics[width=0.5\textwidth,trim=40 100 40 100,clip]{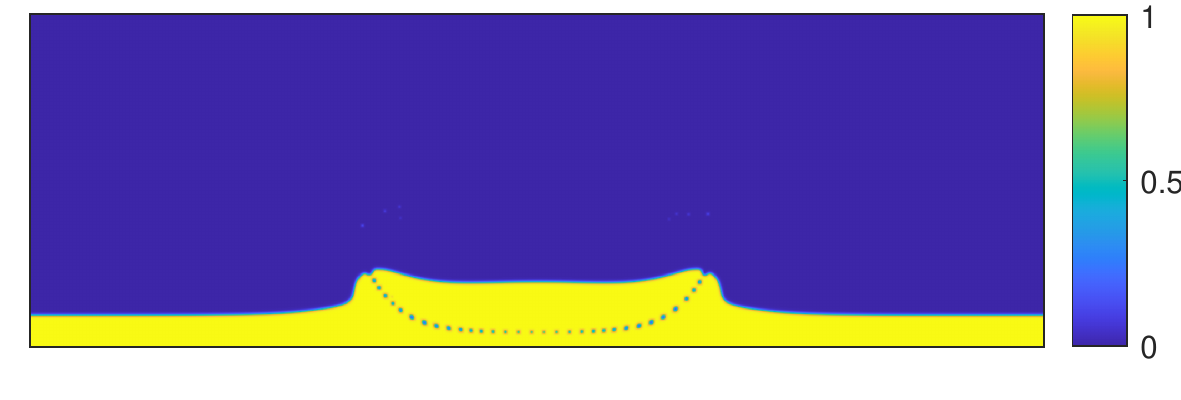}}\\
\caption{Evolution of the instantaneous interface for the droplet splashing on a thin film at $\text{Re}=20$. }
\label{taylor_Re20}
\end{figure}

\begin{figure}[htp]
\centering
\subfigure[$t^*=0.1$]{\includegraphics[width=0.5\textwidth,trim=40 100 40 100,clip]{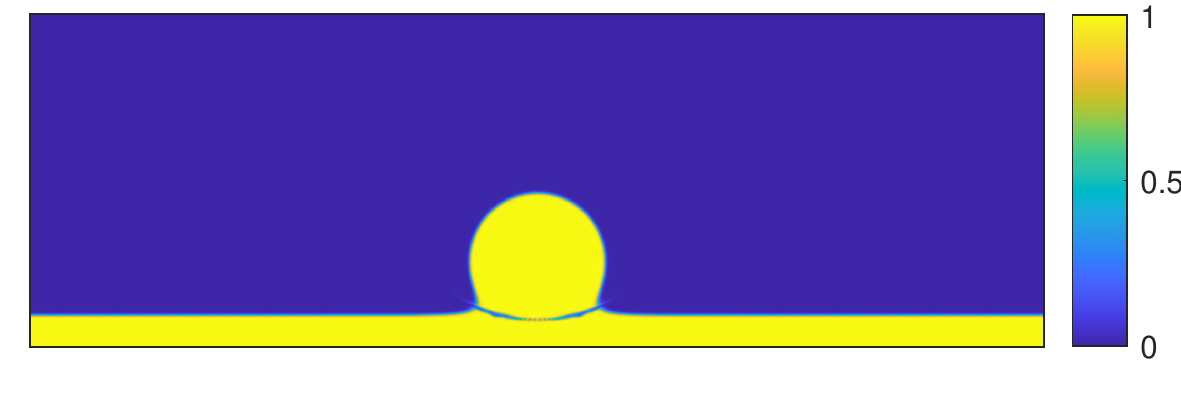}}~
\subfigure[$t^*=0.2$]{\includegraphics[width=0.5\textwidth,trim=40 100 40 100,clip]{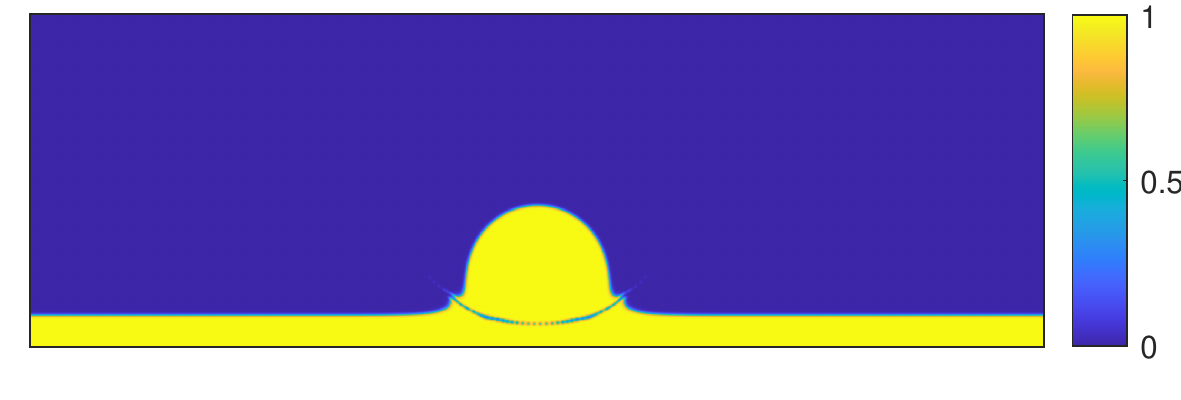}}\\
\subfigure[$t^*=0.4$]{\includegraphics[width=0.5\textwidth,trim=40 100 40 100,clip]{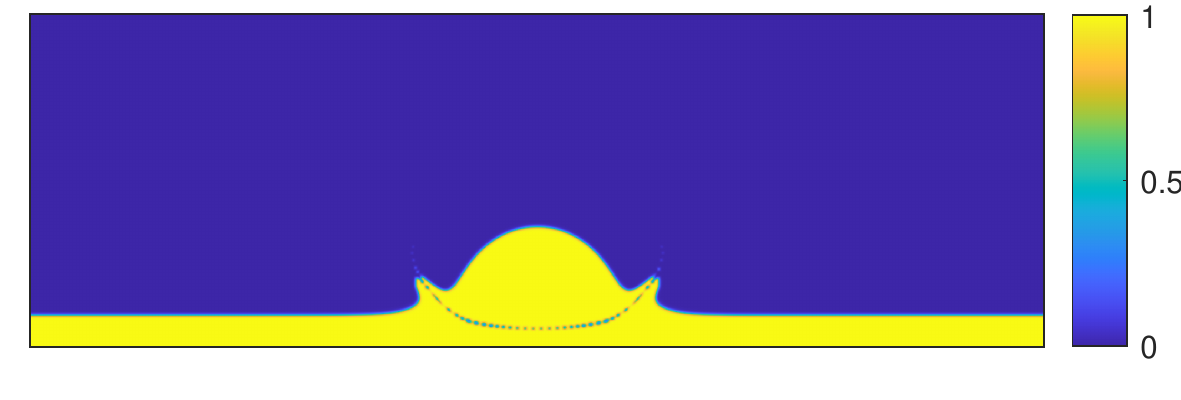}}~
\subfigure[$t^*=0.8$]{\includegraphics[width=0.5\textwidth,trim=40 100 40 100,clip]{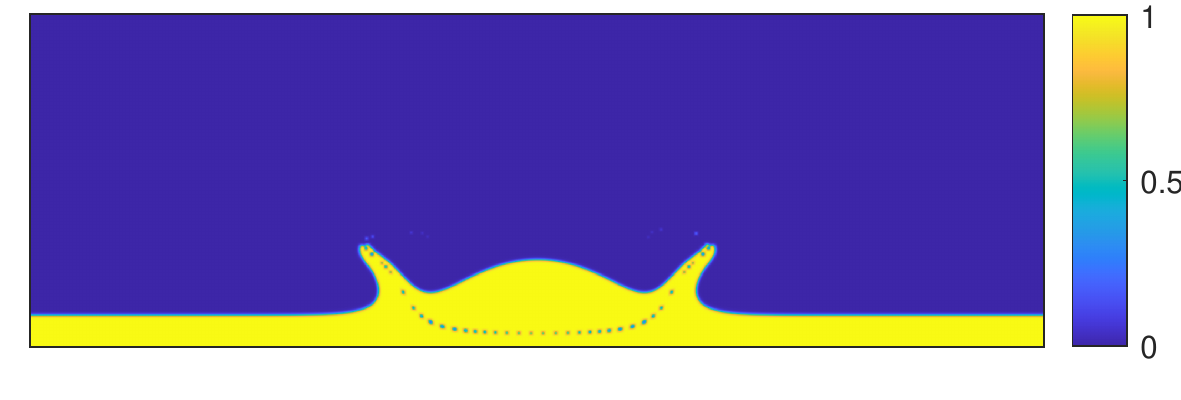}}\\
\subfigure[$t^*=1.2$]{\includegraphics[width=0.5\textwidth,trim=40 100 40 100,clip]{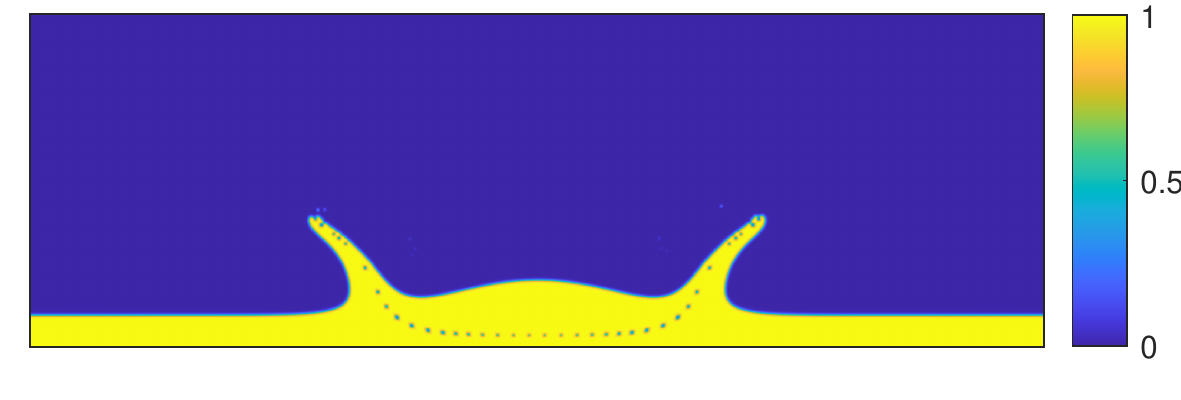}}~
\subfigure[$t^*=1.6$]{\includegraphics[width=0.5\textwidth,trim=40 100 40 100,clip]{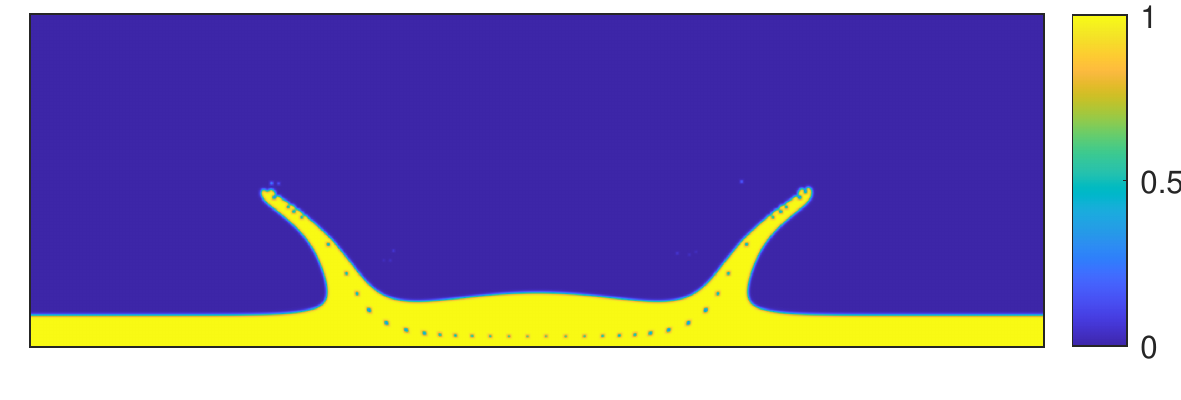}}\\
\caption{Evolution of the instantaneous interface for the droplet splashing on a thin film at $Re=100$. }
\label{taylor_Re100}
\end{figure}

\begin{figure}[htp]
\centering
\subfigure[$t^*=0.1$]{\includegraphics[width=0.5\textwidth,trim=40 100 40 100,clip]{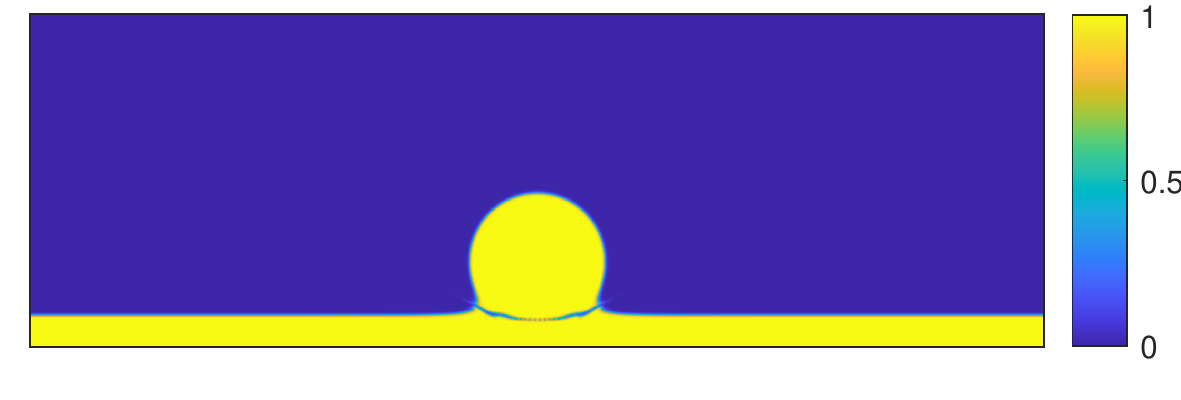}}~
\subfigure[$t^*=0.2$]{\includegraphics[width=0.5\textwidth,trim=40 100 40 100,clip]{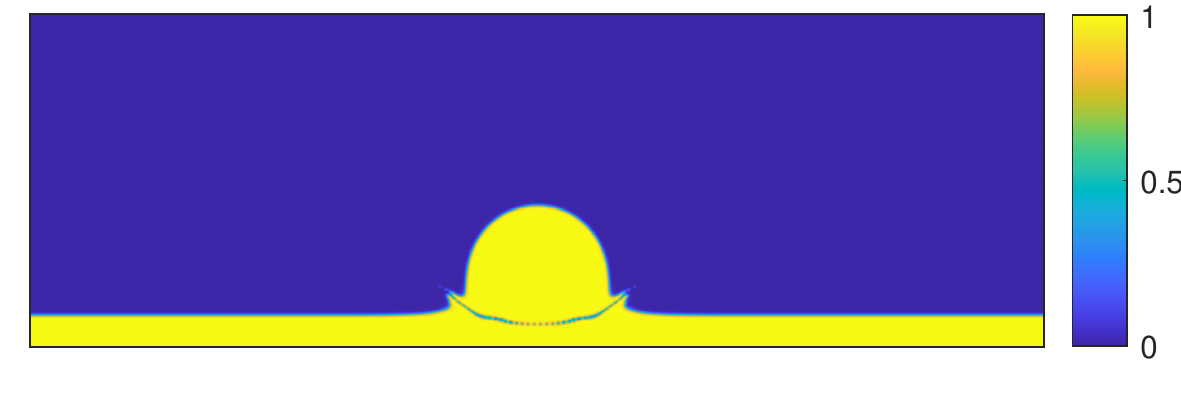}}\\
\subfigure[$t^*=0.4$]{\includegraphics[width=0.5\textwidth,trim=40 100 40 100,clip]{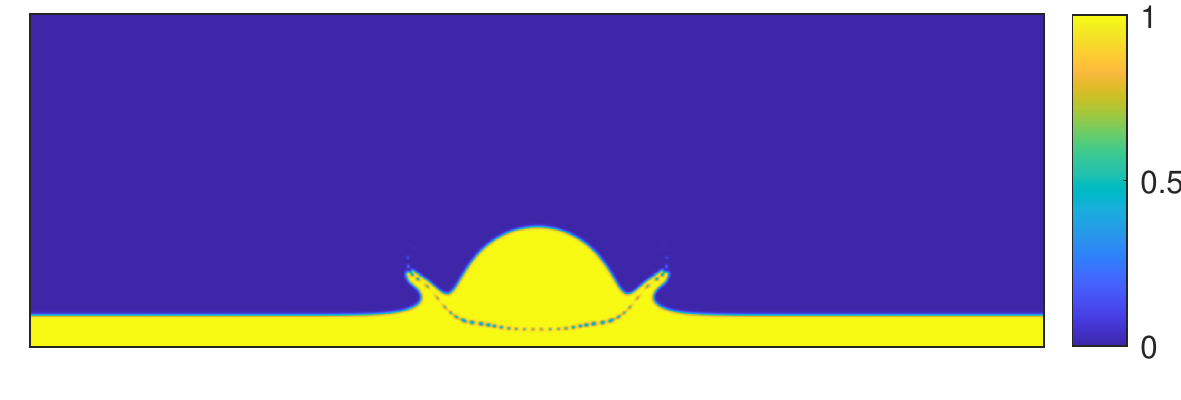}}~
\subfigure[$t^*=0.8$]{\includegraphics[width=0.5\textwidth,trim=40 100 40 100,clip]{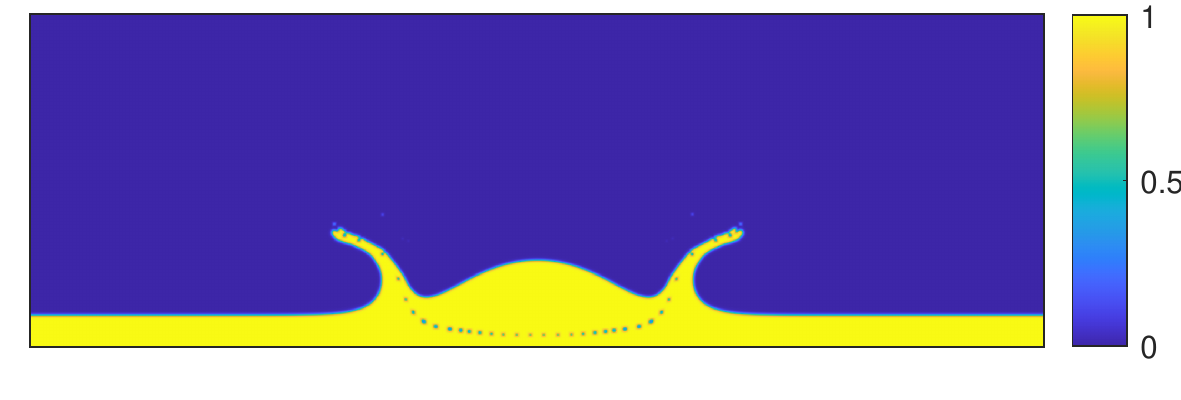}}\\
\subfigure[$t^*=1.2$]{\includegraphics[width=0.5\textwidth,trim=40 100 40 100,clip]{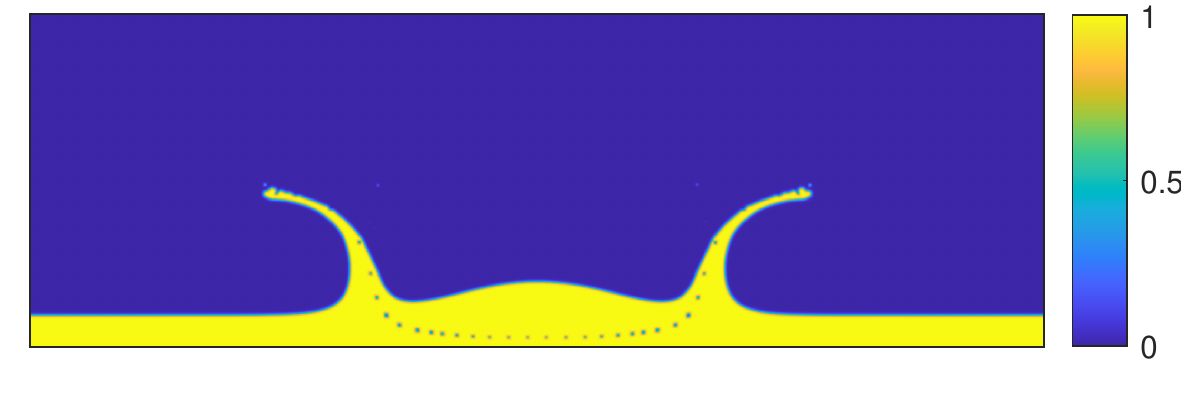}}~
\subfigure[$t^*=1.6$]{\includegraphics[width=0.5\textwidth,trim=40 100 40 100,clip]{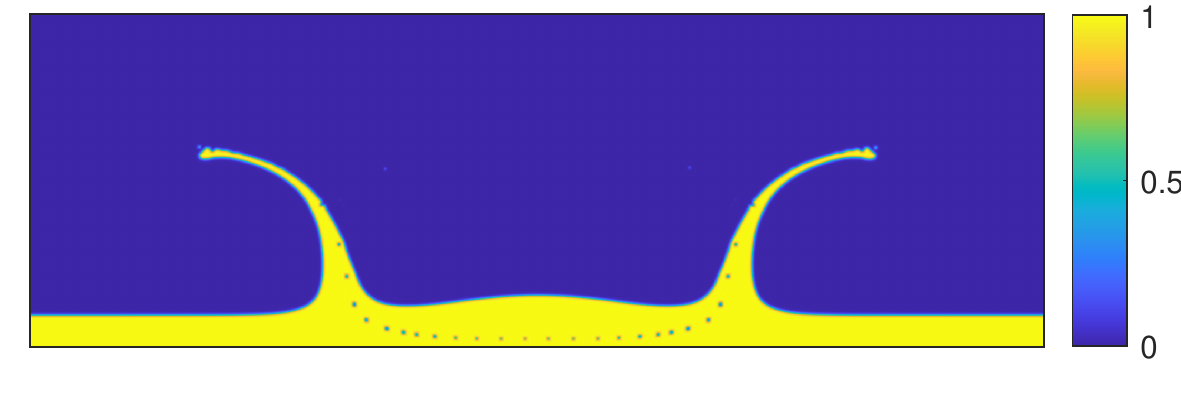}}\\
\caption{Evolution of the instantaneous interface for the droplet splashing on a thin film at $\text{Re}=500$. }
\label{taylor_Re500}
\end{figure}

\begin{figure}[htp]
\centering
\subfigure[$t^*=0.1$]{\includegraphics[width=0.5\textwidth,trim=40 100 40 100,clip]{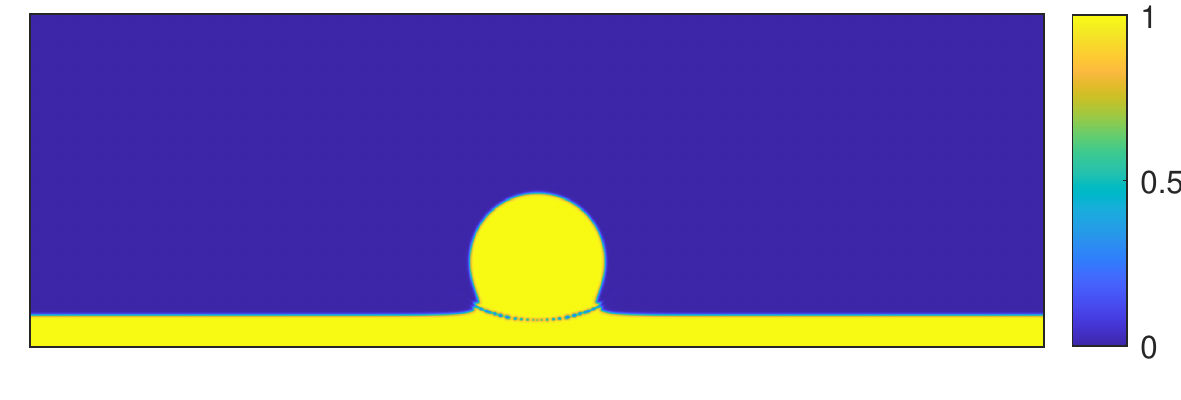}}~
\subfigure[$t^*=0.2$]{\includegraphics[width=0.5\textwidth,trim=40 100 40 100,clip]{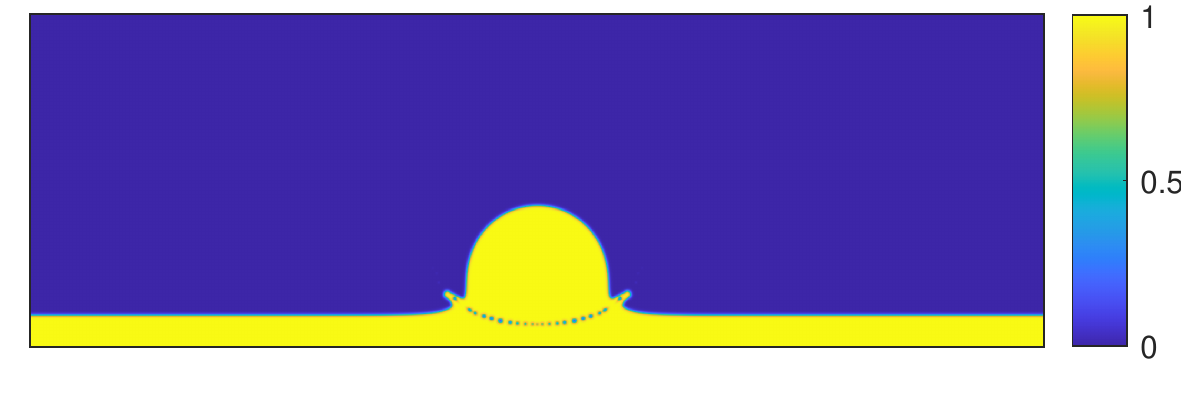}}\\
\subfigure[$t^*=0.4$]{\includegraphics[width=0.5\textwidth,trim=40 100 40 100,clip]{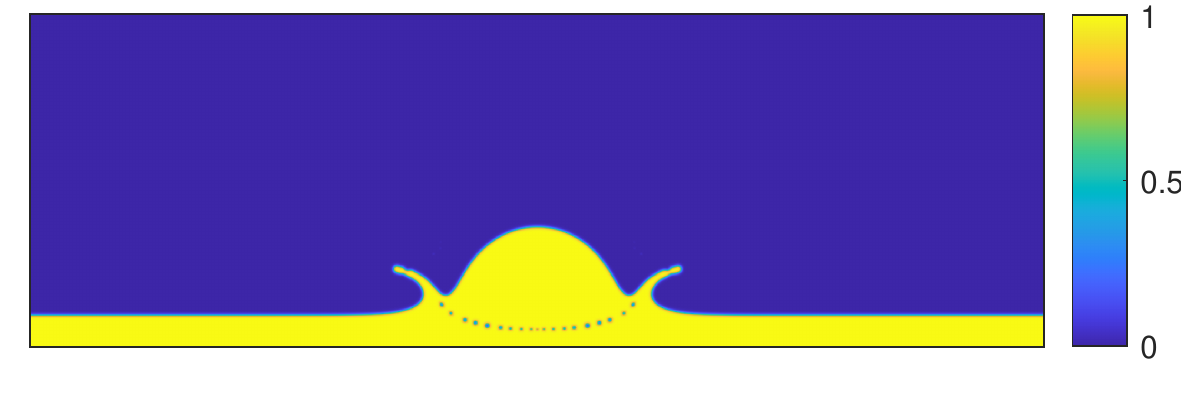}}~
\subfigure[$t^*=0.8$]{\includegraphics[width=0.5\textwidth,trim=40 100 40 100,clip]{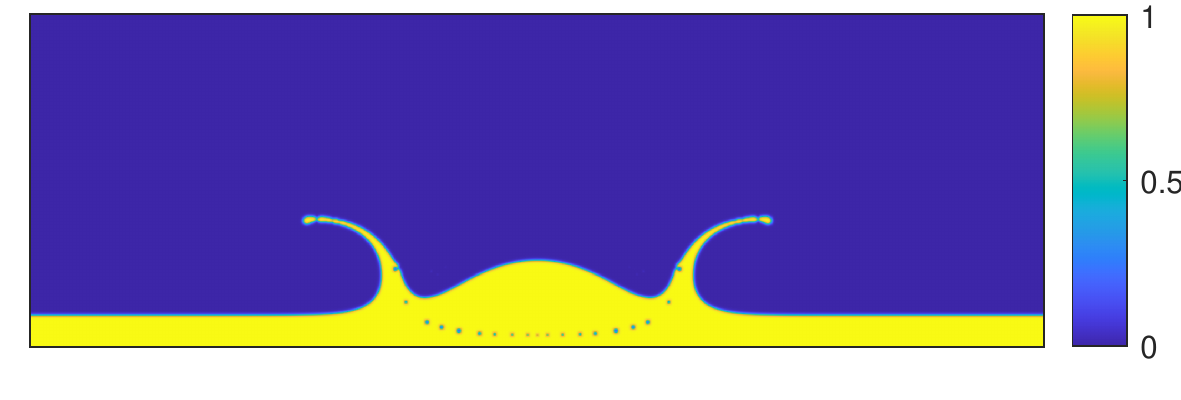}}\\
\subfigure[$t^*=1.2$]{\includegraphics[width=0.5\textwidth,trim=40 100 40 100,clip]{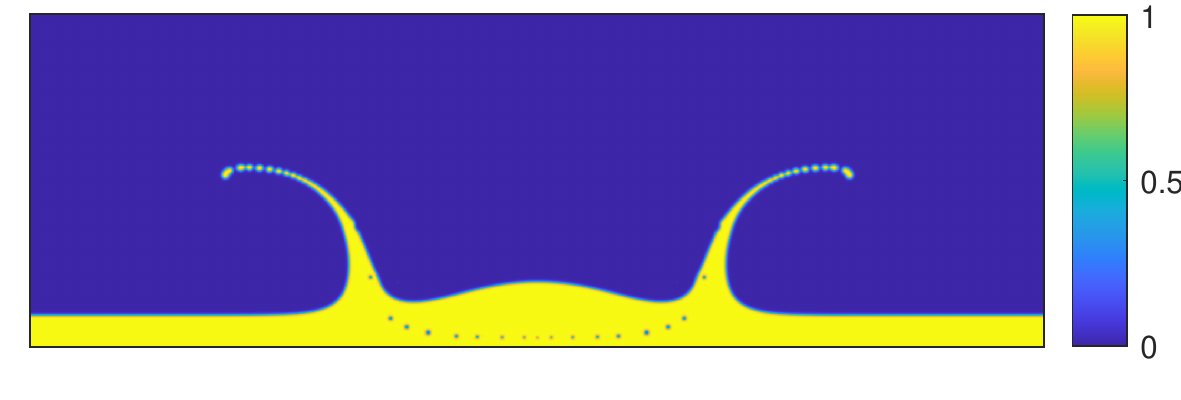}}~
\subfigure[$t^*=1.6$]{\includegraphics[width=0.5\textwidth,trim=40 100 40 100,clip]{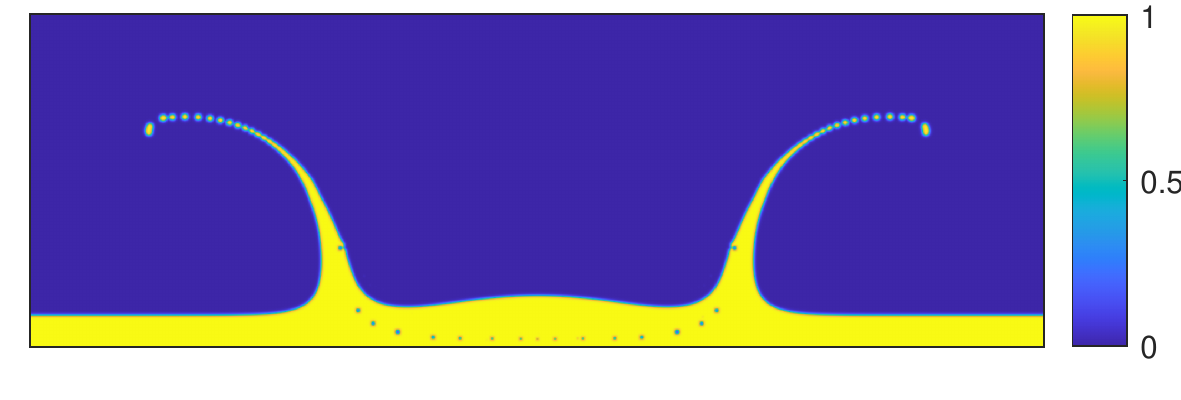}}\\
\caption{Evolution of the instantaneous interface for the droplet splashing on a thin film at $\text{Re}=1000$. }
\label{taylor_Re1000}
\end{figure}

\begin{figure}
  \centering
  \includegraphics[width=0.5\textwidth]{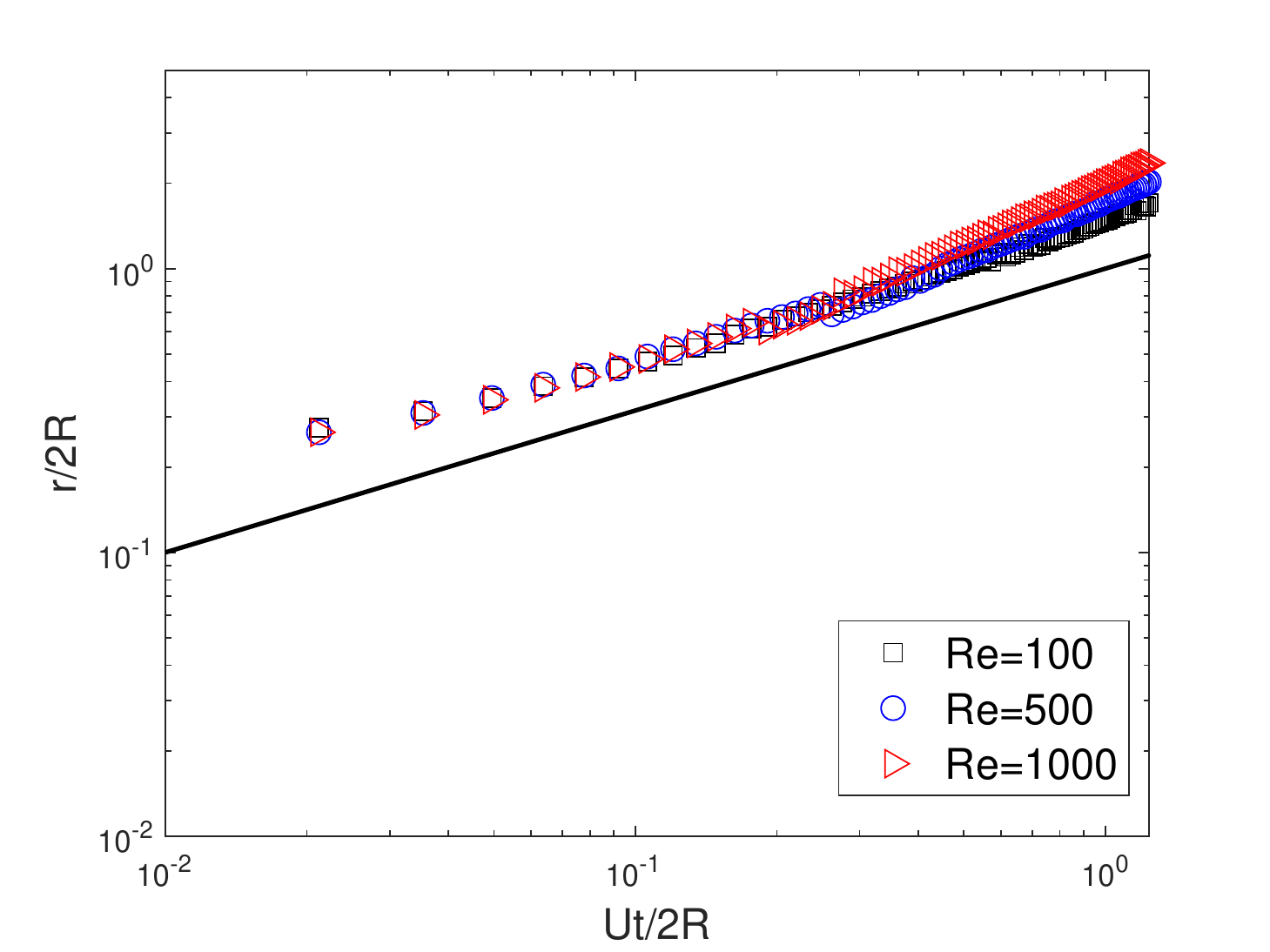}~
  \caption{Log-log plot of the spread factor $r/2R$ as a function of $Ut/2R$ at $We=8000$ and $Re=100, 500$ and $1000$.
  The straight line corresponds to the power law $r=\sqrt{2RUt}$.}\label{spread_factor}
\end{figure}

Figs.~\ref{taylor_Re20}-\ref{taylor_Re500} show the time evolutions of the droplet and the thin liquid film during the impact at three different Reynolds numbers of $20, 100$ and $500$.  The non-dimensional time is measured by $2R/U_0$.
For the case of $\text{Re}=20$, the droplet spreads gently on the surface without splashing, which is  called  deposition process.
An empirical relation has been established between spreading and deposition behaviors through the dimensionless Sommerfeld parameter $\hat{K}=We^{0.5}Re^{0.25}$. When $\hat{K}$ is smaller than a threshold value $\hat{K}_c$ then only deposition is observe while a splash develops for $\hat{K}>\hat{K}_c$.
In the case of $\text{Re}=20$, $\hat{K}=189$ is less than the threshold value  $\hat{K}_c=225$ reported in \cite{josserand2003droplet}.
In the cases of $\text{Re}=100$, as shown in Fig.~\ref{taylor_Re100}, a thin liquid sheet is thrown out radially and vertically outward after the impact. The sheet grows into a corolla. This process is called crown splash.
In the case of $\text{Re}=500$, similar splashing phenomenon can be observed while the  corolla is more obvious.
In \cite{lee2005stable} and \cite{liang2018phase}, the largest Reynold number is $500$. Here, we further simulated  the above case with $\text{Re}=1000$. The time evolutions of the droplet are shown in Fig.~\ref{taylor_Re1000}. It can be observed that
the end rim that grows at the edge of the corolla is unstable and develops fingers of liquid. The fingers eventually breakup into small droplets due to the surface tension effect (also known as the Rayleigh-Plateau instability).
In addition, for all cases, small vapor bubbles are entrapped in the phase interface region and eventually undergo phase change and are absorbed into liquid. Similar results are also found in the previous studies~\cite{liang2018phase}.
Finally, the log-log plot of the spread factor $r/2R$ for $\text{Re}=100, 500$ and $1000$ are plotted in Fig.~\ref{spread_factor}. The straight line denotes the power law $r=\sqrt{2RUt}$ or $r^*=\sqrt{t^*}$.
In previous studies~\cite{liang2018phase,coppola2011insights,li2013lattice}, the growth of the spreading radius is also described by $r^*=C\sqrt{t^*}$,  where $C$ is a fitting coefficient.  For the present study, $C\approx1.45$. Meanwhile, slight deviation from the straight line is  observed at the initial stage, which is probably due to the three-dimensional effect of the process. Overall the numerical results agree well with the prediction of the power law.

\subsection{Dam break flow over dry and wet beds}
\begin{figure}
\centering
\includegraphics[width=0.5\textwidth,trim=10 0 10 10,clip]{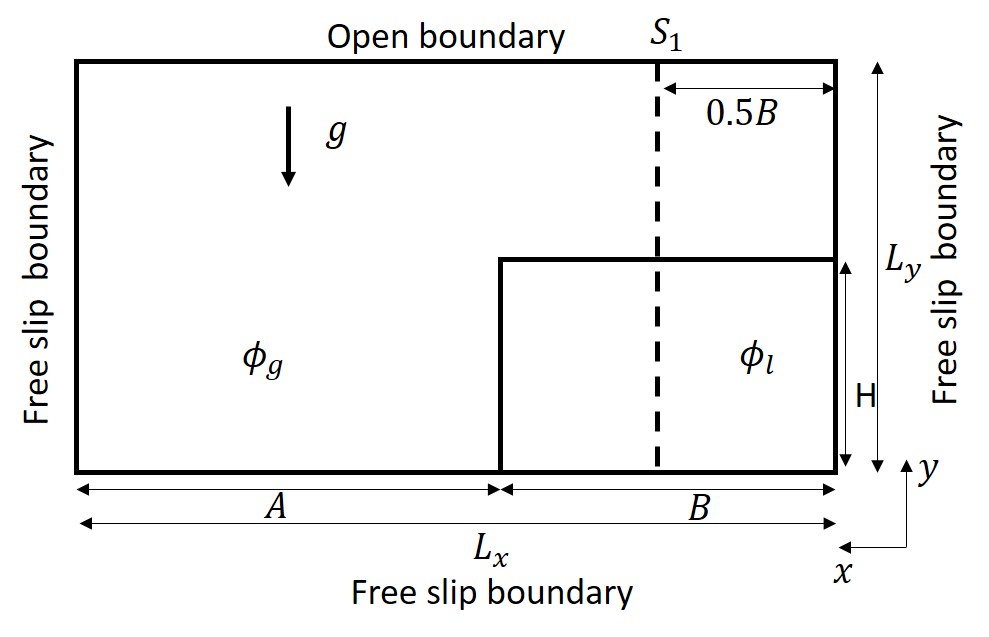}
\caption{Schematic of the initial setup for the dam break over a dry bed.}
\label{fig:dambreakdrysetup}
\end{figure}

The dam break problem is usually generated by a sudden failure of a barrier confining a reservoir filled with fluid and is commonly used to test numerical methods for free surface. The problem  characterized by a high density ratio, high Reynolds number and rapid and complex topological changes has been widely studied by many experiments~\cite{lobovsky2014experimental,hu2010numerical,garoosi2022experimental} and numerical methods. The reliability and performance of the present model is further demonstrated and tested against the experimental data.
The sketch of the computational domain is shown in Fig.~\ref{fig:dambreakdrysetup}.
Free-slip boundary conditions are applied to the left, right and bottom walls and the outflow boundary condition is applied to the top of the computational domain.
Based on the experiment~\cite{lobovsky2014experimental}, the computational domain is described by the length $L_x=1.6m$ and height $L_y=0.6m$. The height and width of the water column are $H=0.3m$ and $B=0.6m$, respectively. The thickness and the removal time of the gate are neglected in simulations.
The physical properties of the water and air are $\rho_h=998 kg/m^3$, $\rho_l=1.2kg/m^3$, $\mu_h=1.0\times 10^{-3} Pa\cdot s$, $\mu_l=1.8\times 10^{-5} Pa.s$, and surface tension coefficient $\sigma=0.07275 N/m$. The gravitational acceleration is $g=9.8m/s^2$. Based on these parameters, the reference velocity, Reynolds number and Weber number are  defined as
\begin{equation}\label{eq:dambreak_dimensionless}
\begin{aligned}
U_{ref} &=\sqrt{gH}=1.7146,\\
\text{Re} &=\frac{\rho_h U_{ref} H}{\mu_h}=5.1245\times 10^{5},\\
\text{We} &=\frac{\rho_h U_{ref}^2 H}{\sigma}=1.2094\times 10^4.
\end{aligned}
\end{equation}
The order parameter is initialized by
\begin{equation}\label{eq}
\phi=\left\{
\begin{array}{ll}
\frac{\phi_h+\phi_l}{2}+\frac{\phi_h-\phi_l}{2}\tanh\frac{2(a-y)}{W},  & x\leq B-W, y\geq H-W  \\
\frac{\phi_h+\phi_l}{2}+\frac{\phi_h-\phi_l}{2}\tanh\frac{2(a-x)}{W}, & x\geq B-W, y\leq H-W \\
\frac{\phi_h+\phi_l}{2}+\frac{\phi_h-\phi_l}{2}\tanh\frac{2(W-\sqrt{(x-B+W)^2+(y-H+W)^2})}{W},&  x\geq B-W, y\geq H-W \\
\phi_h,& \text{otherwise}.
\end{array}\right.
\end{equation}

\begin{figure}[htp]
\centering
\subfigure[$t^*=0.0$]{\includegraphics[width=0.33\textwidth,trim=40 100 40 100,clip]{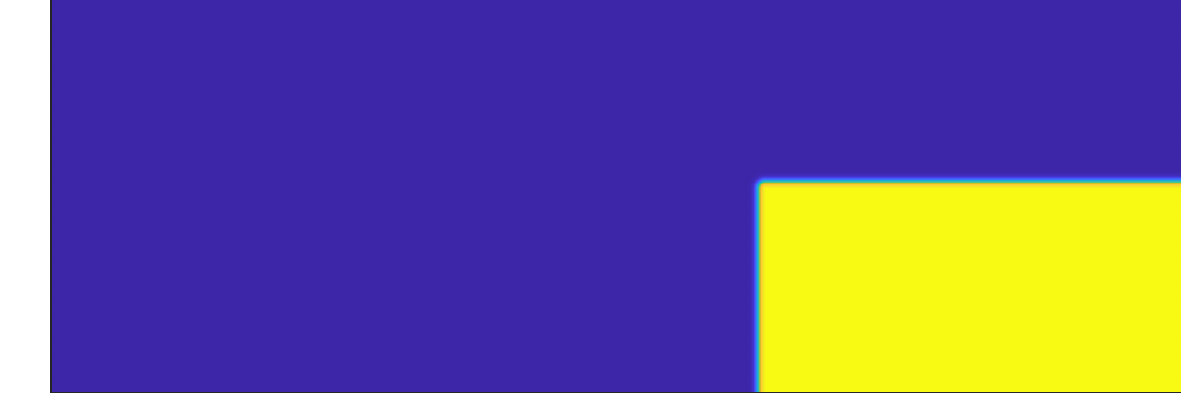}}~
\subfigure[$t^*=0.91$]{\includegraphics[width=0.33\textwidth,trim=40 100 40 100,clip]{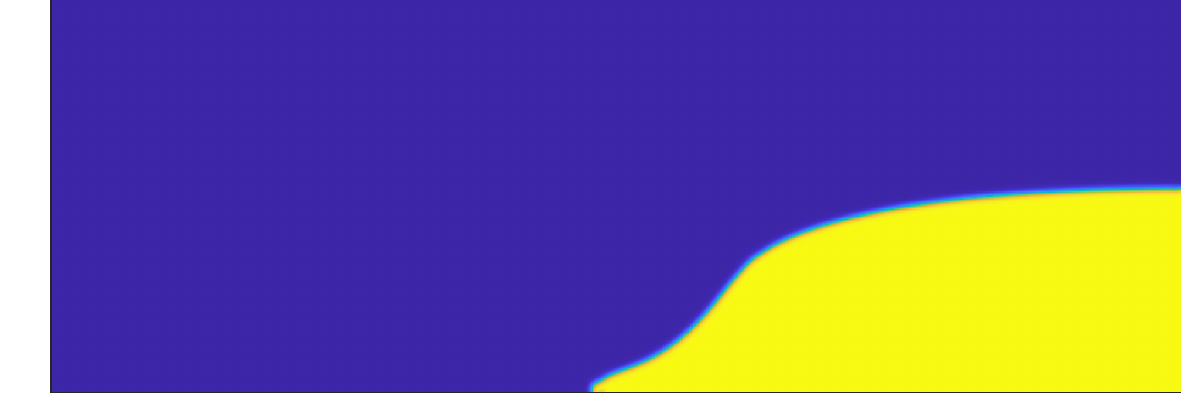}}~
\subfigure[$t^*=1.58$]{\includegraphics[width=0.33\textwidth,trim=40 100 40 100,clip]{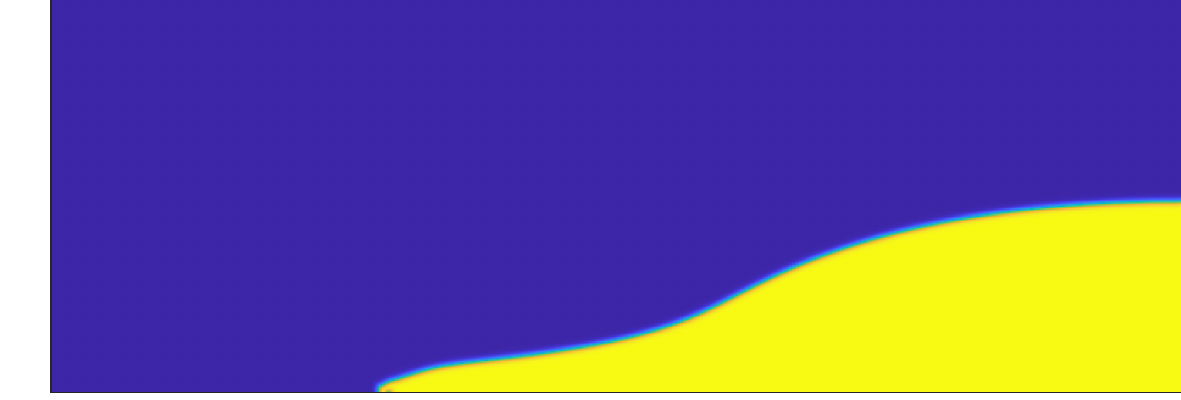}}\\
\subfigure[$t^*=2.13$]{\includegraphics[width=0.33\textwidth,trim=40 100 40 100,clip]{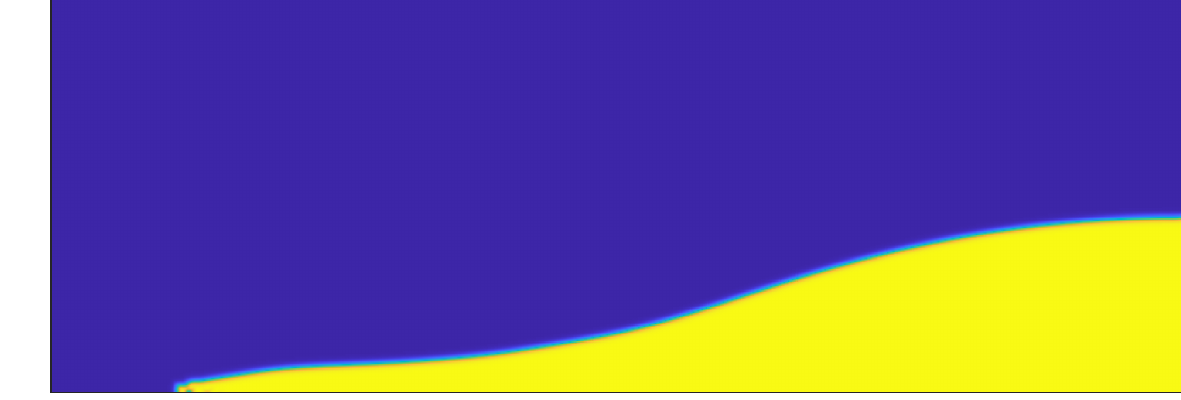}}~
\subfigure[$t^*=2.57$]{\includegraphics[width=0.33\textwidth,trim=40 100 40 100,clip]{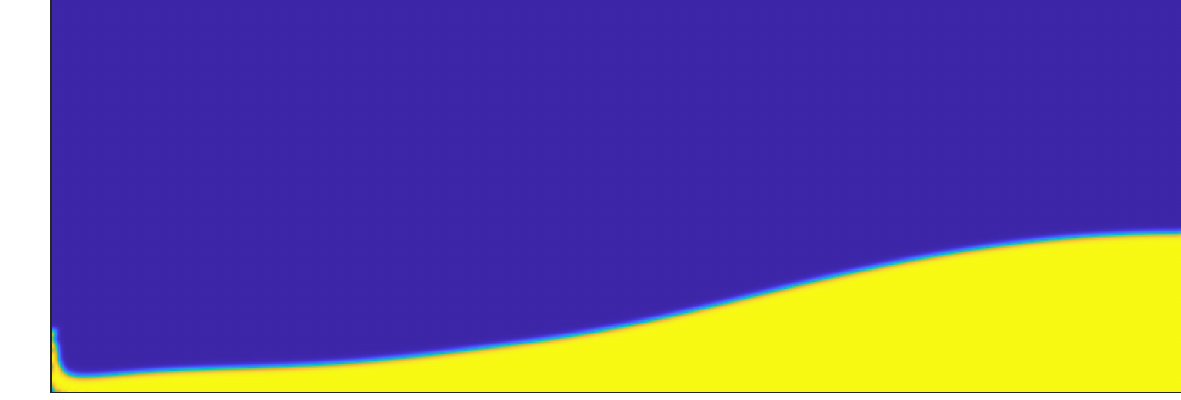}}~
\subfigure[$t^*=3.27$]{\includegraphics[width=0.33\textwidth,trim=40 100 40 100,clip]{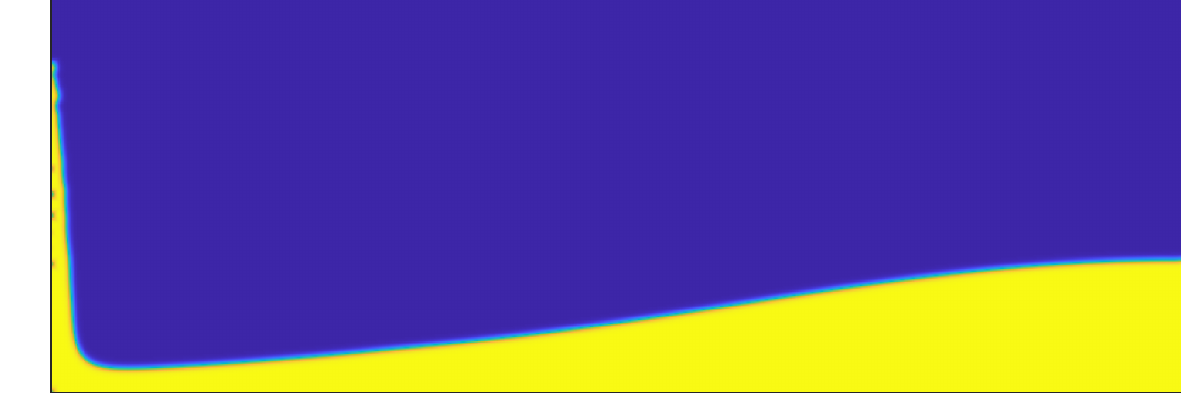}}\\
\subfigure[$t^*=4.93$]{\includegraphics[width=0.33\textwidth,trim=40 100 40 100,clip]{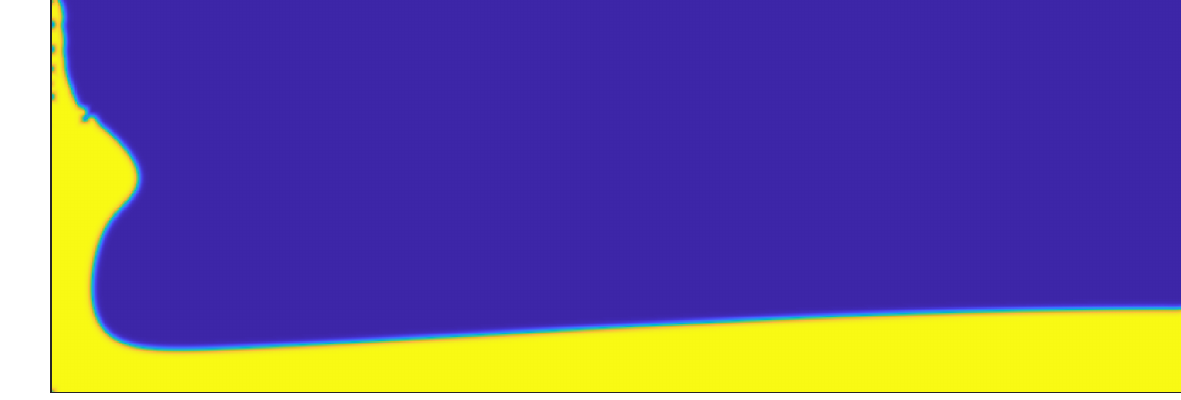}}~
\subfigure[$t^*=5.85$]{\includegraphics[width=0.33\textwidth,trim=40 100 40 100,clip]{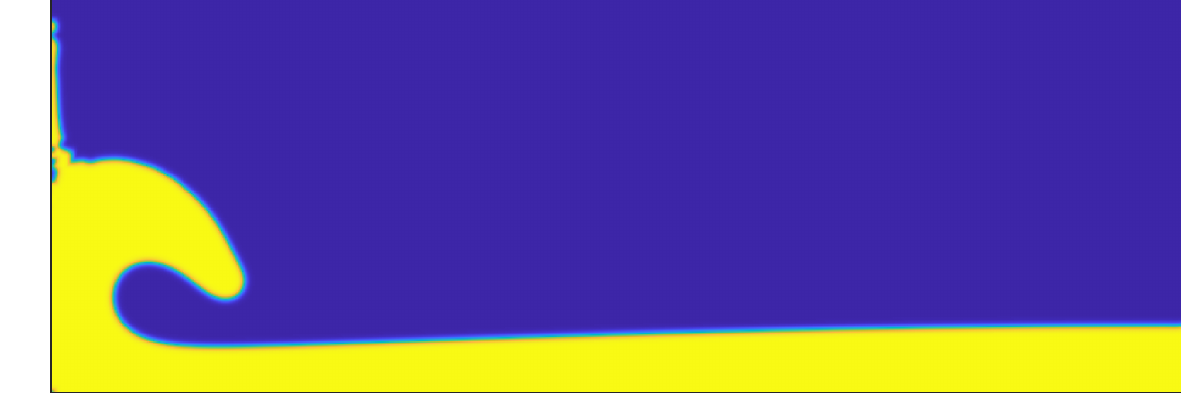}}~
\subfigure[$t^*=6.67$]{\includegraphics[width=0.33\textwidth,trim=40 100 40 100,clip]{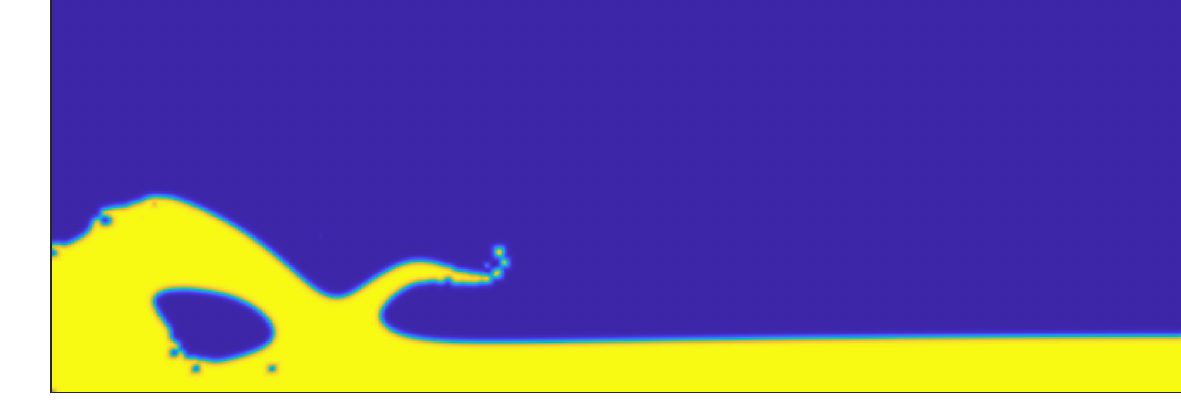}}\\
\subfigure[$t^*=7.54$]{\includegraphics[width=0.33\textwidth,trim=40 100 40 100,clip]{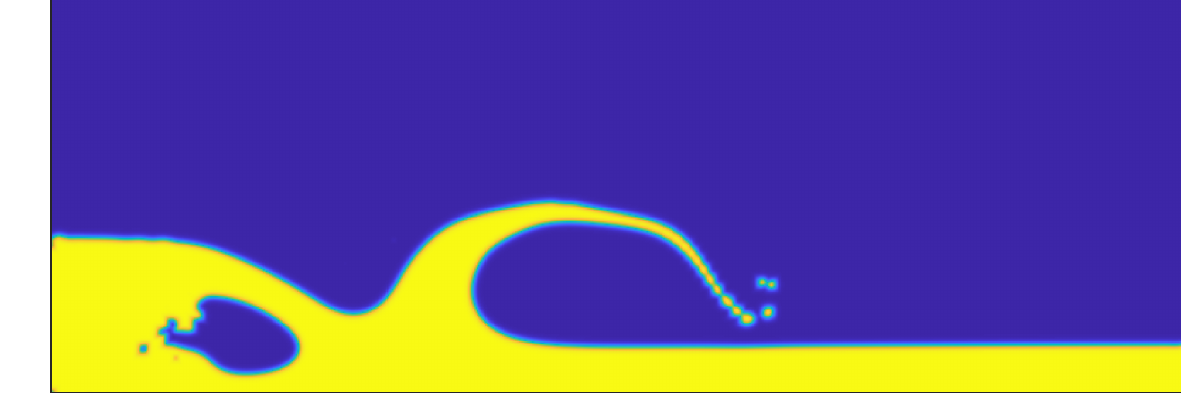}}~
\subfigure[$t^*=8.42$]{\includegraphics[width=0.33\textwidth,trim=40 100 40 100,clip]{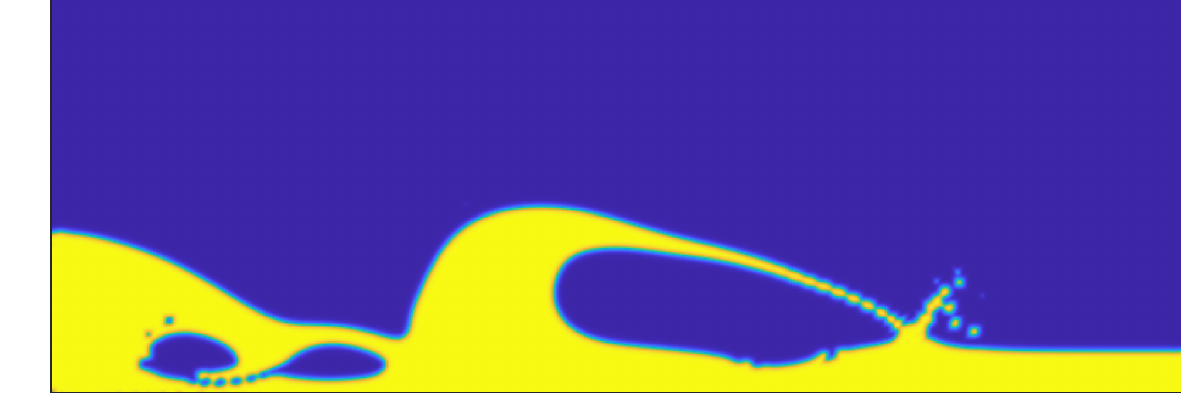}}~
\subfigure[$t^*=9.29$]{\includegraphics[width=0.33\textwidth,trim=40 100 40 100,clip]{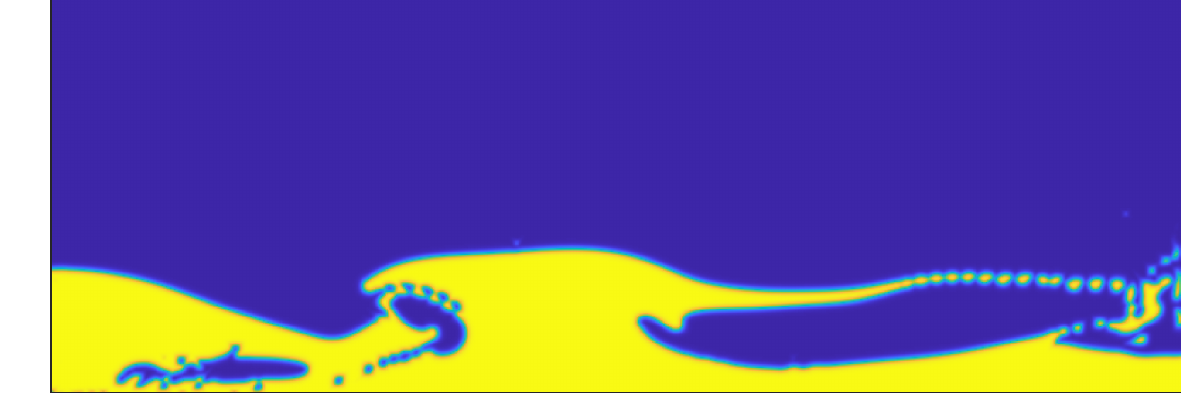}}
\caption{Evolution of the instantaneous interface for dam break flow over the dry bed. }
\label{drybed}
\end{figure}

\begin{figure}
  \centering
\subfigure[]{\includegraphics[width=0.5\textwidth]{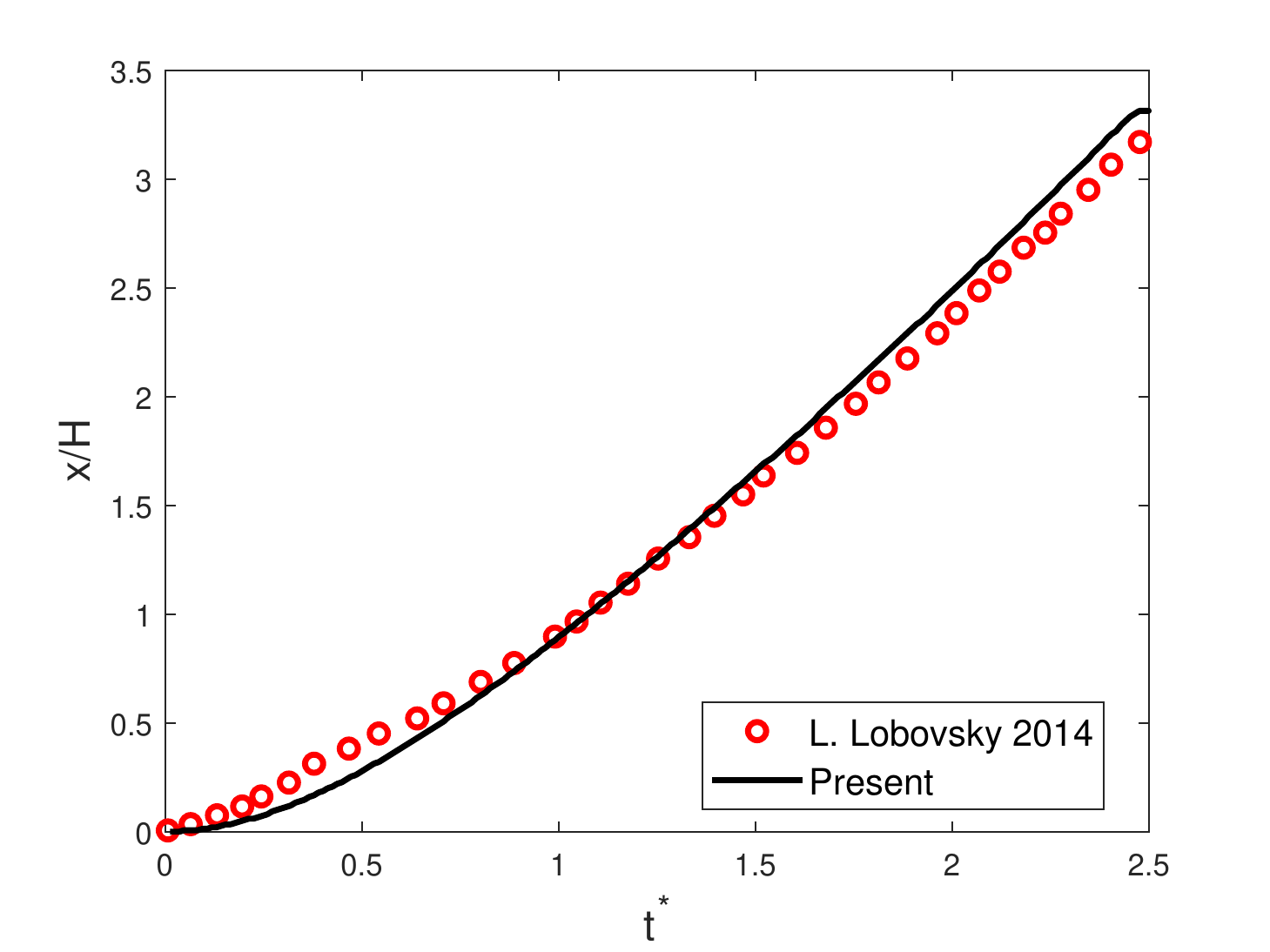}}~
\subfigure[]{\includegraphics[width=0.5\textwidth]{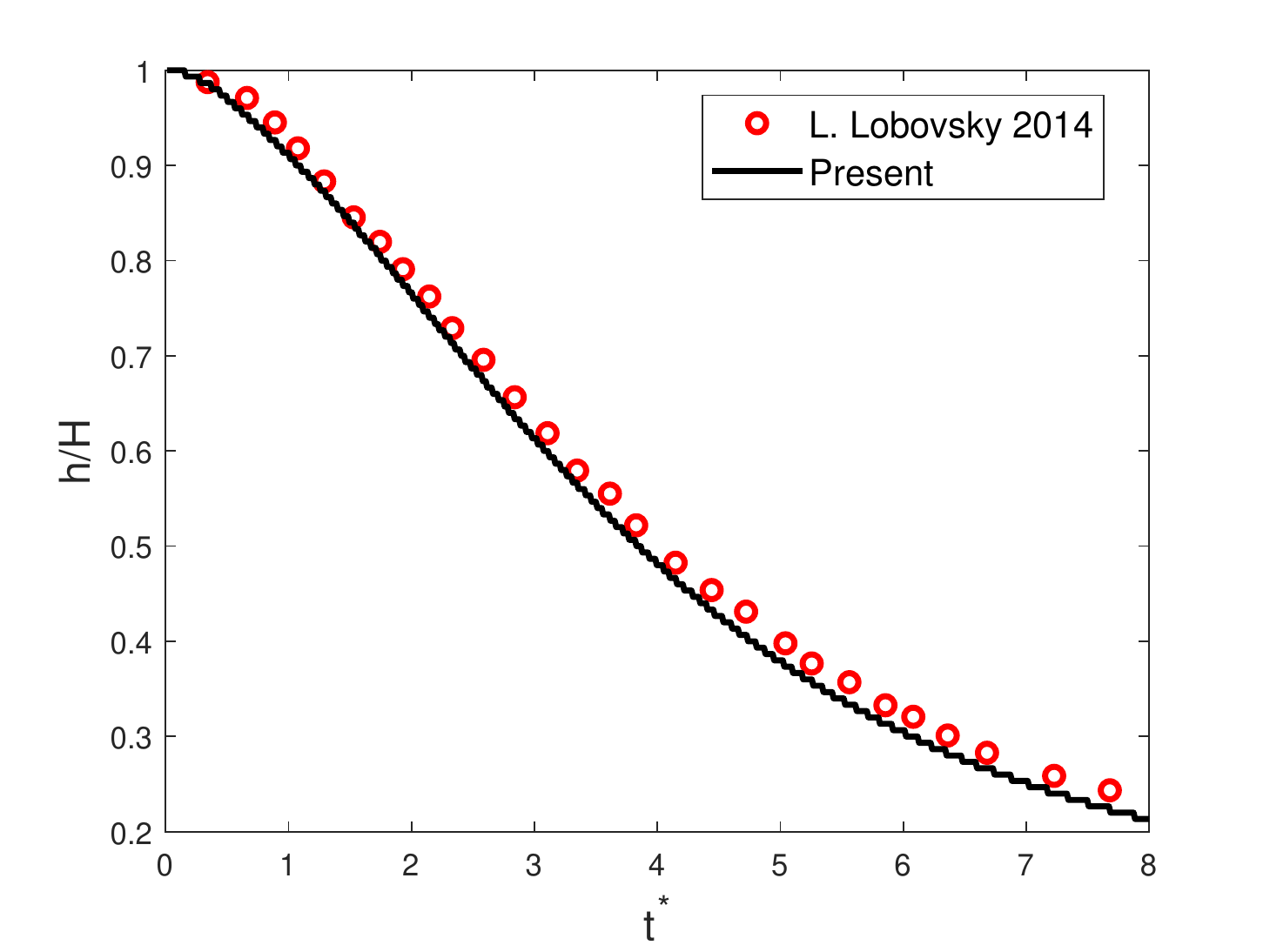}}
  \caption{(a) Evolution of the wave front after dam gate removal and (b) water level elevations at location $S_1$ for dam break flow over a dry bed.}\label{wavefront}
\end{figure}

A comparison between experimental and present numerical results for the dam break flow over a dry bed is shown in Fig.~\ref{drybed}. The time is non-dimensionalized by $t^*=t\sqrt{g/H}$.
It can be seen that the water collapses and travels along the dry bed. Its flow structures seem to be laminar.
When the impacts the downstream vertical wall, a vertical run-up jet is created, then falls onto the underlying fluid.
The interaction between the back flow of the fluid and the fluid yet advancing towards the wall causes a development of a plunging breaker and lots of vorticity.
Compared with the experimental data in~\cite{lobovsky2014experimental},  the initial stages of the dam-break flow scenarios with large free surface deformation correlated very well with the experimental images.
Furthermore, the  temporal evolution of the wave front towards the vertical wall is shown in Fig.~\ref{wavefront}(a) and
the temporal evolution of water levels at the given location $S_1$ is presented in Fig.~\ref{wavefront}(b).
 For comparison, the experimental results are included as well.
At the initial stages of the wave evolution, the position of the wave tip is slightly lower than the experimental data. This may be caused by a limited velocity of the gate's motion in experiment.
After $t^*=1.5$, the positron of the wave tip predicted by the present method is slightly higher than the experimental data. This may be caused by the roughness of the bottom wall. For the entire process, the present results are highly similar to the experimental data. In terms of the water level at location $S_1$, the predicted water levels are slightly higher than the experimental data. This difference is mainly caused by the gate effect. Nevertheless, the curve of the water level predicted by the present model are in good agreement with the experimental results.

\begin{figure}
\centering
\includegraphics[width=0.5\textwidth]{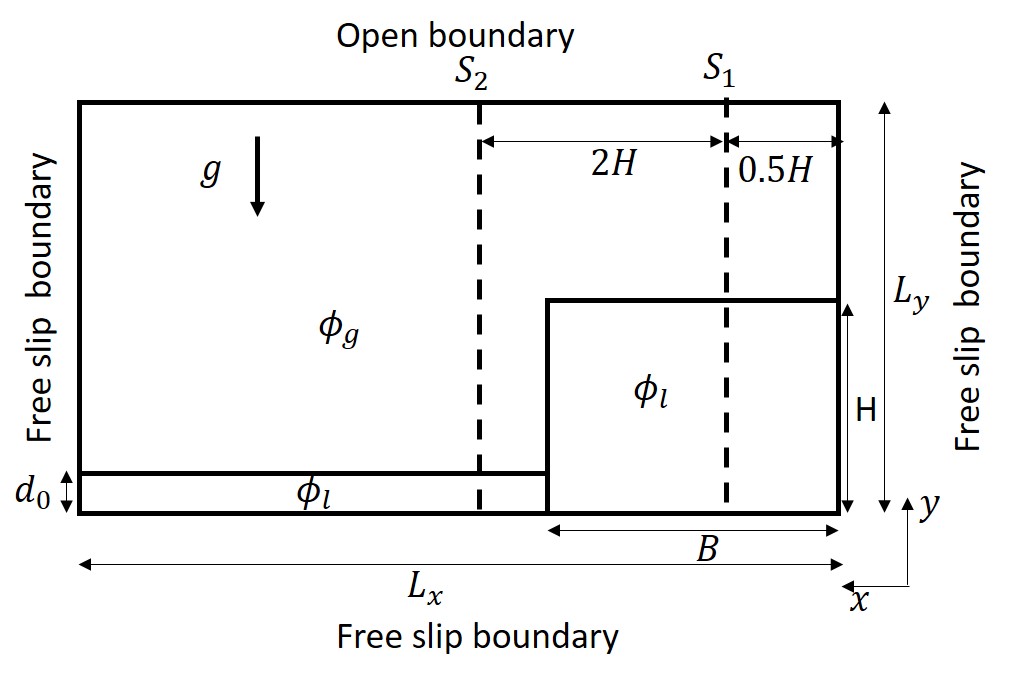}
\caption{Schematic of the initial setup for the dam break over a wet bed.}
\label{fig:dambreak_wetting_setup}
\end{figure}

Finally, we consider a dam break flow over a wet bed, which is recently investigated experimentally and numerically~\cite{garoosi2022experimental}. The sketch of the computational domain is shown in Fig.~\ref{fig:dambreak_wetting_setup}.
The length and height of the computational domain are  $L_x=1m$ and $L_y=0.3m$, respectively. The height and width of the water column are $H=0.2m$ and $B=0.25m$, respectively. The thickness and the removal time of the gate are neglected in simulations.
The boundary condition and the physical properties of the water and air are the same as above.
Based on Eq.(\ref{eq:dambreak_dimensionless}), the reference velocity, Reynolds number and Weber number are $U_{ref}=1.4$, $\text{Re}=2.789\times 10^5$ and $\text{We}=5.375\times 10^3$.
The order parameter is initialized by
\begin{equation}\label{eq}
\phi=\left\{
\begin{array}{ll}
\frac{\phi_h+\phi_l}{2}+\frac{\phi_h-\phi_l}{2}\tanh\frac{2(a-y)}{W},  & x\leq B-W, y\geq H-W  \\
\frac{\phi_h+\phi_l}{2}+\frac{\phi_h-\phi_l}{2}\tanh\frac{2(a-x)}{W}, & x\geq B-W, d_0\leq y\leq H-W \\
\frac{\phi_h+\phi_l}{2}+\frac{\phi_h-\phi_l}{2}\tanh\frac{2(W-\sqrt{(x-B+W)^2+(y-H+W)^2})}{W},&  x\geq B-W, y\geq H-W \\
\phi_h, & y\leq d_0 \\
\phi_h,& \text{otherwise}.
\end{array}\right.
\end{equation}

\begin{figure}[htp]
\centering
\subfigure[$t^*=0.0$]{\includegraphics[width=0.33\textwidth,trim=40 100 40 100,clip]{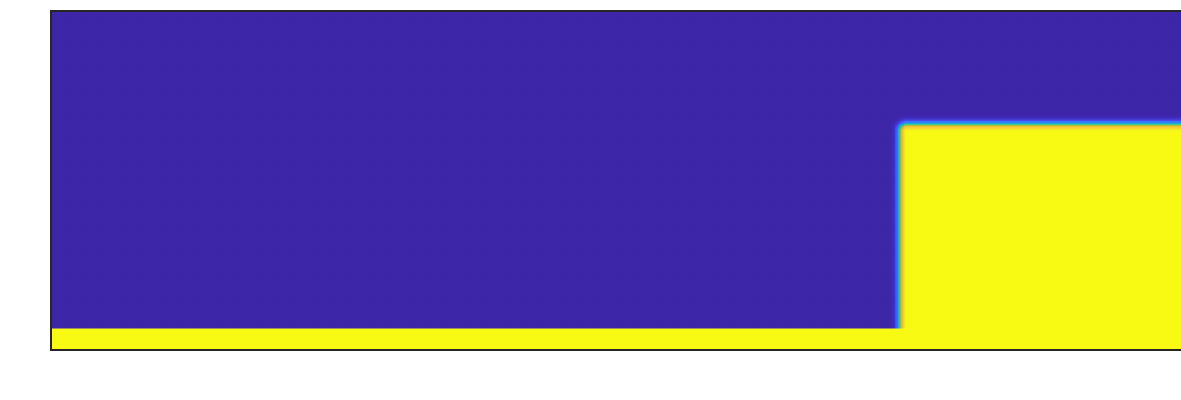}}~
\subfigure[$t^*=0.83$]{\includegraphics[width=0.33\textwidth,trim=40 100 40 100,clip]{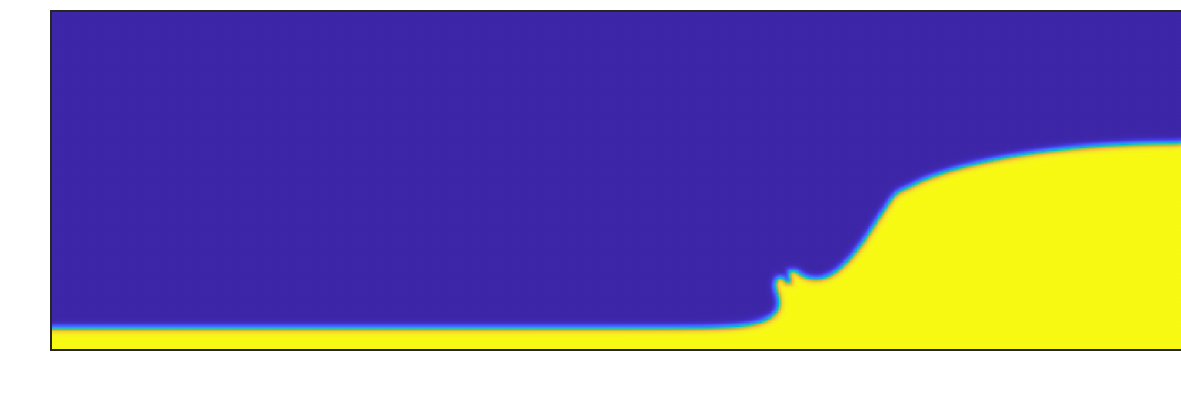}}~
\subfigure[$t^*=1.35$]{\includegraphics[width=0.33\textwidth,trim=40 100 40 100,clip]{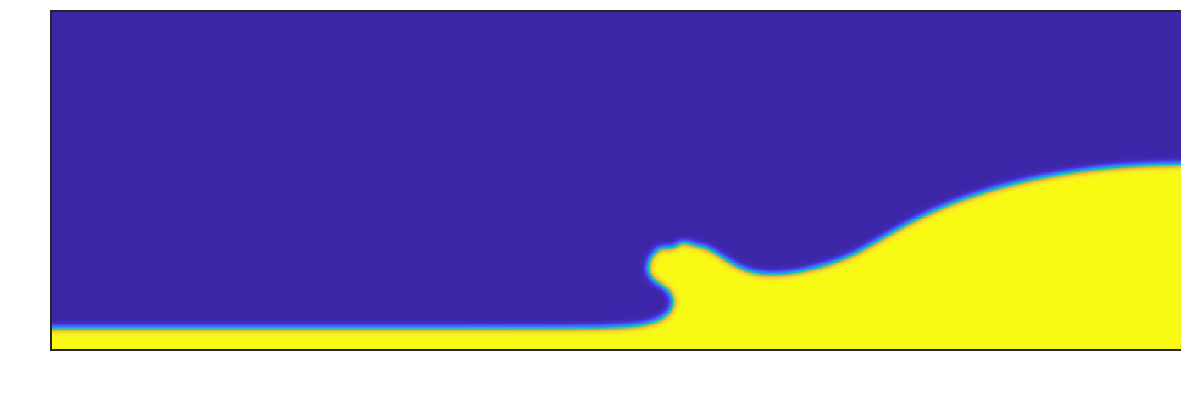}}\\
\subfigure[$t^*=1.765$]{\includegraphics[width=0.33\textwidth,trim=40 100 40 100,clip]{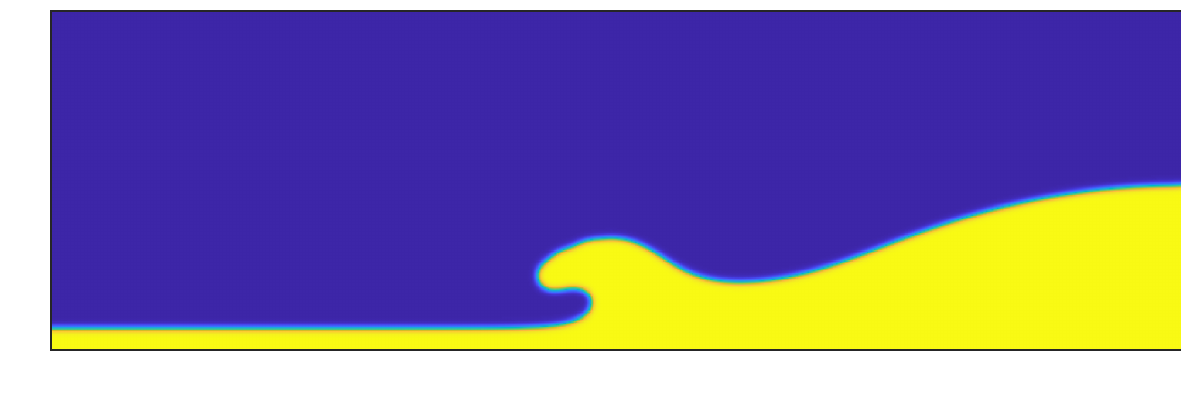}}~
\subfigure[$t^*=1.973$]{\includegraphics[width=0.33\textwidth,trim=40 100 40 100,clip]{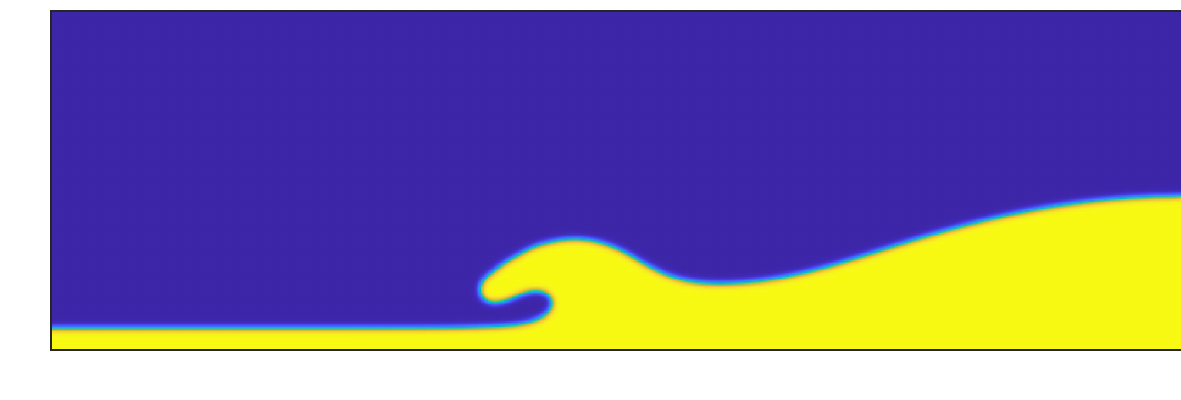}}~
\subfigure[$t^*=2.284$]{\includegraphics[width=0.33\textwidth,trim=40 100 40 100,clip]{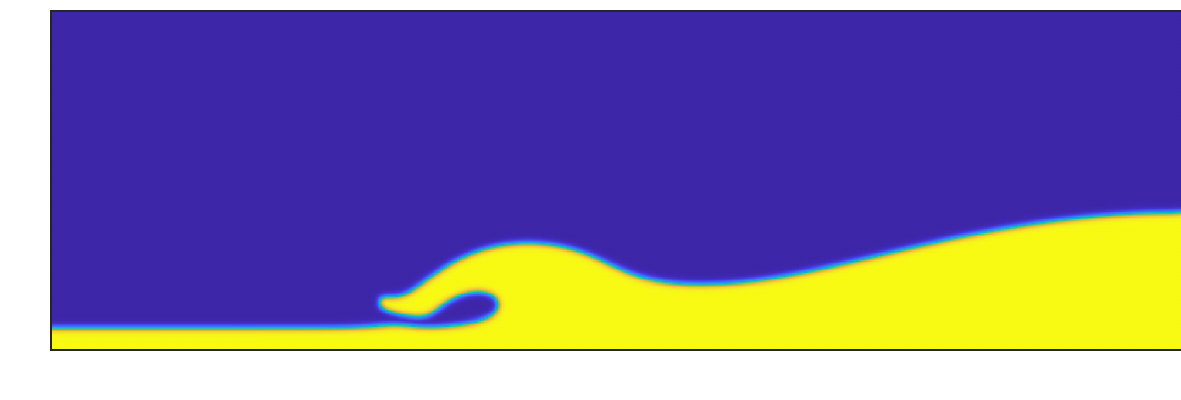}}\\
\subfigure[$t^*=2.596$]{\includegraphics[width=0.33\textwidth,trim=40 100 40 100,clip]{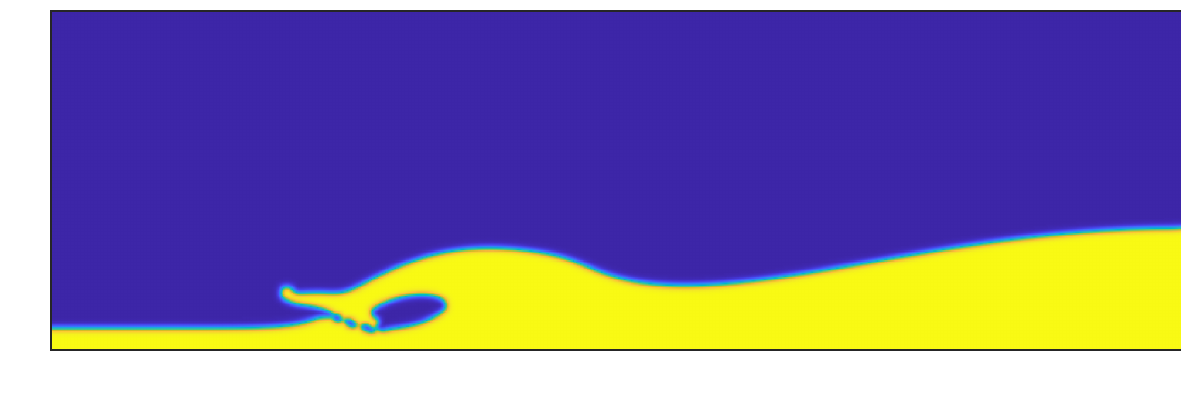}}~
\subfigure[$t^*=2.907$]{\includegraphics[width=0.33\textwidth,trim=40 100 40 100,clip]{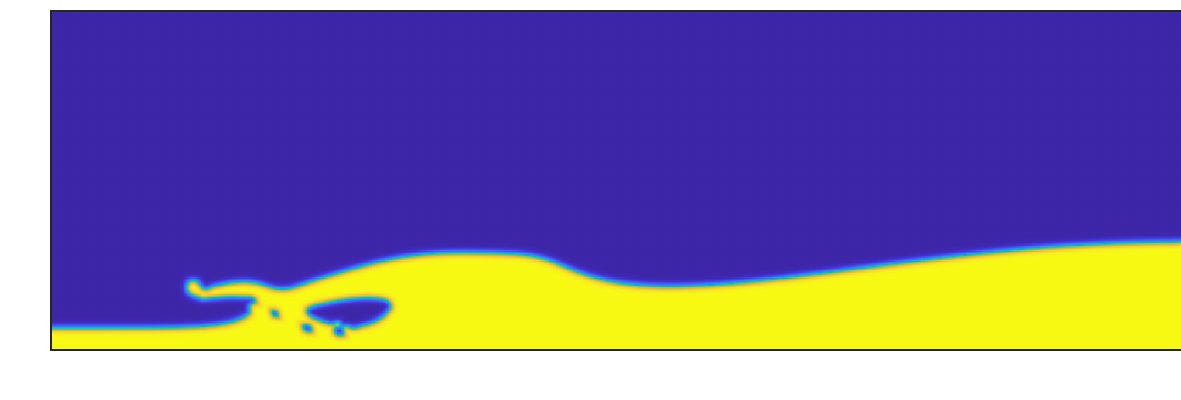}}~
\caption{Evolution of the instantaneous interface for dam break flow over a wet bed. }
\label{dambreak_wet}
\end{figure}

\begin{figure}
  \centering
\subfigure[]{\includegraphics[width=0.5\textwidth]{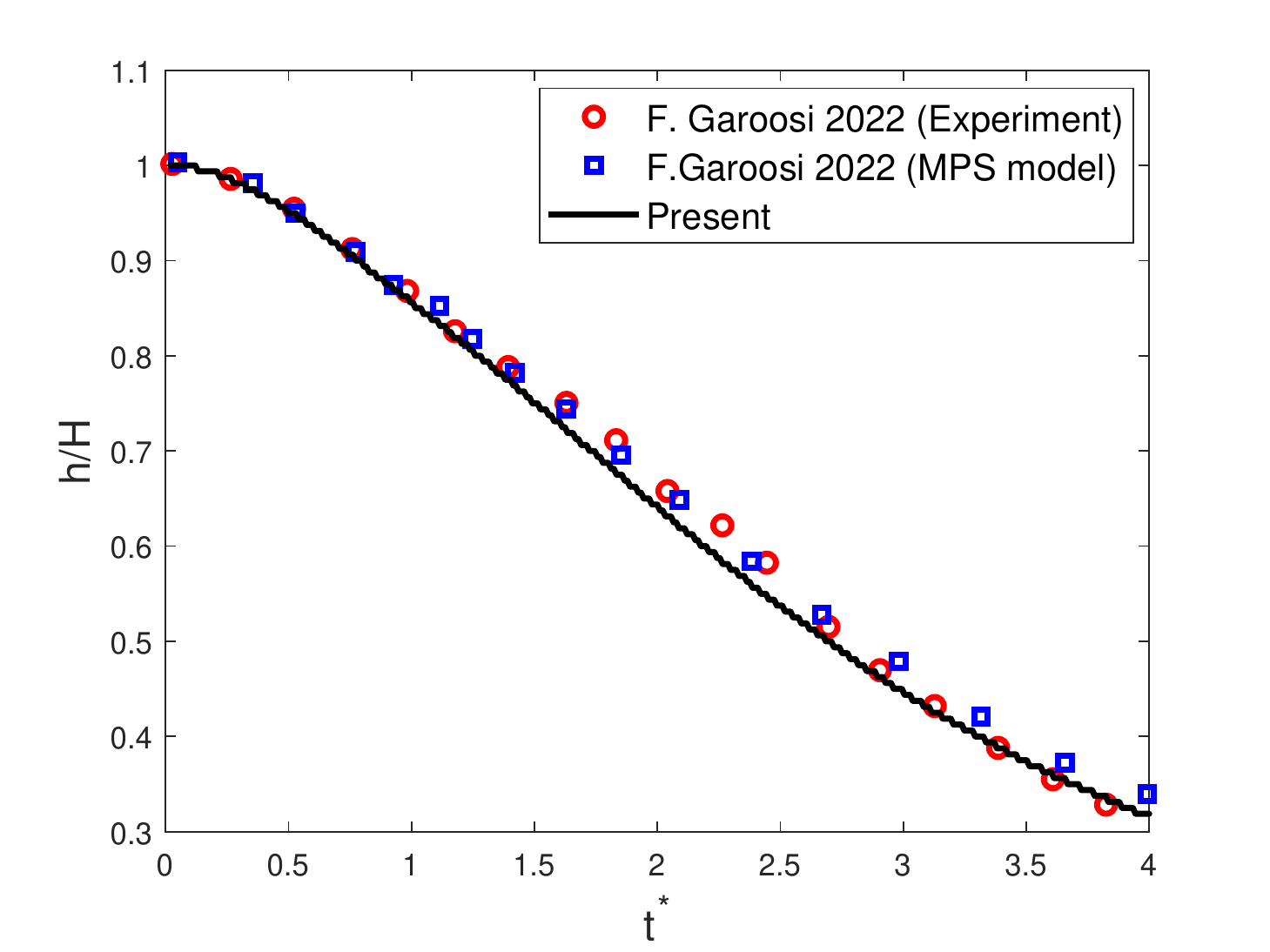}}~
\subfigure[]{\includegraphics[width=0.5\textwidth]{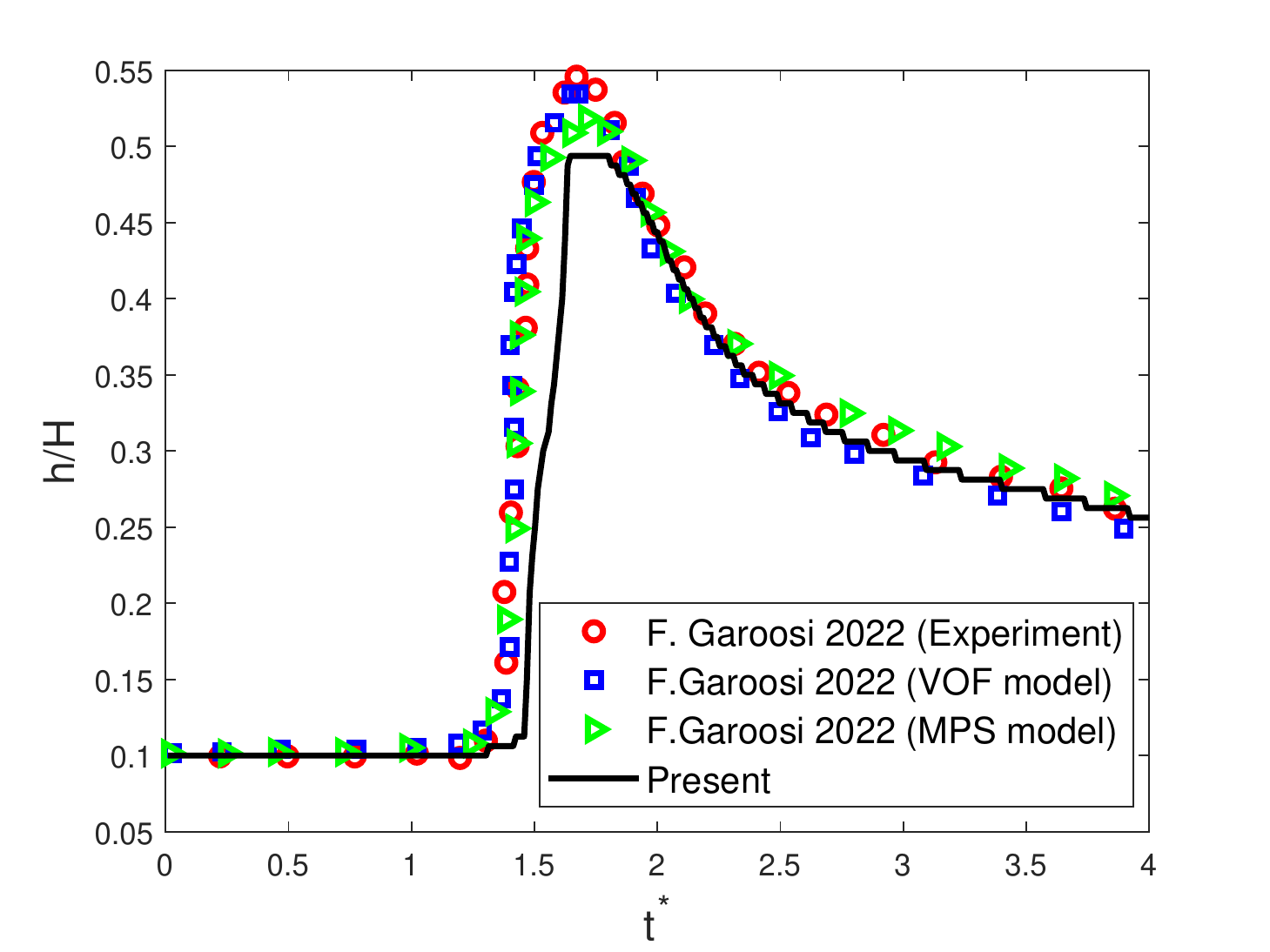}}
  \caption{The time variations of water level height at the locations (a) $S_1$ and (b) $S_2$ for dam break flow over a wet bed.}\label{wavelevel_wet}
\end{figure}

Fig.~\ref{dambreak_wet} shows the downstream wave front propagation along the wet bed during the dam break, including the results of the present study and the experimental data.
In the presence of a continuous fluid layer in the channel, the behavior of the flow becomes remarkably different.
Because the static layer at the bottom resists to a quick replacement,
a mushroom-like jet structure starts to emerge at the junction of the upstream and downstream
water depth. As the mushroom-like jet moves in the x-direction and  decreases in the y-direction as time advances,
lots of air are trapped and wave breaking occurs.
The shapes of the present study is highly similar to the experimental images in ~\cite{garoosi2022experimental}.
The time variations of water level height at the two locations $S_1$ and $S_2$ are shown in Fig.~\ref{wavelevel_wet}.
For quantitative comparison, the experimental results and the numerical results  from \cite{garoosi2022experimental} are added to the diagram.
For both locations, the height of the  water is slightly lower than the  experimental data. This differences may be attributed to the three dimensionality of the experiment.
In spite of this, the predicted water levels at both locations are in good agreement with the previous experimental and  numerical results.

\section{Summary}\label{sec:8}
In this paper  a central moment-based  discrete unified gas-kinetic scheme  has been developed for the accurate and robust phase-field modeling of incompressible two-phase flows with large density ratios and high Reynolds numbers.
In terms of the hydrodynamics equations, the velocity-based distribution function is employed to approximate the formulation of nonconserved momentum equation with surface tension force and an pressure evolution equation.
 The central moment collision operator is adopted  to improve the Galilean invariance, where the continuous form of the Maxwell-Boltzmann equilibrium distribution function is used to calculate the central moments for the hydrodynamic  fields. To further enhance the numerical stability, the Strange-splitting scheme in time is  employed so that the calculations of complex source terms at cell faces is unnecessary.
In terms of interface-capturing, two DUGKS methods with and without force term for the conservative Allen-Cahn equation are presented.

To validate the performance of the present model, the accuracy of both DUGKS methods for interface capturing is first validated through simulating the reversed single vortex. The results from both DUGKS for ACE are nearly identical. Due to computational cost, the DUGKS without force term for ACE is used in  simulations.
Then, the ability to model the surface tension effect of the present solver for the NS equations and the ACE  is demonstrated by two benchmarks problems, that is, static droplet and Oscillating drop. Finally, the present solver is applied  to simulate
 Rayleigh-Taylor instability, rising bubble, droplet splashing and dam break flow over dry and wet beds.
These problems cover a wide a range of density ratios, Reynolds numbers, Weber numbers and topological changes.
The results are  in excellent agreement with the previous computational and experimental results qualitatively and quantitatively.

For the present method, the derivatives of the density and pressure can be evaluated by some advanced techniques such as high-order filter, high-order compact difference or high-order total variation diminishing scheme, then the numerical stability of the present DUGKS can be further enhanced.
In particular, it is possible to eliminate spurious velocities by carefully handling discretization of both the pressure gradient and the surface tension force.
In addition, several artificial pressure  evolution equations modified with a convective term or a diffusive term  can be coupled into the present model to produce a more accurate pressure field~\cite{dupuy2020analysis}. The strategy adopted in the present DUGKS can also be employed to improve the phase-field-based two-phase LBE models. It is also noted that the central-moment-based DUGKS with nine discrete velocities in present work fails  to eliminate all the third order terms of velocity although improvement has been made. This means that the property of Galilean invariance is not completely guaranteed.
An available method to address this is to correct the raw moments of the source term based on the high-order error terms. In addition,  an extension of the present method to three-dimensions with more discrete velocities for simulation of mutliphase flows may also address this deficiency. 
To sum up, the present DUGKS method could be a promise tool to expand applicability of multiphase flows in real-life condition.

\section*{Acknowledgement}
Chunhua Zhang would like to thank Linlin Fei for helpful discussions. This work was supported by the National Numerical Wind Tunnel program, the National Natural Science Foundation of China (Grant No.51836003,11972142, 51806142, 91852205, 91741101, 11972142 and 11961131006), NSFC Basic Science Center Program (Award number 11988102),  Guangdong Provincial Key Laboratory of Turbulence Research and Applications (2019B21203001), Guangdong-Hong Kong-Macao Joint Laboratory for Data-Driven Fluid Mechanics and Engineering Applications (2020B1212030001), and Shenzhen Science and Technology Program (Grant No. KQTD20180411143441009). Computing resources are provided by the Center for Computational Science and Engineering of Southern
University of Science and Technology.

\appendix
\section{ Derivation of  the Naver-stokes equation from the central moment DUGKS}\label{sec:App_A}
Multiplying Eq.~(\ref{eq:NS_kinetic}) two sides with $\bm M$, Eq.~(\ref{eq:NS_kinetic}) can be written in terms of raw moments,
\begin{equation}\label{ap:a:moment}
\partial_t \bm m_f+\widetilde{\bm \xi} \cdot\nabla \bm m_f=-\bm N^{-1}\bm \Lambda^f \bm N(\bm m_f-\bm m_f^{eq})+\bm S_f,
\end{equation}
where $\bm m_f=\bm M\cdot f_i$, $\bm S^f=\bm M\cdot F_i$, and $\widetilde{\bm \xi}=(\bm M \xi_x \bm M^{-1}, \bm M\xi_y \bm M^{-1})$. $\widetilde{\bm \xi}=(\widetilde{\bm \xi}_x,\widetilde{\bm \xi}_y)$ is given by
\begin{equation}\label{eq:apa:mcx}
\widetilde{\bm \xi}_x=
\left[
\begin{matrix}
0& 1& 0&   0&   0& 0& 0& 0& 0  \\
0& 0& 0& 1/2& 1/2& 0& 0& 0& 0  \\
0& 0& 0&   0&   0& 1& 0& 0& 0  \\
0& 1& 0&   0&   0& 0& 0& 1& 0  \\
0& 1& 0&   0&   0& 0& 0&-1& 0  \\
0& 0& 0&   0&   0& 0& 1& 0& 0  \\
0& 0& 0&   0&   0& 1& 0& 0& 0  \\
0& 0& 0&   0&   0& 0& 0& 0& 1  \\
0& 0& 0&   0&   0& 0& 0& 1& 0  \\
\end{matrix}
\right]
\end{equation}
\begin{equation}\label{eq:apa:mcy}
\widetilde{\bm \xi}_y=
\left[
\begin{matrix}
0& 0& 1&   0&   0& 0& 0& 0& 0  \\
0& 0& 0&   0&   0& 1& 0& 0& 0  \\
0& 0& 0& 1/2&-1/2& 0& 0& 0& 0  \\
0& 0& 1&   0&   0& 0& 1& 0& 0  \\
0& 0&-1&   0&   0& 0& 1& 0& 0  \\
0& 0& 0&   0&   0& 0& 0& 1& 0  \\
0& 0& 0&   0&   0& 0& 0& 0& 1  \\
0& 0& 0&   0&   0& 1& 0& 0& 0  \\
0& 0& 0&   0&   0& 0& 1& 0& 0  \\
\end{matrix}
\right]
\end{equation}
Since $\bm N^{-1}\bm \Lambda \bm N$ is invertible, Eq.(~\ref{ap:a:moment}) can be rewritten as
\begin{equation}\label{ap:evolution_mf}
\bm m_f=\bm m_f^{eq}-\bm N^{-1}\bm{\Lambda}_f\bm N(\partial_t \bm m_f+ \widetilde{\bm \xi} \cdot \nabla \bm m_f-\bm S_f),
\end{equation}
So far, we haven't made any approximations for Eq.(\ref{ap:evolution_mf}).
It is noted that each element in $\bm{\Lambda}_f$ is   the order of the Knudsen number ($\text{Kn}$) that is defined as the ratio of mean free path of gas molecules to the characteristic flow length.
In the continuum flow regime where the classical NS equations are valid, we have
\begin{equation}\label{ap:0-order}
\bm m_f=\bm m_f^{eq}+O(\lambda_{f,max}),
\end{equation}
where $\lambda_{f,max}$ is the maximum value in $\bm{\Lambda}_f$.
Inserting Eq.(\ref{ap:0-order}) into (\ref{ap:evolution_mf}) gives
\begin{equation}\label{ap:first-order}
\bm m_f=\bm m_f^{eq}-(\bm N^{-1}\bm \Lambda_f^{-1}\bm N)(\partial_t \bm m^{eq}+\widetilde{\bm \xi}\cdot \nabla\bm  m^{eq}-\bm S_f)+O(\lambda_{f,max}^2).
\end{equation}
This can be used to obtain the Euler equations. Inserting Eq.(\ref{ap:first-order}) into (\ref{ap:evolution_mf}) gives
\begin{equation}\label{eq:apa:second}
\begin{aligned}
\bm m_f=&\bm m_f^{eq}-(\bm N^{-1}\bm \Lambda^{-1}\bm N)  \partial_t \left(\bm m_f^{eq}-(\bm N^{-1}\bm \Lambda^{-1}\bm N)(\partial_t \bm m_f^{eq}+\widetilde{\bm \xi}\cdot\nabla \bm m_f^{eq}-\bm S_f)\right) \\
&-(\bm N^{-1}\bm \Lambda^{-1}\bm N) \left[ \widetilde{\bm \xi}\cdot \nabla\left(
\bm m_f^{eq}-(\bm N^{-1}\bm \Lambda^{-1}\bm N)(\partial_t \bm m_f^{eq}+\widetilde{\bm \xi}\cdot \nabla \bm m_f^{eq}-\bm S_f)+\bm S_f  \right) \right] +O(\lambda_{f,max}^3).
\end{aligned}
\end{equation}
This is sufficient to derive the  Navier-Stokes equations. If the term $\bm m_f$ in the right hand of Eq.(\ref{ap:evolution_mf}) is further replaced by Eq.~\ref{eq:apa:second}, the resulting equation is able to recover governing equations beyond the hydrodynamic regime.
For convenience, Eq.~(\ref{eq:apa:second}) is reformulated as
\begin{equation}\label{eq:apa:second2}
\begin{aligned}
\partial_t \bm m_f^{eq}+\widetilde{\bm \xi}\cdot\nabla \bm m_f^{eq}=&-(\bm N^{-1}\bm{\Lambda}^{-1}\bm N)(\bm m_f-\bm m_f^{eq}) \\
&+\widetilde{\bm \xi}\cdot \nabla ( \bm N^{-1}\bm \Lambda^{-1}\bm N)(\partial_t \bm m_f^{eq}+\widetilde{\bm \xi}\cdot \nabla \bm m_f^{eq}
-\bm S_f)+\bm S_f+O( \lambda_{f,max}^2),
\end{aligned}
\end{equation}
where $\partial_t(\bm N^{-1}\bm {\Lambda}^{-1}\bm N)(\partial_t \bm m_f^{eq}+\widetilde{\bm \xi}\cdot \nabla \bm m_f^{eq}-\bm S_f)$ is ignored.
Substituting Eqs(\ref{eq:rawmoment_NS}),(\ref{eq:apa:mcx}) and (\ref{eq:apa:mcy}) into  Eq.(\ref{ap:first-order}) and letting $i=0,1,2$ lead to
\begin{equation}\label{eq:ap:enler}
\begin{aligned}
\partial_t p+\nabla\cdot (\rho_0 \bm u) &=(\rho_0-c_0 \rho)\nabla\cdot \bm u + O(\lambda_{f,max}),\\
\partial_t (\rho_0 u_x)+ \partial_x (\rho_0 u_x^2) + \partial_y (\rho_0u_x u_y) &=F_x+O(\lambda_{f,max}), \\
\partial_t (\rho_0 u_y)+ \partial_x (\rho_0 u_x u_y) +\partial_y (\rho_0 u_y^2) &=F_y+O(\lambda_{f,max}) ,\\
\end{aligned}
\end{equation}
Substituting Eqs.(\ref{eq:rawmoment_NS}),(\ref{eq:apa:mcx}) and (\ref{eq:apa:mcy}) into  Eq.(\ref{eq:apa:second2}) and letting $i=0,1,2$ lead to
\begin{equation}\label{eq:apa:mass}
\partial_t p +c_0 \rho \nabla\cdot \bm u=0+O(\lambda_{f,max}^2)
\end{equation}
\begin{equation}\label{eq:moment_NS_x}
\begin{aligned}
\partial_t (\rho_0u_x) +\partial_x (\rho_0 u_x^2)+\partial_y (\rho_0 u_x u_y)=&
F_{x}+
\partial_x\left(\frac{\lambda_{f,4} c_f^2}{3} \left(\partial_x u_x-\partial_y u_y \right) +\frac{2\lambda_{f,3}c_f^2}{3}\nabla\cdot \bm u \right)\\
+&
\partial_y\left(\frac{c_f^2\lambda_{f,4}}{3 }\left( \partial_x u_y +\partial_y u_x \right) \right)
+\text{TE}_1+ O(\lambda_{f,max}^2),
\end{aligned}
\end{equation}
\begin{equation}\label{eq:moment_NS_y}
\begin{aligned}
\partial_t (\rho_0u_y) +\partial_x (\rho_0u_x u_y) +\partial_y  (\rho_0u_y^2) =&
F_{y}+\partial_x\left(\frac{c_f^2\lambda_{f,3}}{3}\rho_0 \left(\partial_x u_y +\partial_y u_x \right)  \right)\\
+&
\partial_y\left(\frac{c_f^2\lambda_{f,4}}{3}\rho_0\left( \partial_y u_y -\partial_x u_x \right)
+\frac{2c_f^2\lambda_{f,3}}{3}\rho_0 \nabla\cdot \bm u \right)+\text{TE}_2+O(\lambda_{f,max}^2),
\end{aligned}
\end{equation}
where the formulations in Eq.(\ref{eq:ap:enler}) have been used.
$\text{TE}_1$ and $\text{TE}_2$ are error terms, defined as
\begin{equation}\label{eq}
\begin{aligned}
\text{TE}_1=&-\partial_y\left(u_x u_y\nabla\cdot (\rho_0\bm u) \right) \\
&+\partial_x\left[
-\lambda_{f,3}\left(
(2u_x^2+\frac{u_y^2}{2})\partial_x(\rho_0 u_x)
+(2u_y^2+\frac{u_x^2}{2})\partial_y(\rho_0 u_y)\right)\right.\\
&+\left.\lambda_{f,4}\left(
(\frac{1}{2}u_y^2-2u_x^2)\partial_x(\rho_0 u_x)
+(2u_y^2-\frac{u_x^2}{2})\partial_y(\rho_0 u_y)
\right)
\right]
\end{aligned}
\end{equation}
\begin{equation}\label{eq}
\begin{aligned}
\text{TE}_2=&-\partial_x \lambda_{f,4} u_x u_y\nabla\cdot (\rho_0\bm u) \\
&+\partial_y\left[
\lambda_{f,4}
\left(
(2u_x^2-\frac{u_y^2}{2})\partial_x(\rho_0 u_x)+
(\frac{u_x^2}{2}-2u_y^2)\partial_y(\rho_0 u_y)
\right)\right.   \\
&-\left.\lambda_{f,3}
\left(
(2u_x^2+\frac{u_y^2}{2})\partial_x (\rho_0 u_x)
+(\frac{u_x^2}{2}+2u_y^2)\partial_y(\rho_0 u_y)
\right)
 \right]
\end{aligned}
\end{equation}
It can be found that both $\text{TE}_1$ and $\text{TE}_2$ are of the order $O(\lambda_{g,max}\bm u^3)$ or $O(\text{Kn} \text{Ma}^3)$, where $\text{Ma}$ represents the Mach number. For multiphase flow with low Mach numbers, these terms can be neglected.
To recover the correct momentum equations, the kinematic and bulk viscosities should be defined as
\begin{equation}\label{eq}
\nu=\frac{c_f^2}{3}\lambda_{f,4}=\frac{c_f^2}{3}\lambda_{f,5}, \quad \eta=\frac{2c_f^2}{3}\lambda_{f,3},
\end{equation}

\section{ Derivation of   the Allen-Cahn equation from the central moment DUGKS}\label{sec:App_B}
The derivation of the conservative Allen-Cahn equation from the DUGKS is similar to the previous process of the derivation of NSE.
From Eq.(\ref{eq:apa:second}), we can obtain the equations of the moment of distribution function $g$ by replacing $\bm m_f$  with $\bm m_g$ and $\bm S_f$ with $\bm S_g$,
\begin{equation}\label{eq:apb:gsecond2}
\begin{aligned}
\partial_t \bm m_g^{eq}+\widetilde{\bm \xi}\cdot \nabla \bm m_g^{eq}=& -(\bm N^{-1}\bm \Lambda_g^{-1}\bm N)(\bm m_g-\bm m_g^{eq})
+\nabla \cdot ( \bm N^{-1}\bm \Lambda_g^{-1}\bm N)(\partial_t \bm m_g^{eq}+\widetilde{\bm \xi}\cdot\nabla \bm m_g^{eq} -\bm S_g)\\
&+\bm S_g+O( \bm \lambda_{g,max}^2),
\end{aligned}
\end{equation}
where $\bm m_g=\bm M g_i$ and $\bm S_g=\bm MF^{g}$.
Inserting Eqs.~(\ref{eq:rawmoments_continuum_AC}), ~(\ref{eq:apa:mcx}) and ~(\ref{eq:apa:mcy}) into Eq.~(\ref{eq:apb:gsecond2}), then letting $i=0$, we have
\begin{equation}\label{eq:ap:AC}
\begin{aligned}
\partial_t\phi +\partial_x(\phi u_x)+\partial_y(\phi u_y)
 &= \partial_x\left(
\underline{ \left(\lambda_{g,0}-\lambda_{g,1} \right)u_x\left( \partial_t\phi +\partial_x(\phi u_x)+\partial_y(\phi u_y)\right)}+
c_{s,g}^2\lambda_{g,1}(\partial_x \phi -n_x\Theta)
\right)
\\
& +
\partial_y\left(
\underline{\left( \lambda_{g,0}-\lambda_{g,1} \right) u_y\left( \partial_t\phi +\partial_x(\phi u_x)+\partial_y(\phi u_y)\right)}+c_{s,g}^2\lambda_{g,2}(\partial_y \phi -n_y\Theta)
\right)
\end{aligned}
\end{equation}
The underlined terms are of the order $O(\lambda_{g,max}^2)$ and can be neglected. Then Eq.(\ref{eq:ap:AC}) can be written in vector form
\begin{equation}\label{eq}
\partial_t \phi+\nabla\cdot (\phi\bm u)=\nabla\cdot(M(\nabla\phi-\Theta\bm n)),
\end{equation}
where the mobility $M$ is defined as
\begin{equation}\label{eq:apb-mobility}
M=c_{s,g}^2\lambda_{g,1}=c_{s,g}^2\lambda_{g,2}.
\end{equation}

For the DUGKS without force term for ACE, Eq.(\ref{eq:apb:gsecond2}) is reduced to
\begin{equation}\label{apb:gwithoutforce}
\begin{aligned}
\partial_t \bm m_g^{eq}+\widetilde{\bm \xi}\cdot \nabla \bm m_g^{eq}& =-(\bm N^{-1}\bm \Lambda_g^{-1}\bm N)(\bm m_g-\bm m_g^{eq}) \\
&+\nabla \cdot ( \bm N^{-1}\bm \Lambda_g^{-1}\bm N)\left(\partial_t \bm m_g^{eq}+\widetilde{\bm \xi}\cdot\nabla \bm m_g^{eq}\right) +O( \lambda_{g,max}^2),
\end{aligned}
\end{equation}
Inserting Eqs.~(\ref{eq:rawmoments_continuum_AC}), ~(\ref{eq:apa:mcx}) and ~(\ref{eq:apa:mcy}) into Eq.~(\ref{eq:apb:gsecond2}), and letting $i=0$ yield
\begin{equation}\label{eq:ap:dugkswithoutforce}
\begin{aligned}
\partial_t\phi +\partial_x(\phi u_x+c_{s,g}^2\Theta n_x)+\partial_y(\phi u_y+c_{s,g}^2\Theta n_y)
 = \partial_x\left( c_{s,g}^2\lambda_{g,1}\partial_x \phi \right)   +\partial_y\left(c_{s,g}^2\lambda_{g,2}\partial_y \phi\right)
+O( \lambda_{g,max}^2),
\end{aligned}
\end{equation}
Similarly, Eq.(\ref{eq:ap:dugkswithoutforce}) can be reexpressed as follows
\begin{equation}\label{eq}
\partial_t \phi+\nabla\cdot (\phi\bm u)=\nabla\cdot M(\nabla\phi-\Theta\bm n),
\end{equation}
where the mobility is the same to Eq.(\ref{eq:apb-mobility}).
\bibliographystyle{unsrt}
\bibliography{mybibfile}

\end{document}